%% file: SandersonHelmi2012rev.tex
\documentclass[useAMS,usenatbib,a4]{mn2e}

\pdfpagewidth=\paperwidth 
\pdfpageheight=\paperheight

\newcommand{\sub}[2]{\ensuremath{#1_{\mathrm{#2}}}}
\newcommand{\super}[2]{\ensuremath{#1^{\mathrm{#2}}}}

\newcommand{\unit}[2]{\ensuremath{\textrm{#1}^{#2}}}
\newcommand{\half}{\ensuremath{\frac{1}{2}}}

\newcommand{\zmin}{\ensuremath{z_{\mathrm{min}}}}
\newcommand{\zmax}{\ensuremath{z_{\mathrm{max}}}}

\newcommand{\cosmax}{\ensuremath{\left(\cos \vartheta\right)_{\mathrm{max}}}}
\newcommand{\sinmax}{\ensuremath{\left(\sin \vartheta\right)_{\mathrm{max}}}}
\newcommand{\cosmin}{\ensuremath{\left(\cos \vartheta\right)_{\mathrm{min}}}}
\newcommand{\sinmin}{\ensuremath{\left(\sin \vartheta\right)_{\mathrm{min}}}}

\newcommand{\cosdiffmax}{\ensuremath{\left[\cos(\vartheta-\phi_s) \right]_{\mathrm{max}}}}
\newcommand{\sindiffmax}{\ensuremath{\left[\sin(\vartheta-\phi_s) \right]_{\mathrm{max}}}}
\newcommand{\cosdiffmin}{\ensuremath{\left[\cos(\vartheta-\phi_s) \right]_{\mathrm{min}}}}
\newcommand{\sindiffmin}{\ensuremath{\left[\sin(\vartheta-\phi_s) \right]_{\mathrm{min}}}}

\newcommand{\hmumin}{\ensuremath{(h-u)_{\mathrm{min}}}}
\newcommand{\hmumax}{\ensuremath{(h-u)_{\mathrm{max}}}}

\newcommand{\rmax}{\ensuremath{r_{\mathrm{max}}}}


\usepackage{graphicx} 
\usepackage{epstopdf} 
\usepackage[intlimits]{amsmath}
\usepackage{amssymb}
\usepackage{multirow}

\begin{document}

\title{An analytical phase-space model for tidal caustics}
\author[R. E. Sanderson, A. Helmi]{Robyn E. Sanderson, Amina Helmi${}^{1}$\\
${}^{1}$Kapteyn Astronomical Institute, P.O. Box 800, 9700 AV Groningen, The Netherlands}

\maketitle

\begin{abstract}
The class of tidal features around galaxies known as ``shells'' or ``umbrellas" comprises debris that has arisen from high-mass-ratio mergers with low impact parameter; the nearly radial orbits of the debris give rise to a unique morphology, a universal density profile, and a tight correlation between positions and velocities of the material. As such they are accessible to analytical treatment, and can provide a relatively clean system for probing the gravitational potential of the host galaxy. In this work we present a simple analytical model that describes the density profile, phase-space distribution, and geometry of a shell, and whose parameters are directly related to physical characteristics of the interacting galaxies. The model makes three assumptions: that their orbit is radial, that the potential of the host is spherical at the shell radii, and that the satellite galaxy had a Maxwellian velocity distribution. We quantify the error introduced by the first two assumptions and show that selecting shells by their appearance on the sky is a sufficient basis to assume that these simplifications are valid. We further demonstrate that (1) given only an image of a shell, the radial gravitational force at the shell edge and the phase-space density of the satellite are jointly constrained, (2) that combining the image with measurements of either point line-of-sight velocities or integrated-light spectra will yield an independent estimate of the gravitational force at a shell, and (3) that an independent measurement of this force is obtained for each shell observed around a given galaxy, potentially enabling a determination of the galactic mass distribution. 
\end{abstract}


\section{Introduction}
Recently, large-scale sky surveys and deep follow-up images have been used to discover a wealth of tidal debris around our Galaxy and others nearby \citep[e.g.,][]{2001Natur.412...49I,2009Natur.461...66M,Trujillo2009,Martinez-Delgado2010,Radburn-Smith2011}.  This tidal debris comes in many shapes and sizes, from the huge tidal arms created by interacting, roughly equal-mass galaxies to the fainter features observed around nearby galaxies.  These smaller-scale features are thought to come from interactions of the large galaxy with much smaller satellites, which are called minor mergers.  Evidence that minor mergers occur in nature is an important link to our cosmological history, since cosmological simulations of dark matter indicate that about half the mass in the Milky Way's outer regions was accreted in this way \citep[for example,][]{2010arXiv1010.2491M,Wang2011}.  Minor mergers are also useful as a way to constrain the shape and mass of the large galaxy, since tidally stripped material from the smaller satellite galaxy behaves as test particles in the relatively undisturbed potential of the larger host galaxy \citep{2001Natur.412...49I,2004ApJ...610L..97H,2005ApJ...619..800J,Eyre2009,2010ApJ...714..229L}. The remnants of these mergers give us a way to measure the characteristics of the dark components of galaxies, which are predicted with great accuracy by cosmological models and simulations.

Some minor mergers create patterns of tidal debris that look like shells or umbrellas.  The first such debris was identified around elliptical galaxies by \citet{malin:1983aa}; these galaxies were called ``shell galaxies" because of these distinctive features.  \citet{quinn:1986aa,hernquist:1988aa,hernquist:1989aa} showed that the shells were probably created by a minor merger on a nearly radial orbit.  This explains the alternate spacing of the shells on either side of the host galaxy since they are formed as material from the satellite piles up approximately at turning points: the satellite initially had a distribution of energies that is reflected in the different radii of the shells. More recently, similar features have been discovered around nearby disk galaxies  \citep{2001Natur.412...49I,2009Natur.461...66M,Martinez-Delgado2010}.  The vast improvements in imaging since shell galaxies were first identified, and the relative proximity of these objects, have revealed more of their structure than had previously been observed.  In some cases \citep{Fardal2012,Romanowsky2012}, the objects are even close or bright enough that velocity information could be obtained.  

Shells from nearly radial mergers are particularly special because there is a direct correlation between the kinematic properties of debris and its location relative to the host galaxy.  Thanks to the near-symmetry of the encounter, the system can be considered in a two-dimensional projection $(r, v_r)$ of the full six-dimensional phase space without much loss of information.  In this two-dimensional projection, the initially cold satellite material, once unbound, forms a thin stream that winds through phase space, so that for any spatial location $r$ there are a small, finite number of streams with different characteristic $v_r$.  \citet{Merrifield1998} pointed out that this correlation could be used to estimate the mass of the galaxy hosting a shell, if the line-of-sight velocity of the material could be measured at different points in the shell.

The dynamics governing the formation and shape of this stream are closely related to earlier work on spherically symmetric secondary infall of matter accreting onto dark matter halos.  As shown in \citet{fillmore:1984aa} and \citet{bertschinger:1985aa} radial accretion of gravitating, cold, collisionless matter forms a series of infinite-density peaks at successive radii, known as caustics.  \citet{2006MNRAS.366.1217M} further showed that for warm matter with a finite velocity dispersion, the peaks take on a finite width and height, but although they are no longer caustics in the mathematically rigorous sense they retain many of the same properties.   This work, motivated by cosmological simulations, considered the radial collapse of spherically distributed matter, but \cite{1999MNRAS.307..495H} showed that caustics are also produced when a small, initially self-bound satellite falls  into the (assumedly) static potential of a larger host galaxy. In fact, caustics are a universal product of dynamical systems with turning points, regardless of the type of symmetry \citep{tremaine:1999aa,2001PhRvD..64f3515H}. This work has been recently confirmed with state-of-the-art computer simulations using realistic galactic potentials by \cite{vogelsberger:2007aa}, and systems resembling shell galaxies are produced in cosmological simulations with semi-analytic stellar components \citep{Cooper2011}.  \citet{2010ApJ...725.1652S} demonstrated the connection between the shape of the density profile of the caustic and the physical characteristics of the interacting galaxies. For these reasons, in this work we will refer to shells as ``tidal caustics", a term which emphasizes their high degree of symmetry and the correlation of positions and velocities.

\citet{EbrovaIvana2012} recently discussed the possibility of using integrated-light spectra to measure the gravitational potential in shell galaxies by obtaining the line-of-sight velocity profiles of the shell debris. Their results are similar to some of those given in this work, with a few important differences. First, they consider only spherical shells from satellites on perfectly radial orbits, not the effect of relaxing these assumptions to include angular momentum, potential flattening, or projection effects. In this work we explore how all three of these things affect the line-of sight velocity distribution. Furthermore, the line-of-sight velocity profiles derived by \citeauthor{EbrovaIvana2012} do not account for the nonzero thickness of the shell. This causes the peaks of the profile in their model to be thinner than and slightly offset from the peaks in their simulated shells; we show in this work that including the shell thickness solves both problems.  Finally, they do not discuss how the surface-brightness profile of the shell is related to the kinematic profile, whereas we show that the two can be used together to simplify the fitting process by constraining the shell's geometry independently prior to modelling its kinematics.

In this paper we present a simple analytical model for the density and phase space distributions of tidal caustics (Sections \ref{subsec:udp} and \ref{subsec:PDF}) and show how its parameters are related to the radial component of the gravitational force $g_s$ exerted by the host galaxy at the radius of a tidal caustic. We project the model phase-space distribution into the space of observables to calculate the surface-brightness distribution (Section \ref{sec:images}) and the line-of-sight velocity distribution for point sources in the shell (Section \ref{sec:vzmax}). In Section \ref{sec:sims} we use simulations to test various aspects of the model: the assumptions of radial orbits and spherical symmetry (Section \ref{subsec:assumptionTests}) and its ability to reproduce the phase-space distribution (Section \ref{subsec:phaseSpaceComparison}), surface brightness (Section \ref{subsec:imageComparison}) and line-of-sight velocity distribution (Section \ref{subsec:DLOScomparison}). In Section \ref{sec:spectra} we discuss an additional application: the calculation of the velocity profile for integrated-light spectra, which we show is also sensitive to $g_s$. In Section \ref{sec:conclusion} we summarize.

\section{The model}
\label{sec:model}
\subsection{Density profile}
\label{subsec:udp}

\begin{figure*}
\begin{center}
\begin{tabular}{cc}
\includegraphics[width=0.4\textwidth]{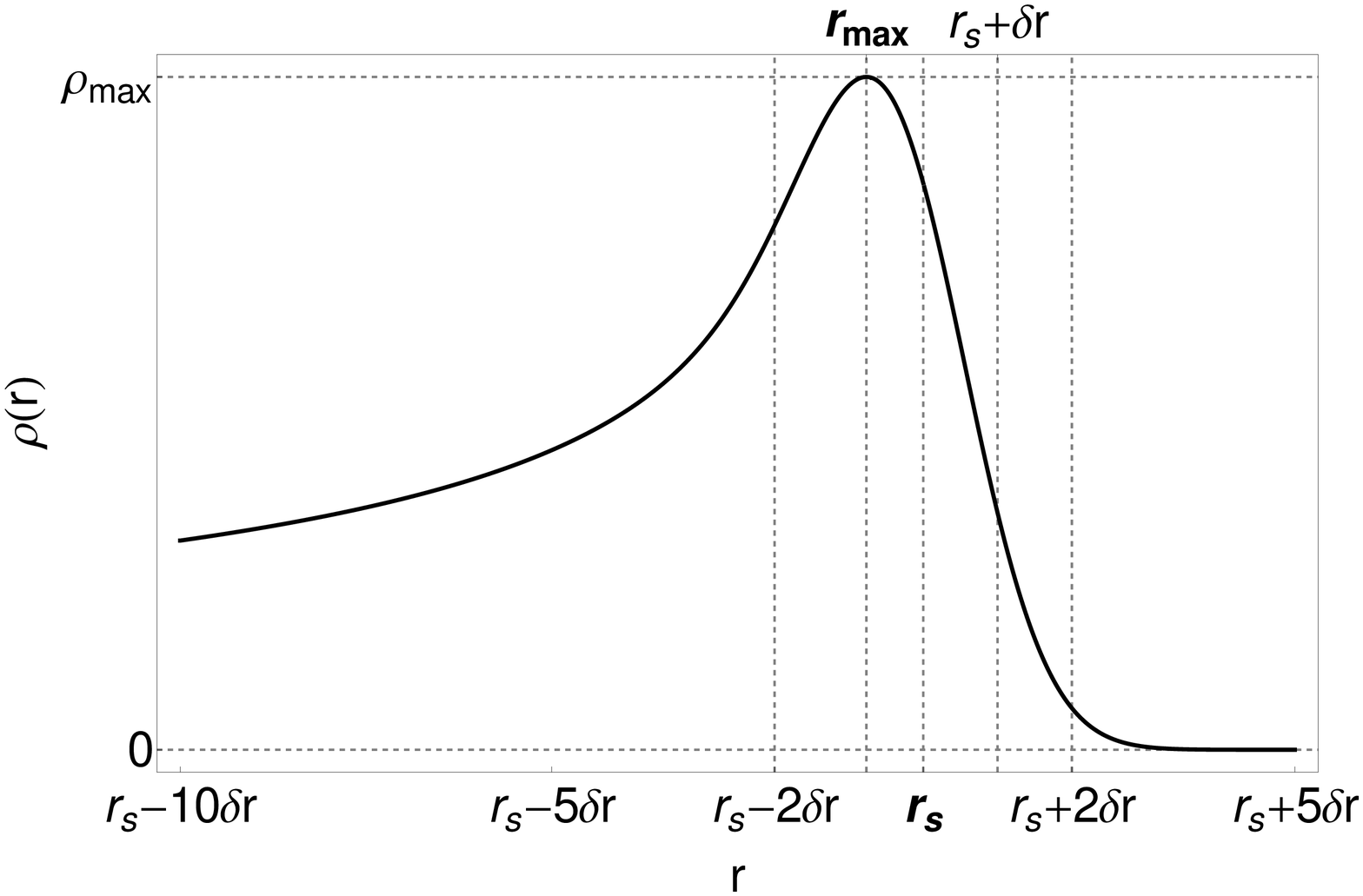} &\includegraphics[width=0.4\textwidth]{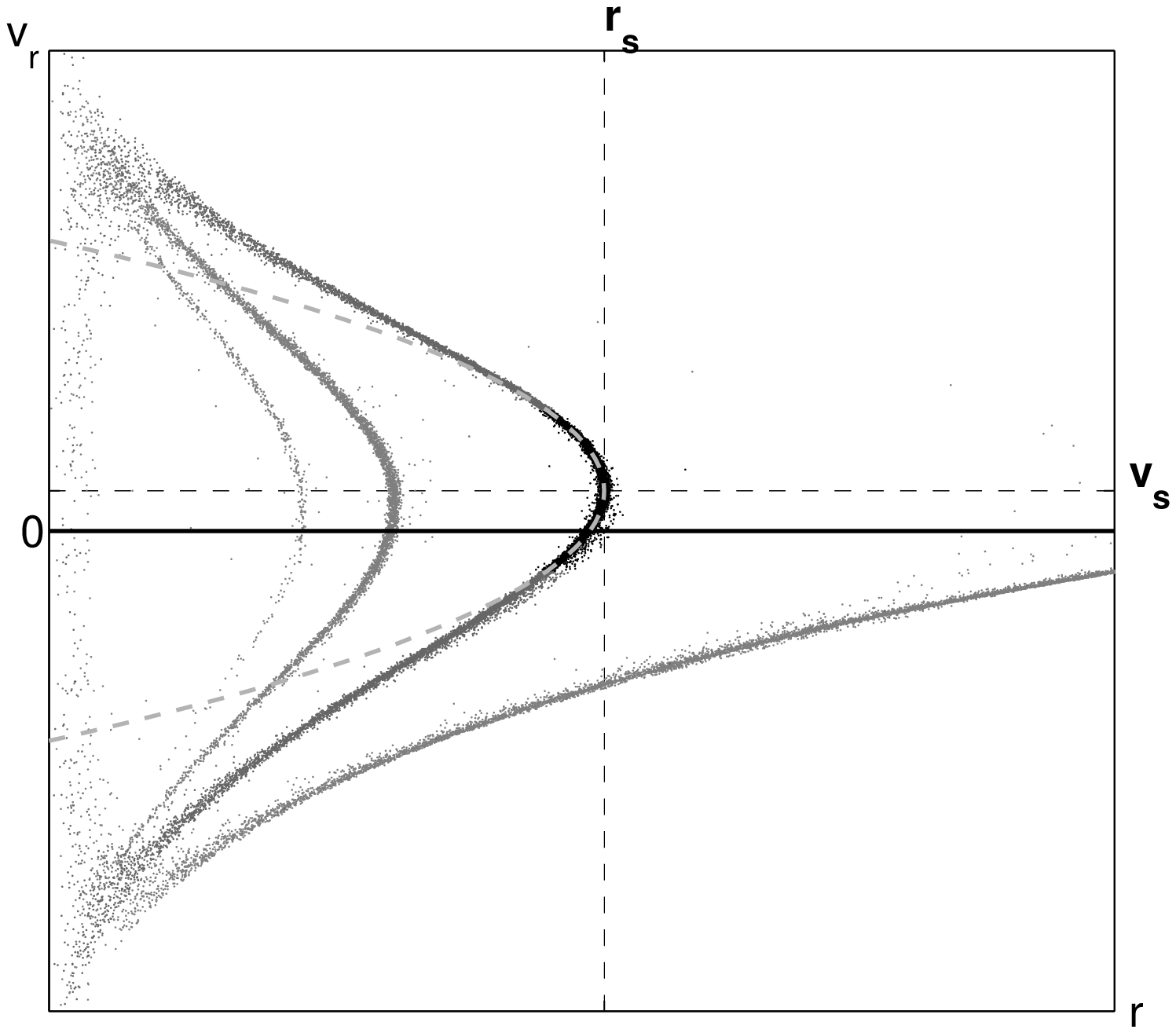}
\end{tabular}
\caption{Left: Universal caustic form for radial infall from a population with initial Gaussian velocity dispersion.  $r_s$ and $\delta_r$ are defined in Section \ref{subsec:udp} and formulae for the peak radius $\sub{r}{max}$ and density $\sub{\rho}{max}$ are given in Equations \eqref{eq:rmax} and \eqref{eq:rhomax}, respectively. Right: Phase space distribution taken from simulation D. Near the peak of one of the caustics (black), the phase space distribution follows the curve given by Equation \eqref{eq:rvrcurve} (dashed line), where $v_s$ is the radial velocity of particles at the caustic surface.  $\kappa$ is defined in Equations \eqref{eq:kappaderiv} and \eqref{eq:kappa}.}  
\label{fig:exampleCaustic}
\end{center}
\end{figure*}

In previous work \citep{2010ApJ...725.1652S}  we derived a simple analytical form for a one-dimensional caustic formed from a system with initial random velocities drawn from a Maxwellian distribution (Figure \ref{fig:exampleCaustic}, left panel).  In the case where this system is a small, gravitationally self-bound satellite galaxy falling radially into the center of a static host galaxy, the one-dimensional density profile of the caustic, as a function of the galactocentric radius $r$, can be written in terms of four physical parameters $\delta_r$, $r_s$, $\kappa$, and $f_0$:
\begin{equation}
\label{eq:generalDP}
\rho(r) = \frac{f_0}{\sqrt{2\pi \kappa}} \sqrt{\left|r-r_s\right|}\ e^{-\left(r-r_s\right)^2/4\delta_r^2}\ \mathcal{B}\left[\frac{(r-r_s)^2}{4 \delta_r^2}\right]
\end{equation}
where $\mathcal{B}$ is a piecewise combination of modified Bessel functions of the first kind:
\begin{equation}
\mathcal{B}(u) = 
\left\{
\begin{array}{cc}
\frac{\pi}{2} \left[ \mathcal{I}_{-1/4}(u) +  \mathcal{I}_{1/4}(u) \right]  & r \le r_s  \\
\frac{\pi}{2} \left[ \mathcal{I}_{-1/4}(u) -  \mathcal{I}_{1/4}(u) \right]  &  r > r_s 
\end{array}
\right.
\end{equation}

In Equation \eqref{eq:generalDP}, $\delta_r$ is the characteristic width of the caustic surface, which depends on the phase space distribution  of the dwarf galaxy that created the caustic.  A perfectly ``cold" distribution, in which all particles had the same initial energy, would give rise to a true caustic with zero width and infinite density.  We refer to caustics in a less rigorous sense: they have finite density and width but still exhibit a large local density enhancement.  In practice, determining $\delta_r$ from first principles is not straigthforward \citep[see, e.g.,][]{1999MNRAS.307..495H} Since we wish to remain as model-independent as possible, we will treat $\delta_r$ as a free parameter in the remainder of this paper.

We take the radius of the caustic surface, $r_s$, to be the location in the $(r,v_r)$ projection where the stream is locally vertical, as shown in the right panel of Figure \ref{fig:exampleCaustic}.  This radius is close to the radius of peak density, $r_{\mathrm{max}}$, but not equal to it because of the nonzero thickness of the caustic. The peak location $\sub{r}{max}$ and the peak density $\sub{\rho}{max}$ may be determined by solving $d\rho/dr=0$ numerically (necessary because of the Bessel functions). In this way we find that $r_{\mathrm{max}}$ is related to $r_s$ by
\begin{equation}
\label{eq:rmax}
r_{\mathrm{max}} = r_s - 0.765\ \delta_r, 
\end{equation}
and the density at this location is
\begin{equation}
\label{eq:rhomax}
\rho_{\mathrm{max}} = 1.021\ f_0 \sqrt{\frac{\delta_r}{\kappa}}.
\end{equation}

Expanding around $r_s$, the shape of the stream in $(r,v_r)$ can be approximated by a quadratic function:
\begin{equation}
\label{eq:rvrcurve}
 r = r_s - \kappa(v_r - v_s)^2,
\end{equation}
where the curvature $\kappa$ measures the shape of the stream near $r_s$:
\begin{equation}
\label{eq:kappaderiv}
\kappa = -\half \left. \frac{d^2r}{dv_r^2}\right|_{r_s},
\end{equation} 
and $v_s$ is the radial velocity at the caustic surface. Although all the material forming the caustic is near apocenter, $v_s$ is not zero because stars in the caustic do not all have the same energy, but are ordered by energy along the stream from lowest (most bound) to highest (least bound). As time goes on, the energy (and therefore apocenter) of stars passing through the caustic gradually increases, so the caustic gradually moves outward with time---in other words, $v_s$ is always positive.

We can relate $\kappa$ to the gravitational force at $r_s$ by differentiating the energy equation.  For a purely radial orbit in a spherical galactic potential $\Phi(r)$, we have
\begin{equation}
\label{eq:radialEnergy}
\frac{d}{dr} \left( E = \half v_r^2 + \Phi(r) \right) \to v_r \frac{dv_r}{dr} = -\frac{d\Phi}{dr} = g(r)
\end{equation}
for the first derivative.  If we invert the derivative in this expression and differentiate with respect to $v_r$, then express the first derivative in terms of $g$, we find that
\begin{equation}
\label{eq:kappa}
\kappa = -\frac{1}{2g(r_s)} \left[ 1 - \left .\frac{v_s^2}{g(r_s)^2} \frac{dg}{dr}\right|_{r_s} \right].
\end{equation}
 We can suppress the dependence on the tide, $dg/dr$, by using Equation \eqref{eq:kappa} to obtain a lower limit on $\kappa$ instead of an equality, under certain assumptions about the mass distribution.  It can be shown from the Poisson equation that $dg/dr$ is negative for any spherical density distribution that satisfies the relation  
\begin{equation}
\frac{M(<r)}{\frac{4}{3}\pi r^3} \equiv \bar{\rho}(r) < \frac{3}{2} \rho(r),
\end{equation}
at the radius of interest. For a power-law density distribution, $\rho \propto r^\gamma$, this condition is satisfied for all $\gamma<-1$. Observational evidence and simulations of dark halos both indicate that the density almost certainly falls off faster than this at most radii, especially in the outer regions where shells are usually found. If we assume $dg/dr<0$, then the second term in Equation \eqref{eq:kappa} is always positive, so that the quantity in brackets is always larger than 1.  The Poisson equation also requires that $g<0$ at all $r$ for a spherical potential, since there is no negative mass, implying that $\kappa$ is positive.  Thus neglecting the term involving the derivative of $g$ turns Equation \eqref{eq:kappa} into a lower limit on $\kappa$:
\begin{equation}
\kappa \geq \frac{1}{2 |g(r_s)|}
\end{equation}
For most reasonable models of the halo the neglected term is much less than 1, given that $v_s$ is close to zero and the tidal forces are weak at large radius (where shells are most easily observed). Neglecting the gravity gradient term gives 
\begin{equation}
\label{eq:kappaApprox}
\kappa \approx \frac{1}{2|g(r_s)|},
\end{equation}
relating the curvature of the phase-space stream directly to the gravitational force exerted by the host galaxy. 

\subsubsection{The effect of nonzero angular momentum and deviations from spherical symmetry}
\label{subsec:nonzeroLandTides}
If the satellite galaxy had some initial orbital angular momentum $L$, the energy equation  \eqref{eq:radialEnergy} is no longer correct.  In this case, following the same derivation for $\kappa$ while including nonzero $L$ (but still assuming a spherical potential) leads to the relation
\begin{eqnarray}
\label{eq:kappaWithL}
\kappa &=& -\frac{1}{2 g_s} \left(1 + \frac{L^2}{r_s^3 g_s}\right)^{-3} \left[1 - \frac{v_s^2}{g_s^2} \left.\frac{dg}{dr}\right|_{r_s} \right. \nonumber \\
&& \qquad \qquad \qquad + \left. \frac{L^2}{r_s^3 g_s}\left(2 + \frac{3v_s^2}{r_s g_s} + \frac{L^2}{r_s^3 g_s} \right) \right],
\end{eqnarray}
where $g_s\equiv g(r_s)$. The overall $1/2g$ dependence is preserved, and the first two terms in the square brackets correspond to those in Equation \eqref{eq:kappa}. However, there are now new terms that all depend on the dimensionless quantity
\begin{equation}
\label{eq:smallLdef}
-\frac{L^2}{r_s^3 g_s} \equiv \epsilon_L = \frac{L^2}{r_s^2 v_c^2},
\end{equation} 
which compares the orbital angular momentum to that of a circular orbit at $r_s$, since $v_c(r_s) = \sqrt{r_s |g_s|}$ is the circular velocity at $r_s$. The minus sign is included in the definition to ensure a positive $\epsilon_L$, since $g_s$ is negative. When the orbit is purely radial, $\epsilon_L=0$; when it is circular, $\epsilon_L=1$. 

Solving Equation \eqref{eq:kappaWithL} for $g$ gives a complicated dependence on the angular momentum that can be simplified by expanding in terms of $\epsilon_L$ and its counterpart for the tidal force,
\begin{equation}
 \epsilon_T \equiv -\frac{v_s^2}{g_s^2} \left.\frac{dg}{dr}\right|_{r_s} = -\left(\frac{v_s}{v_c}\right)^2 \left(\frac{r_s}{v_c}\right)^2 \left.\frac{dg}{dr}\right|_{r_s},
\end{equation} 
and requiring that both be much less than 1. To first order in both quantities,
\begin{equation}
\label{eq:gWithLandTides}
 g(r_s) \approx -\frac{1}{2\kappa} \left[ 1 + \epsilon_T - \epsilon_L \left( 1 + \frac{6 \kappa v_s^2}{r_s}\right)\right] + \mathcal{O}(\epsilon_L \epsilon_T).
\end{equation} 
The leading-order change in $\kappa$ due to the angular momentum is at the $L^2$ level, so we expect that Equation \eqref{eq:kappaApprox} will remain a decent approximation even for satellites on fairly non-radial orbits. Furthermore, the corrections for tides and angular momentum have opposite signs, so at intermediate $L$ we may expect help from competing error terms.

\subsection{Phase-space distribution}
\label{subsec:PDF}

Caustics have a nearly one-dimensional structure in phase space as well as a simple density profile, thanks to the fact that the material in them was initially compact in phase space.  The preservation of this initial small phase-space volume means that stripped material will spread out in a thin stream through phase space. In the case of the caustics formed by a nearly radial encounter in a spherical potential, the stream flows mainly along two of the six available coordinate directions: galactocentric radius $r$ and radial velocity $v_r$. This is why, following Equation \eqref{eq:rvrcurve}, we can construct a model of the caustic phase space distribution as a one-dimensional function of the six-dimensional phase space coordinates $(\vec{x},\vec{v})$:
\begin{equation}
 f(\vec{x},\vec{v}) \propto \delta\left[r_s - r - \kappa (v_r - v_s)^2\right].
\label{eq:phaseSpacePropto}
\end{equation}
Normalizing this distribution so that the integral over a shell is unity results in the expression
\begin{equation}
\label{eq:phaseSpaceSingleE}
  f(r, v_r) = \frac{15\sqrt{\kappa}}{16 r_s^{5/2} \Omega_s}\delta\left[r_s - r - \kappa \left(v_r - v_s\right)^2\right],
\end{equation}
where $\Omega_s$ is the solid angle spanned by the shell. The derivation of this expression is given in Appendix \ref{appx:phaseSpaceNorm}.

Unlike the model of the density profile, this model for the phase space distribution does not self-consistently take the energy spread into account, since this will result in a spread of the phase space distribution around this line. For example, if Equation \eqref{eq:phaseSpaceSingleE} is integrated over $v_r$, one obtains the limit of Equation \eqref{eq:generalDP} as $\delta_r \to 0$: a piecewise function proportional to $1/\sqrt{r_s-r}$ for $r<r_s$ and zero beyond the caustic radius.  However, we can obtain a profile consistent with Equation \eqref{eq:generalDP} by assuming that the caustic is made up of many particles with slightly different caustic radii, normally distributed around the average $r_s$. This leads to a Gaussian form for the distribution function in $r$ and $v_r$,
\begin{equation}
 \label{eq:phaseSpaceGaussian}
f(r,v_r) = \frac{15}{16 r_s^{5/2} \Omega_s} \sqrt{\frac{\kappa}{2\pi \delta_r^2}} \exp\left\{\frac{-\left[r_s - r - \kappa (v_r-v_s)^2\right]^2}{2 \delta_r^2}\right\}
\end{equation}
In Appendix \ref{appx:phaseSpaceGaussian} we give a derivation of this form, and show that integrating over all $v_r$ retrieves the functional form of Equation \eqref{eq:generalDP}.

\section{Direct applications of the model: images and line-of-sight velocities}

\subsection{Images of caustics}
\label{sec:images}

Making a few additional assumptions about the geometry of a shell allows us to project the radial density profile given in Equation \eqref{eq:generalDP} onto the plane of the sky to produce an image of a caustic. We assume the material in the caustic is distributed evenly in angle, over some solid angle $\Omega_s$.  The edge of the caustic can then be approximated as a spherical segment that spans this solid angle, and the angular extent of the debris can be modeled as a cone with its base at the center of the host galaxy.  Far from the caustic this is not a good approximation, but within a few $\delta_r$ of $r_s$ it appears to be adequate. 

\begin{figure*}
\begin{center}
\begin{tabular}{cc}
\includegraphics[width=0.43\textwidth]{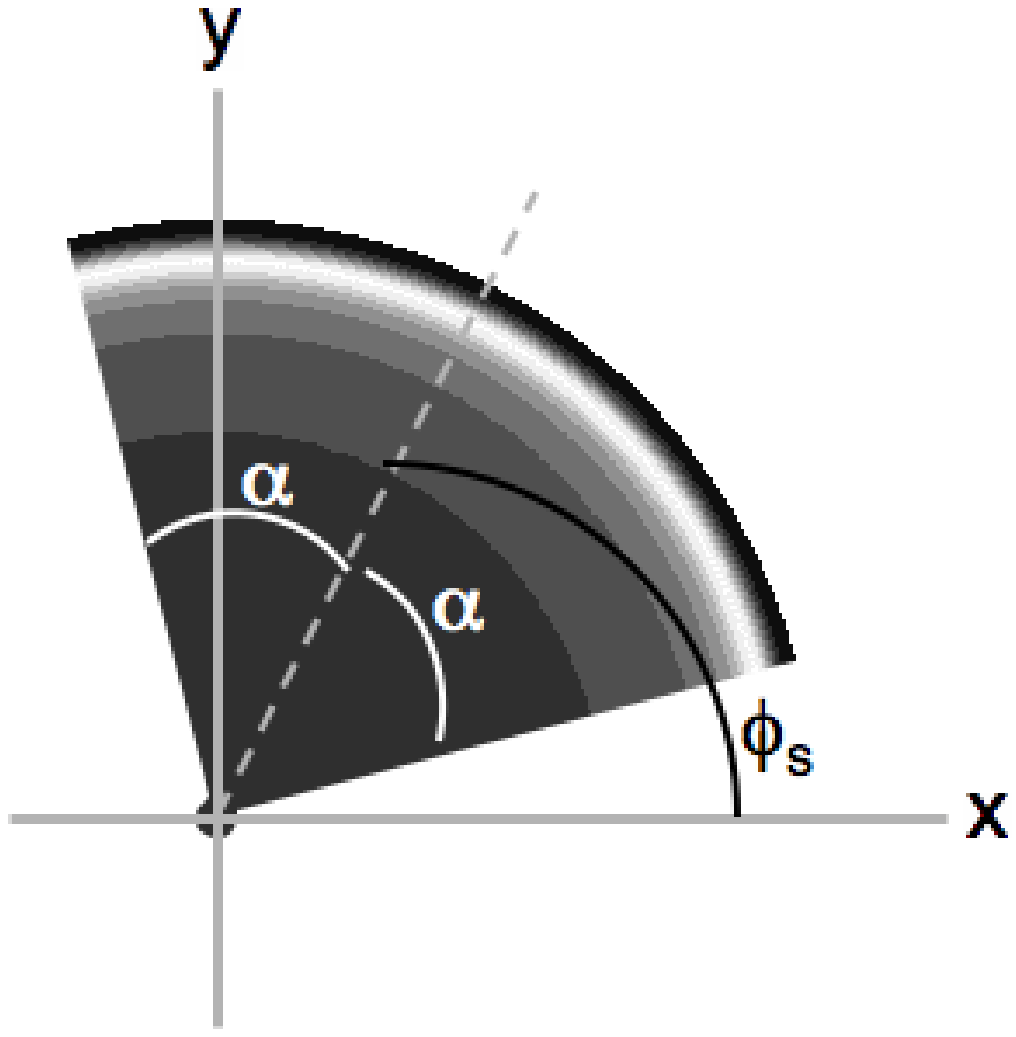} & \includegraphics[width=0.43\textwidth]{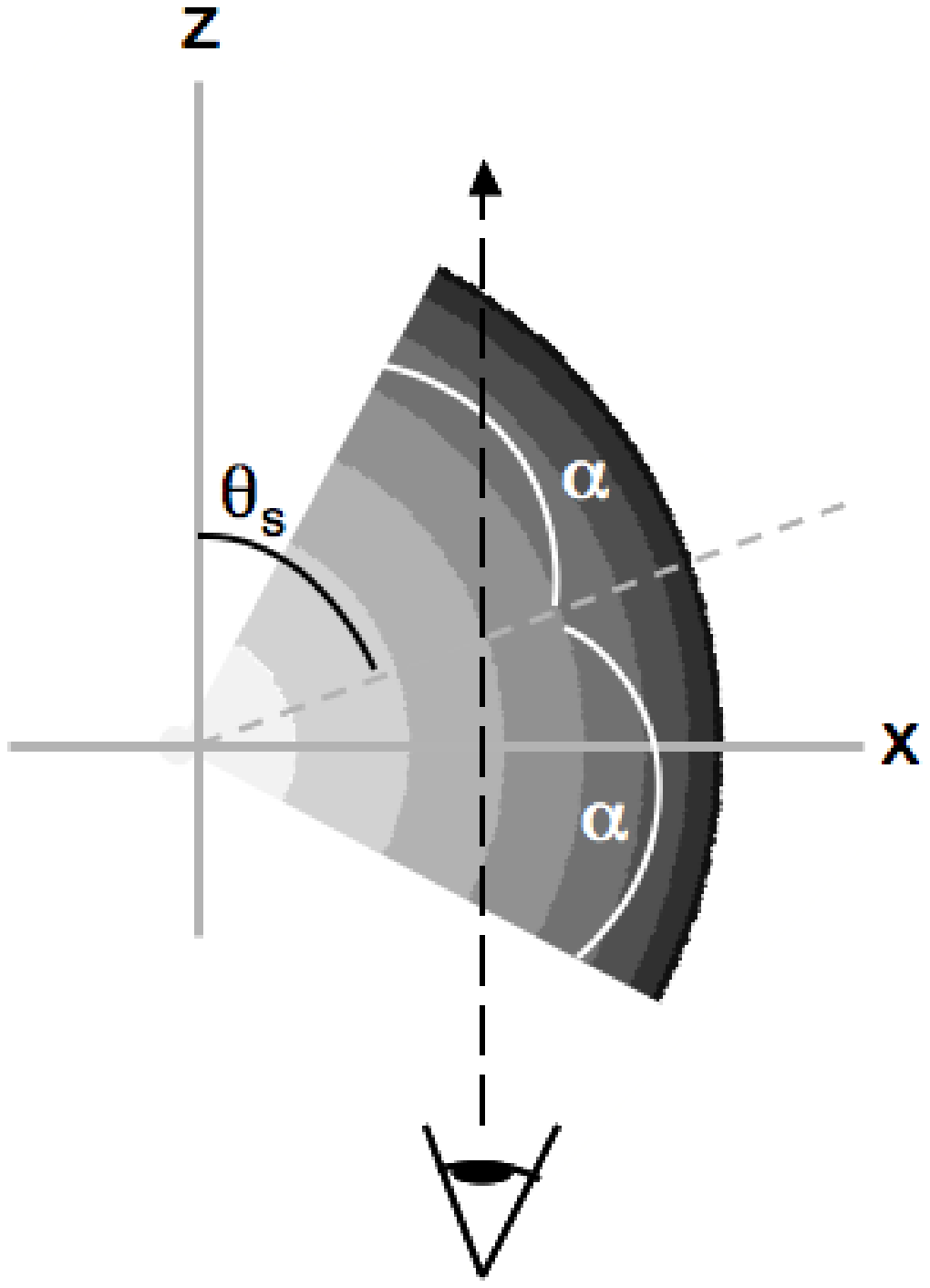}
\end{tabular}
\end{center}
\caption{Left: the geometry of the shell and cone in the $x$--$y$ plane (the plane of the sky) in a slice through $z$ = 0. The background shading is proportional to the density (lighter=denser), for a shell with a thickness of 1/50th its radius. Superimposed are the angular limit of the shell $\alpha$ (white) and the orientation of the shell in the sky plane $\phi_s$ (black). Right: the geometry of the shell in the $x$--$z$ plane (along the line of sight), sliced at $y$ = 0. Here the background shading is proportional to the radial velocity reltaive to the shell edge, $|v_r ? v_s|$ (lighter=higher value); at a given radius the radial velocity is double-valued at $\pm |v_r ? v_s|$. The angular limit $\alpha$ (white) and the inclination $\theta_s$ with respect to the line of sight (black) are superimposed. A line of sight (black dashed line) with a given projected radius $R$, equal to $x$ in this slice, probes a range of radial velocities and therefore line-of-sight velocities $v_z$. The range of $v_z$ is limited by the values of $v_s$ and $\kappa$, but also by the angular extent of the shell and its orientation with respect to the observer.}  
\label{fig:angles}
\end{figure*}

We take a coordinate system in which $z$ is the direction along the line of sight, centered on the host galaxy. Thus the $x$--$y$ plane is effectively the plane of the sky, and will be used to represent the observers' view. For this work we assume that the angular size of the galaxy is small enough that we can ignore the difference between the $z$ coordinate and the radial line of sight (spherical projection effects). The shape and orientation of the debris in the caustic are then described by three parameters: the angle $\theta_s$ of the cone relative to the line of sight $z$, the angle $\phi_s$ of the cone symmetry axis on the sky, and the opening angle $\alpha$ of the cone (Figure \ref{fig:angles}).  The solid angle enclosed by the cone is $\Omega_s = 2\pi (1-\cos \alpha)$.  The angles are defined such that $(\theta_s, \phi_s)=(0,0)$ corresponds to a cone opening directly away from the observer along the line of sight.  If the caustic has a sharp edge in projection, this means that $\theta_s$ must be close to $\pi/2$, in which case $\alpha$ is close to the value measured in the plane of the sky.  

For a given sky position $(x,y)$, we obtain the surface density $\Sigma$ by integrating the density distribution of Equation \eqref{eq:generalDP} along the line of sight $z$, for all $z$ inside the cone:
\begin{eqnarray}
\label{eq:sbxy}
\Sigma(x,y) &=& \mathcal{A} \int_{\zmin(\theta_s,\phi_s,\alpha;x,y)}^{\zmax(\theta_s,\phi_s,\alpha;x,y)} dz \sqrt{\left|\sqrt{x^2+y^2+z^2}-r_s\right|}\nonumber \\
&& \times\ e^{-\left(\sqrt{x^2+y^2+z^2}-r_s\right)^2/4\delta_r^2}\nonumber\\
&&\times\ \mathcal{B}\left[\frac{(\sqrt{x^2+y^2+z^2}-r_s)^2}{4 \delta_r^2}\right], 
\label{eq:losIntegral}
\end{eqnarray}
where we have defined $\mathcal{A}\equiv f_0/\sqrt{2\pi \kappa}$. The limits of the integration depend on the geometric parameters of the shell and the sky position.  The standard cone equations lead to analytic expressions for these limits in terms of the cone angles and sky position, which are derived in Appendix \ref{appx:rotatedCone}.  The integral along the line of sight must be done numerically, but since the expressions for $\zmin$ and $\zmax$ are analytic in $\theta_s$, $\phi_s$, and $\alpha$, standard minimization routines can be used to find best fit values by comparing the calculated and measured profiles. To this end we also provide derivatives of the integration limits with respect to the parameters in Appendix \ref{appx:rotatedCone}.

The normalization $\mathcal{A}$ involves the phase-space density $f_0$ and the gravitational force at the caustic $g_s$. If the model is fit to a surface-brightness map, the normalization obtained by the fit must be scaled by the estimated mass-to-light ratio of the debris to obtain the normalization of the mass-density profile.  Using Equation \eqref{eq:kappaApprox} as an approximation for $\kappa$ and guessing a reasonable value (or range) of $f_0$, we can in principle obtain an estimate of $g_s$:
\begin{equation}
 g_s \approx -\frac{\pi \mathcal{A}^2}{f_0^2}.
\end{equation}   
However, due to the strong inverse scaling of $g_s$ with $f_0$, the estimate of the phase-space density has to be within better than a factor of about 3 just to get the right order of magnitude for the gravitational force. Furthermore $f_0$ refers to the fine-grained density, not the coarse-grained density sampled by the stars, so it is unlikely that this approach will yield a reasonable constraint on $g_s$. We conclude that realistically, images alone can only constrain the combination $f_0\sqrt{g_s}$. As we discuss in the next section, kinematic data can separate this constraint into independent measurements.

Although it cannot produce an independent measurement of $g_s$, the surface-brightness fit plays a crucial role in the interpretation of kinematic data, which are likely far more difficult to obtain than an image. Fitting the image determines $r_s$, $\alpha$, $\theta_s$, $\phi_s$, and the normalization $\mathcal{A}$, so that the velocity data is only required to determine the remaining unknowns $\kappa$ and $v_s$. Fitting for only two parameters instead of seven will substantially reduce the amount of velocity information needed to measure $g_s$.

\subsection{Discrete line-of-sight velocity measurements of caustics}
\label{sec:vzmax}

In some shells, the line-of-sight velocities of point sources can be measured. These measurements probe the projection of the phase space distribution of Equation \eqref{eq:phaseSpaceGaussian} onto the space of projected galactocentric distance $R$ and line-of-sight velocity $v_z$. As shown by \citet{Merrifield1998}, this projected distribution has a distinctive, roughly triangular shape for shells with $\theta_s \approx \pi/2$.  In this projection the maximum magnitude of the line-of-sight velocity at a given projected radius, $v_{z,\mathrm{max}}(R)$, depends on $g_s$. We now derive an expression for $v_{z,\mathrm{max}}(R)$, similar to the derivation in \citet{Merrifield1998} but properly including the nonzero outward velocity of the caustic. 

\label{subsec:envelope}
A given projected radius $R$ samples different spherical radii $r$ that correspond, via the phase space distribution, to different line-of-sight velocities $v_z$.  For a satellite on a purely radial orbit,
\begin{equation}
 v_z = v_r \frac{z}{r},
\end{equation}
so that for a given $r$ and $R$, the line-of-sight velocity at the midline of the stream is described by
\begin{equation}
 v_z = \left(v_s \pm \sqrt{\frac{r_s-r}{\kappa}}\right)\frac{\sqrt{r^2-R^2}}{r}.
\end{equation}
To find the maximum $v_z$ at a given projected radius $R$ we need to maximize this expression with respect to $r$, to find the spherical radius $r_e$ that contributes the highest line-of-sight velocities. Technically this results in a 6th-order polynomial expression for $r_e$; however the approximation used by \citeauthor{Merrifield1998}, $r_e \approx (r_s+R)/2$, works quite well for the region near the shell.

Using the linear approximation for $r_e$ and the first-order approximation for $\kappa$ gives the formula for the maximum line-of-sight velocity as a function of projected radius alone, with $g(r_s)$ and $v_s$ as parameters:
\begin{equation}
\label{eq:vlosmax}
 v_{z,\mathrm{max}} = \frac{\sqrt{r_s^2 + 2 R r_s - 3 R^2}\left(\sqrt{g(r_s)(r_s-R)}+v_s\right)}{r_s+R}
\end{equation}
If we take the limit $v_s\to 0$ and keep only first-order terms in an expansion in $R$ about $r_s$, we recover the expression obtained by \citeauthor{Merrifield1998}:
\begin{equation}
 \label{eq:vlosMK}
\super{\sub{v}{z, max}}{MK} = \sqrt{\frac{g_s}{r_s}} (r_s - R).
\end{equation} 

If enough line-of-sight velocity measurements of point sources are available to fully probe a shell's density in the projected phase plane, $f(R,v_z)$, then Equation \eqref{eq:vlosmax} can be fitted to their maximum-velocity envelope to measure $g_s$. However, in order to know how many point sources are required, we need to know what $f(R,v_z)$ looks like. We can calculate it by converting Equation \eqref{eq:phaseSpaceGaussian} to cylindrical coordinates and integrating over everything but $R$ and $v_z$. This means we must take the geometry of the shell into account, using the cone model derived in the previous section. The full phase space distribution is limited to the cone by some function $\Theta(\theta_s,\phi_s,\alpha; \theta,\phi)$ in spherical coordinates.  Since we are projecting onto the plane of the sky, we can separate the $\phi$-dependence so that the cone limits are
\begin{equation}
 \Theta(\theta_s,\phi_s,\alpha;R,z) \Theta(|\phi-\phi_s|\le \alpha).
\end{equation} 
We have assumed the velocity of the material is purely radial, and so $v_\theta$ and $v_\phi$ are both zero. In cylindrical coordinates this becomes
\begin{equation}
 \delta(v_\phi=0) \delta(v_R = v_z R/z).
\end{equation} 
Then the full phase-space distribution is
\begin{eqnarray}
 f(\vec{x},\vec{p}) &=& \mathcal{F}_0 \exp\left\{\frac{-\left[r_s -  \sqrt{R^2 + z^2} - \kappa (\frac{R v_R}{r} + \frac{z v_z}{r}-v_s)^2\right]^2}{2 \delta_r^2}\right\} \nonumber \\
&&\times \Theta(\theta_s,\phi_s,\alpha;R,z) \Theta(|\phi-\phi_s|\le \alpha) \nonumber\\
&&\times \delta(v_\phi=0)  \delta(v_R = v_z R/z)
\end{eqnarray} 
where
\begin{equation}
\mathcal{F}_0 \equiv \frac{15}{32 \pi r_s^{5/2}(1-\cos\alpha) } \sqrt{\frac{\kappa}{2\pi \delta_r^2}}
\end{equation} 

Now we must integrate over $v_\phi$, $v_R$, $\phi$, and $z$. All of these but the $z$ integral are delta functions or step functions, and can be trivially evaluated:
\begin{eqnarray}
 f(R,v_z) = & 2 \mathcal{F}_0 \alpha \int_{-\infty}^{\infty} dz \Theta(\theta_s,\phi_s,\alpha;R,z) \times \nonumber & \\
& \exp\left\{\frac{-\left[r_s -  \sqrt{R^2 + z^2} - \kappa (\frac{R^2 v_z}{r z} + \frac{z v_z}{r}-v_s)^2\right]^2}{2 \delta_r^2}\right\} \nonumber & \\
\end{eqnarray}
The remaining step function limits the line-of-sight integral to points inside the spherical cone used to describe the shell's geometry; thus we can rewrite the integral by replacing the limits of integration with the functions of $R$ derived in Appendix \ref{appx:rotatedCone} (suppressing the parameter dependence for simplicity), just as we did to calculate surface brightness in Section \ref{sec:images}:
 \begin{equation}
 \label{eq:projectedPhaseDensity}
  f(R,v_z) = 2 \mathcal{F}_0 \alpha \int_{\zmin(R)}^{\zmax(R)} dz e^{-\left[r_s -  \sqrt{R^2 + z^2} - \kappa (\frac{R^2 v_z}{r z} + \frac{z v_z}{r}-v_s)^2\right]^2/2 \delta_r^2}.
 \end{equation} 
The line-of-sight integral is then performed numerically to obtain the distribution for a given $R$ and $v_z$. We will discuss the specific form that results from this integral further in Section \ref{subsec:DLOScomparison}.

\section{Comparisons with simulations}
\label{sec:sims}

To check that the analytical model represents the salient features of caustics, we will compare it with a set of simulations intended to gradually relax the assumptions of spherical symmetry and radial orbits.  The simulations are labeled A through D in order of decreasing symmetry; series B and C gradually increase the angular momentum in two different potentials.  All the simulations presented here follow the disruption of an N-body Plummer satellite in a static potential representing the host galaxy. Integrations were performed using a serial implementation of a leapfrog symplectic integrator (kick-drift-kick) written in Scheme and C by Will Farr; the Barnes-Hut tree method \citep{1986Natur.324..446B} is used to calculate interparticle forces with Plummer softening $\epsilon=100$ pc and tree opening angle $0.8$ radians. The code uses a single adaptive timestep whose length is determined by the maximum particle acceleration: $\Delta t= \eta \sqrt{2\epsilon/|\sub{a}{max}|}$, with constant of proportionality $\eta=0.1$. With these settings energy is conserved to about 1 part in $10^4$ or better. The host-galaxy forces were computed analytically except in the case of simulation D, which uses the multipole-expansion code {\tt GalPot} \citep[public release via][]{1998MNRAS.294..429D} to calculate forces from a combination of axisymmetric and spherical potentials. Table \ref{tbl:sims} summarizes the parameters of all the simulations.  

For simulations A and B, the initial position and velocity vectors for the satellite galaxy lie in the sky plane, so that the resulting shells are symmetric with respect to $z=0$. This is the orientation in which the shell edges appear sharpest and the kinematic signature is least complicated. In simulations C and D we relax this assumption. 

\begin{figure*}
 \includegraphics[width=0.95\textwidth]{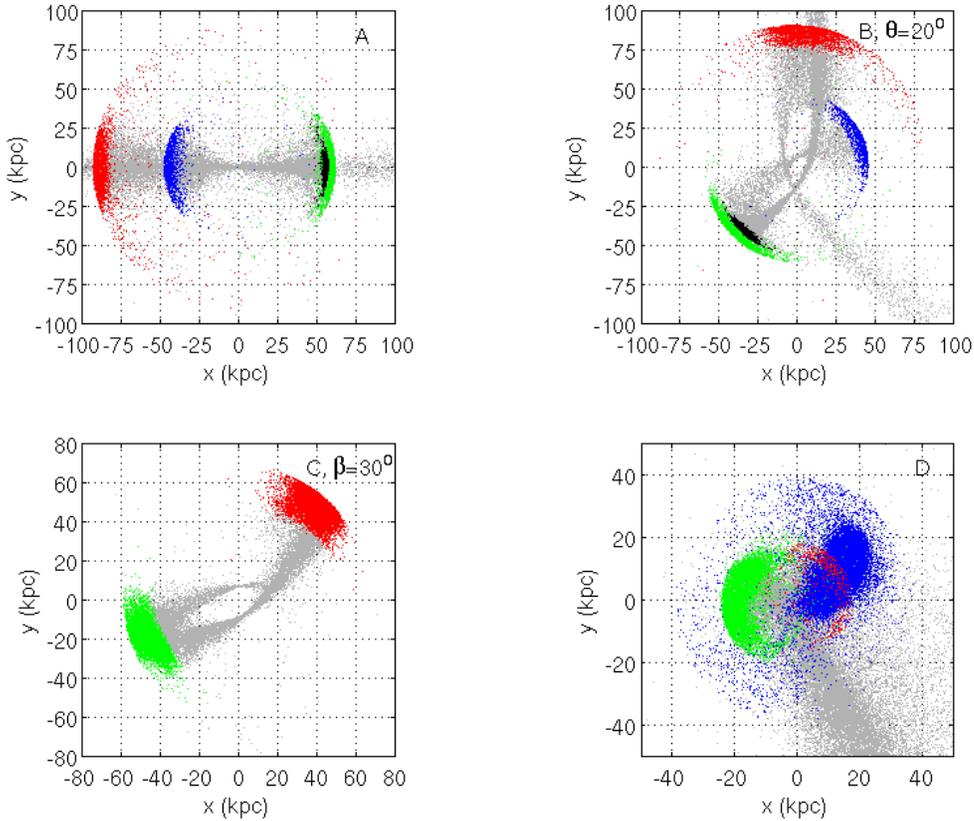}
 \caption{Views of the sky ($x$--$y$) plane for the four types of simulations used in this work. In each simulation, caustics selected in $r-v_r$ space (shown in different colors) reflect the shell features seen in the plane of the sky. Simulations A-C have the same mass ratio of about $10^{-4}$; simulation $D$ has a mass ratio of about $3 \times 10^{-3}$ (see Table \ref{tbl:sims} for more information). For B and C, one simulation in each series is shown with the given angle $\theta$ or $\beta$, as described in Section \ref{sec:sims} and Table \ref{tbl:sims}. For C, this value of $\beta$ corresponds to a starting $R=54.9$ kpc (see Appendix \ref{appx:potls}. In all panels, half of the particles are plotted.}
\label{fig:simsxy}
\end{figure*}

Simulation A (Figure \ref{fig:simsxy}, top left) is the base case that fulfills all the assumptions in the derivation presented in Section \ref{sec:model}: the satellite evolves in a static, spherical potential on a purely radial orbit ($\sub{L}{COM}=0$). Simulations B use the same potential but gradually increase the orbital angular momentum, parameterized by the angle $\theta$ between the initial position and velocity vectors of the center of mass.  The orbital energy of the center of mass is kept constant by maintaining $|\sub{\mathbf{v}}{COM}|=v_c(\sub{r}{COM})$ and starting all the orbits from the same initial position, so that the angular momentum is $\sub{L}{COM}=\sub{L}{circ}\sin \theta$, where $\sub{L}{circ} \equiv \sub{r}{COM} v_c$. In this way $\theta=0$ corresponds to a radial orbit and $\theta=90^\circ$ is a circular orbit. An example with a moderate amount of angular momentum, $\theta = 20^{\circ}$, is shown in the top right panel of Figure \ref{fig:simsxy}.

Series C uses a highly flattened, axisymmetric cored logarithmic potential intended to exaggerate the effect of departures from spherical symmetry, thus stretching the capabilities of the model to the limit. The scale parameters of the potential (core radius and asymptotic circular velocity) were determined by fitting the rotation curve of this potential in the symmetry plane ($z=0$) to that of the isochrone potential used in Simulations A and B (see Appendix \ref{appx:potls}). The flattening $q$ was set to $q=0.78$, which is highly flattened but still has positive density everywhere. This is much more flattened than we expect dark halos to be: at a few times the halo scale radius, where shells tend to be located, $q$ is generally between 0.9 and 1 in cosmological Milky-Way-sized dark halos \citep{2007MNRAS.377...50H}. We chose a set of orbits whose parent spherical orbit has the same radius as Simulations B, but which have $\sub{z}{max}=25$ kpc to probe the potential well out of the plane of symmetry. This requirement results in a maximum $z$ angular momentum of $\sub{L}{z,max} = 0.71\sub{L}{circ}$, relative to the planar circular orbit. We varied $L_z$ in 10 even steps between zero and this maximum, parameterizing the fraction of $\sub{L}{z,max}$ in an orbit using an angle $\beta$, defined such that $L_z = \sub{L}{z,max} \sin \beta$. This angle is roughly analogous to $\theta$ in Simulations B. More information about the generation of the initial conditions for this series is given in Appendix \ref{appx:potls}.

Simulation D (Figure \ref{fig:simsxy}, bottom right) is an N-body model of a real set of caustics in a realistic, though still static, galactic potential (a mass model of M31). This model, obtained by \citet{fardal:2007aa}, includes two caustics produced by the nearly-radial encounter of an intermediate-mass satellite galaxy. This simulation is included so we can compare our analytic expressions to more realistic conditions.

\begin{table*}
\caption{Information about the four simulations whose results are presented in this work: the name by which the simulations are referred to in this work, the type and parameters of the potential used to represent the host galaxy in each simulation, and the mass and initial radius ($\sub{\mathbf{r}}{COM}$) and velocity ($\sub{\mathbf{v}}{COM}$) of the satellite whose disruption creates shells. For Series B, $\theta$ is the angle between $\sub{\mathbf{r}}{COM}$ and $\sub{\mathbf{v}}{COM}$; for Series C, $\beta$ is defined in the introduction to Section \ref{sec:sims}. Expressions for the potential types in the table are given in Appendix \ref{appx:potls}.}
\begin{center}
\begin{tabular}{lp{1.1in}p{1.2in}p{0.7in}lp{1.3in}p{0.7in}}
Name & Potential type & Potential parameters & Satellite mass ($\times 10^8\ M_{\odot}$) & Initial $\sub{\mathbf{r}}{COM}$ (kpc) & Initial $\sub{\mathbf{v}}{COM}$ (km \unit{s}{-1}) $v_c \equiv v_c(\sub{r}{COM})$ & Integration time (Myr)\\
\hline
A & Spherical Isochrone & $M = 2.7 \times 10^{12}\ M_{\odot}$, $b = 8.0$ kpc & $2.2$& (-40, 0, 0) & ($v_c$, 0, 0) &  1000 \\
Series B  & Spherical Isochrone & $M = 2.7 \times 10^{12}\ M_{\odot}$, $b = 8.0$ kpc  & $2.2$& (-40, 0, 0)  & ($v_c \cos \theta$, $v_c \sin \theta$, 0), $\theta \in \{10,20,\ldots, 90\}^{\circ}$ & 1000 \\
Series C  & Flattened Cored Logarithmic & $v_c = 411\ \textrm{km}\ \unit{s}{-1}$, $r_0 = 5.0\ \textrm{kpc}$, $q = 0.78$, {$\Phi_0 = -1.22\ ($kpc \unit{Myr}{-1}$)^2$} & $2.2$& ($-R$, 0, 25) & $(0, 11.9 \sin \beta/(R/\textrm{kpc}), 0)$, $37.5 \leq R(\beta) \leq 57.4$ kpc, on zero-velocity curve, $0\leq \beta \leq 90^\circ$ as defined in Section \ref{sec:sims} & 3000 \\
D & Hernquist bulge, exponential disk, spherical NFW halo \citep[see][]{Geehan2006}& $\sub{M}{M31}(< 125$ kpc$) = 7.3 \times 10 ^{11} M_{\odot}$, $\sub{r}{halo} = 7.63$ kpc; see Table 2 of \cite{fardal:2007aa} for complete parameters  & $22$& (-34.75, 19.37, -13.99)  & (67.34, -26.12, 13.50)  & 840 
\end{tabular}
\end{center}
\label{tbl:sims}
\end{table*}%
\subsection{Tests of the assumptions of radial orbits and spherical symmetry}
\label{subsec:assumptionTests}
In Section \ref{subsec:nonzeroLandTides} we determined how the shape of a caustic in the $r$-$v_r$ plane depends on the tide at the caustic surface and the angular momentum. In analyzing real shells we cannot correct for these unknown quantities, but with numerical experiments we can test how the assumption that both are zero will affect estimates of $g_s$. To do this, we obtain $\kappa$ from fits to the projected $(r,v_r)$ distribution for shells in the simulations of Series B. We first estimate $g$ using Equation \eqref{eq:kappaApprox}, then progressively correct this estimate at first order according to Equation \eqref{eq:gWithLandTides}, taking first tides, then angular momentum into account. The simulations in the series form four caustics at different radii, and we obtain an independent estimate of $g_s$ from each. 

\begin{figure*}
 \begin{tabular}{ccc}
\includegraphics[width=0.39\textwidth]{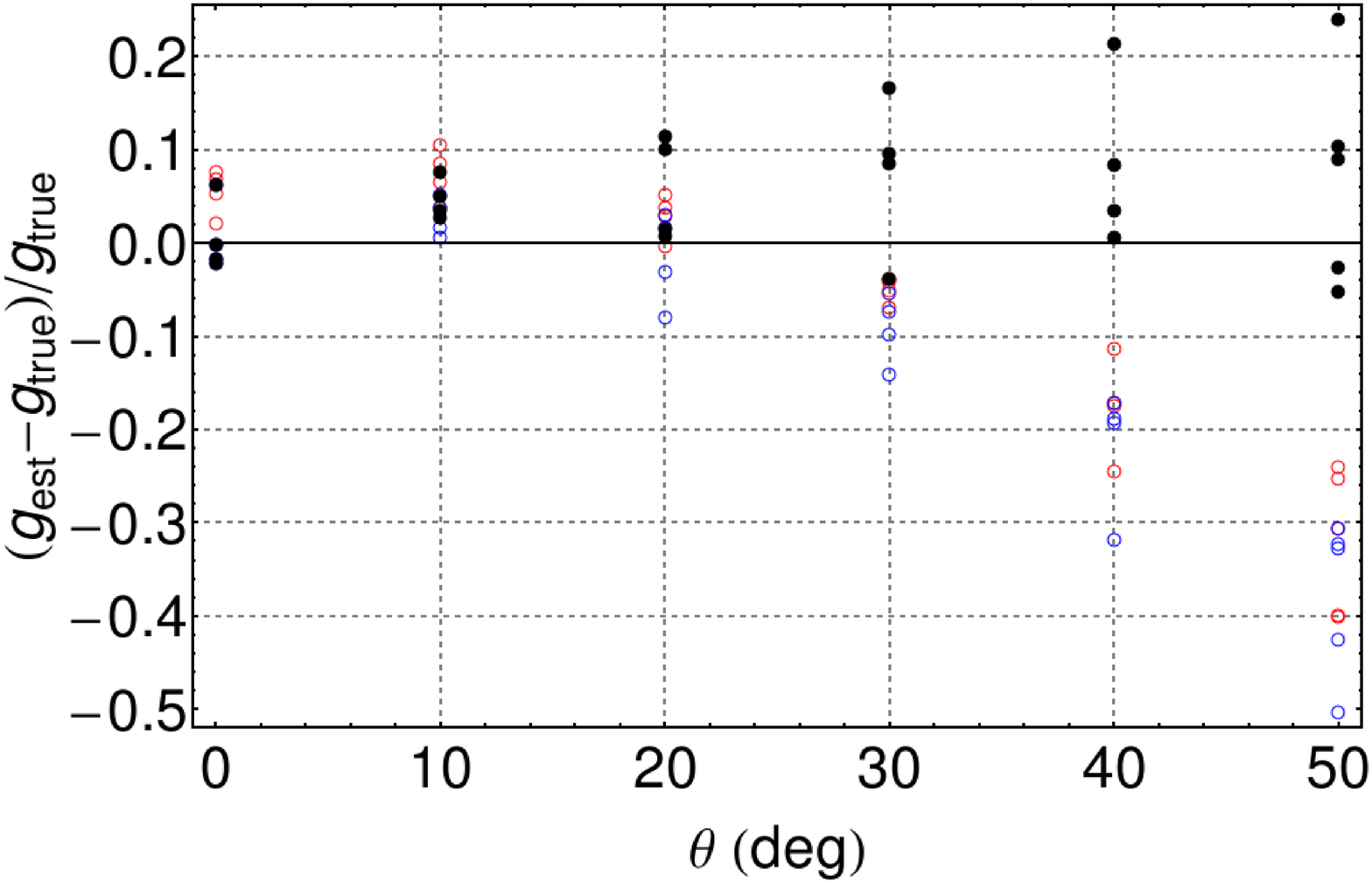}   & \includegraphics[width=0.3\textwidth]{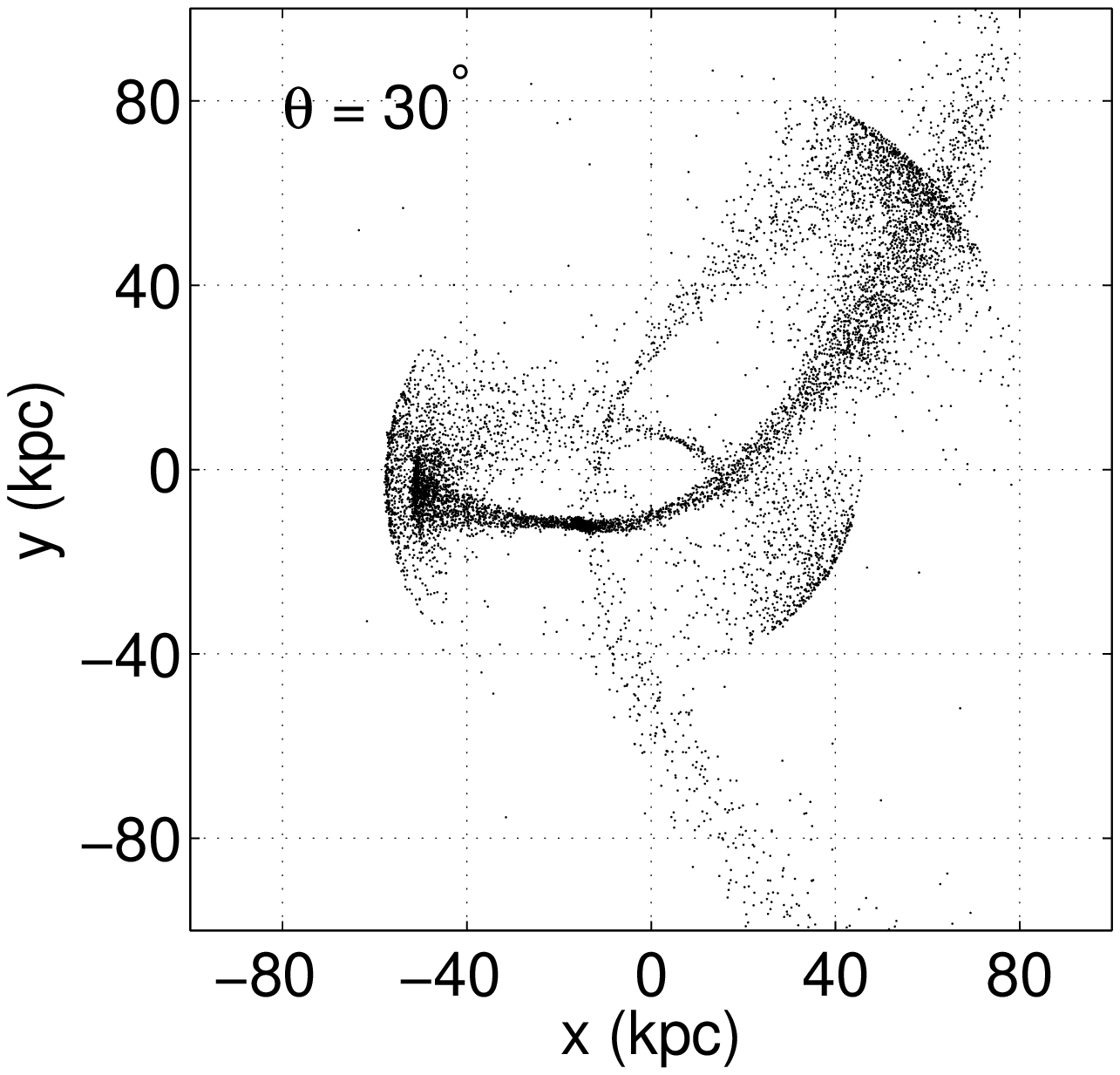} & \includegraphics[width=0.3\textwidth]{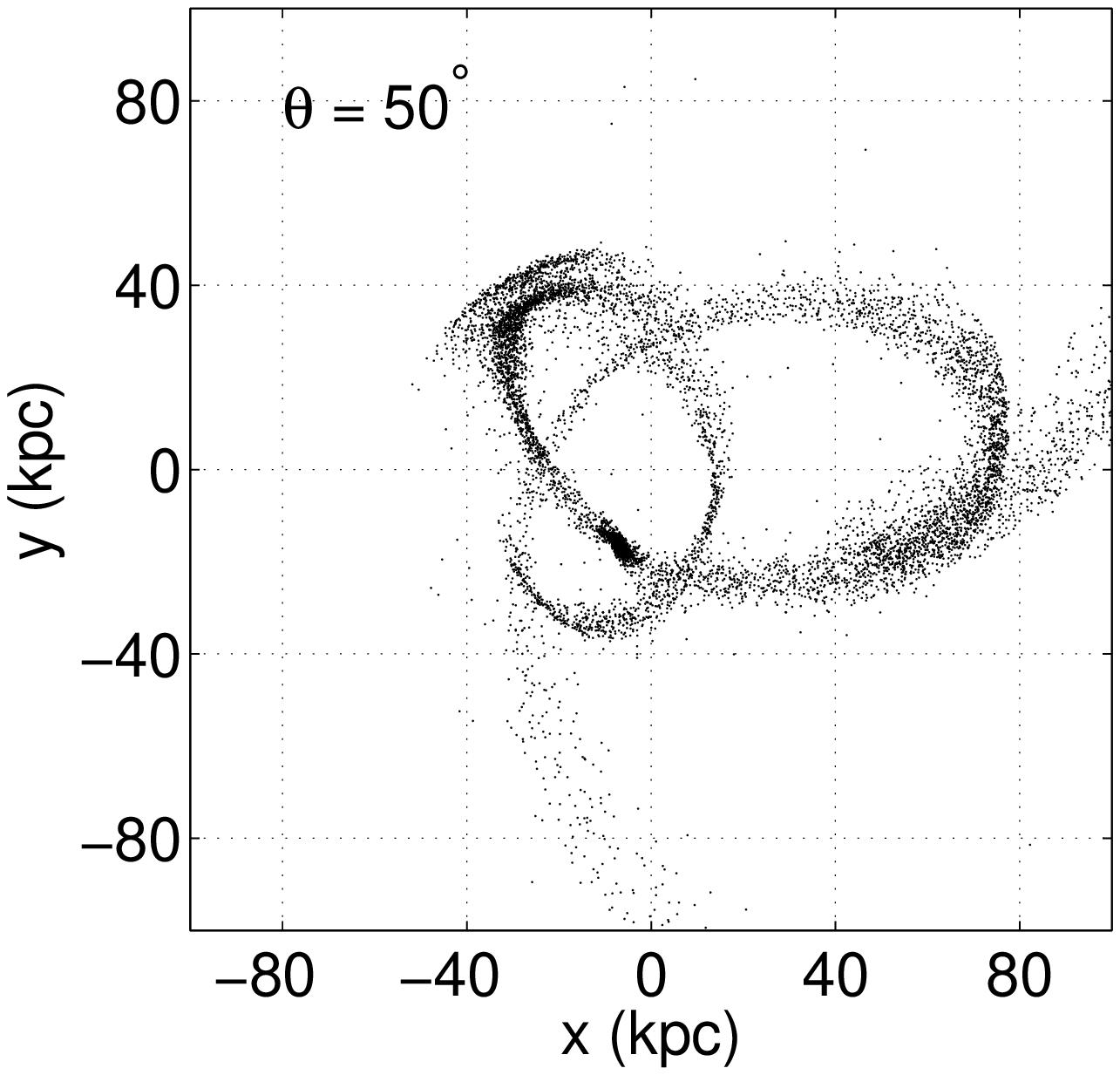}
 \end{tabular}
\caption{Left: the fractional error in recovering the radial gravitational force $g$ (Equation \ref{eq:radialEnergy}), calculated for the spherical potential of Simulations B which have increasing angular momentum parameterized by $\theta$, as discussed in the introduction to Section \ref{sec:sims}. The error remains small even at fairly large values of the orbital angular momentum, which is proportional to $\sin\theta$. The blue open circles show the relative error from using Equation \eqref{eq:kappaApprox} to estimate $g$. The red open circles include the first-order correction for tidal forces from Equation \eqref{eq:gWithLandTides}; the solid black points correct for both tides and angular momentum at first order using the same equation. Identical symbols at the same incident angle represent the different caustics in each simulation. Center and right: the tidal features at $\theta=50^\circ$ (right), when the error on $g_s$ exceeds about 20 percent, still exhibit a sharp outer edge but are morphologically distinct from fan-shaped shells with lower angular momentum (e.g. at $\theta = 30^\circ$, center). }
\label{fig:gRecoverySpherical}
\end{figure*} 

As seen in the left panel of Figure \ref{fig:gRecoverySpherical}, the spherical approximation still recovers the gravitational force to within about 20 percent, even for relatively large amounts of angular momentum (up to half the maximum value). This is due to the quadratic dependence of $\epsilon_L$ on $L$ and the partial cancellation of $\epsilon_T$ and $\epsilon_L$. In some cases the combined first-order corrections for tides and angular momentum are sufficient to eliminate nearly all the error, while in others the correction is less good, especially at larger $L$. This is because the average $L$ of material in each shell is not necessarily equal to $\sub{L}{COM}$, which was used to calculate the correction. Since $\epsilon_L$ is quadratic in $L$ this difference is magnified at larger $L$.

We also see that the difference between a system that has less than 20 percent error using the radial approximation (Figure \ref{fig:gRecoverySpherical}, center panel) and one that has much more (Figure \ref{fig:gRecoverySpherical}, right panel) coincides with the difference between a shell-like and a stream-like morphology.  The example with more angular momentum still has a sharp edge at the apocenter of each caustic, but the morphology of the debris is otherwise more stream-like than shell-like. Projection effects will increase these differences; the two systems shown here are at an ideal viewing angle where the shells look the sharpest. We conclude that identifying shells (consisting of a fan shape and a sharp outer edge) by eye as low-$L$ systems is a sufficiently strict selection criterion for applying the radial-orbit approximation to obtain estimates of $g_s$.

\begin{figure}
 \includegraphics[width=0.48\textwidth]{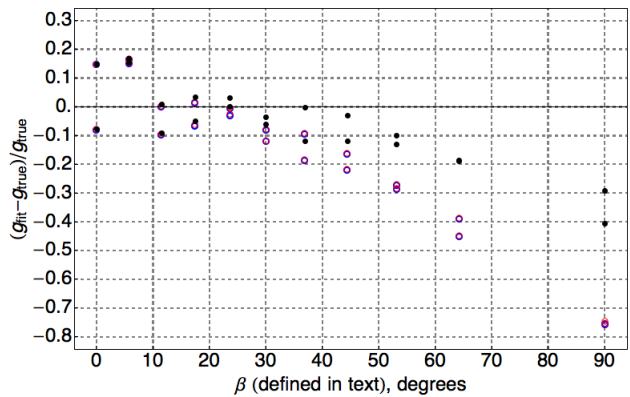}
\caption{As in the left panel of Figure \ref{fig:gRecoverySpherical}, but for simulations in series C (a highly flattened potential). In this case there are two caustics per simulation. The parameter $\beta$, which measures the fraction of maximum $z$ angular momentum, is defined in the introduction to Section \ref{sec:sims} and further discussed in Appendix \ref{appx:potls}. The errors after correction for the angular momentum, shown as black points, are reduced but do not vanish since precession is not taken into account. }
\label{fig:gRecoveryAxi}
\end{figure}

Deviations from spherical symmetry in the potential can also affect the ability to recover $g_s$. Out of the plane of symmetry, orbits in an axisymmetric potential will precess, thickening the phase space stream out of the $(r,v_r)$ plane. Furthermore, the spherical radius $r$ and its conjugate velocity $v_r$ are not symmetry coordinates in an axisymmetric potential, so there is no guarantee that the analysis developed here will still be applicable to non-spherically-symmetric systems. To see how well we can do for a significantly flattened system, we repeat the process used to obtain Figure \ref{fig:gRecoverySpherical} on Series C, with one change. Because the caustics are now quite thick, we first fit the density profile using Equation \eqref{eq:generalDP} to obtain $r_s$ and $\delta_r$. Then we force $r_s$ to $\sub{r}{max}$ when fitting Equation \eqref{eq:rvrcurve} to the $(r,v_r)$ distribution to obtain $g_s$ and $v_s$. Otherwise the value of $r_s$ obtained from fitting Equation \eqref{eq:rvrcurve} with equal weights for all the N-body points tends to underestimate $r_s$, which leads to extra error in determining $g_s$. We compare the recovered $g_s$ with the derivative of the potential with respect to $r$ in the $z=0$ plane, evaluated at $r_s$. The results are shown in Figure \ref{fig:gRecoveryAxi}. Equation \eqref{eq:gWithLandTides} does not include the effect of precession, so the error prediction from the spherical analysis does not fully correct the estimate, but we can still recover $g$ within 30 percent for $\beta$ up to ~50 degrees. At small $\beta$ there tend to be more cases where there is substructure within the stream in the $(r,v_r)$ plane because the material is actually passing through caustics in both symmetry coordinates, $R$ and $z$. This leads to bad fits because we are trying to fit a model for a first-order (fold) caustic to a higher-order caustic---the intersection of two folds---that is much sharper. This is a reflection of the highly flattened potential, and causes the relatively large error at small $\beta$. In addition, at small $\beta$ most of the material is at high $z$, where the force differs most from that in the plane. At larger $\beta$ the material tends to be in only an $R$-caustic, so our density model is appropriate and easier to fit; the best performance is therefore in the transition between these two regions: where angular momentum is not yet too large for a good fit, but there is enough that the $R$ and $z$ periods are quite dissimilar. In the case of the best estimate, at $\beta=23.6^\circ$, the material in the caustic is distributed fairly symmetrically around the $z=0$ plane.  We conclude that although we expect galactic potentials to be flattened, the inaccuracy from assuming spherical symmetry is not large enough to prevent us from measuring $g$.

\subsection{Comparisons of model and simulated phase-space distributions}
\label{subsec:phaseSpaceComparison}
\begin{figure*}
\begin{tabular}{ccc}
  \includegraphics[width=0.3\textwidth]{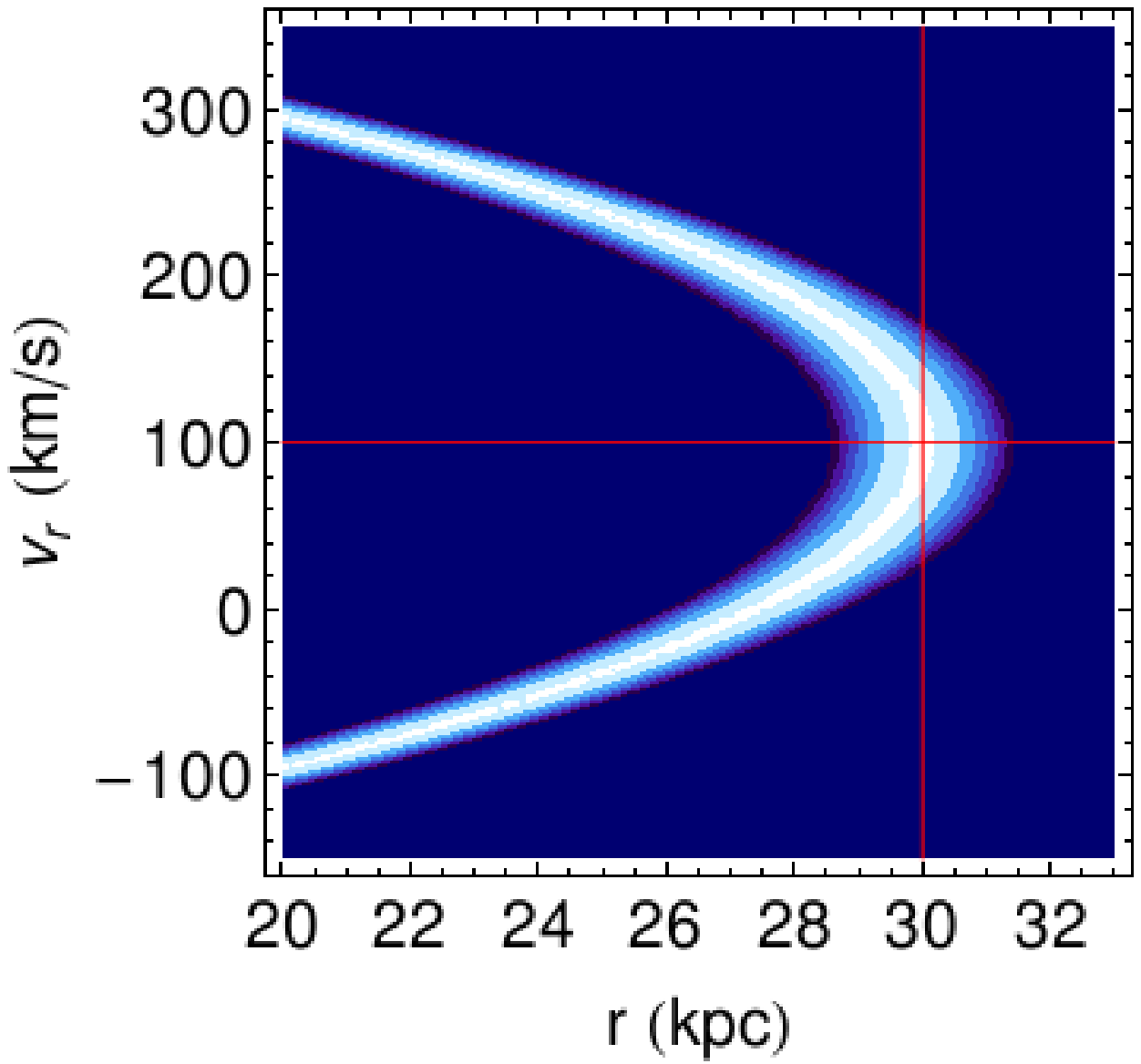} &   \includegraphics[width=0.3\textwidth]{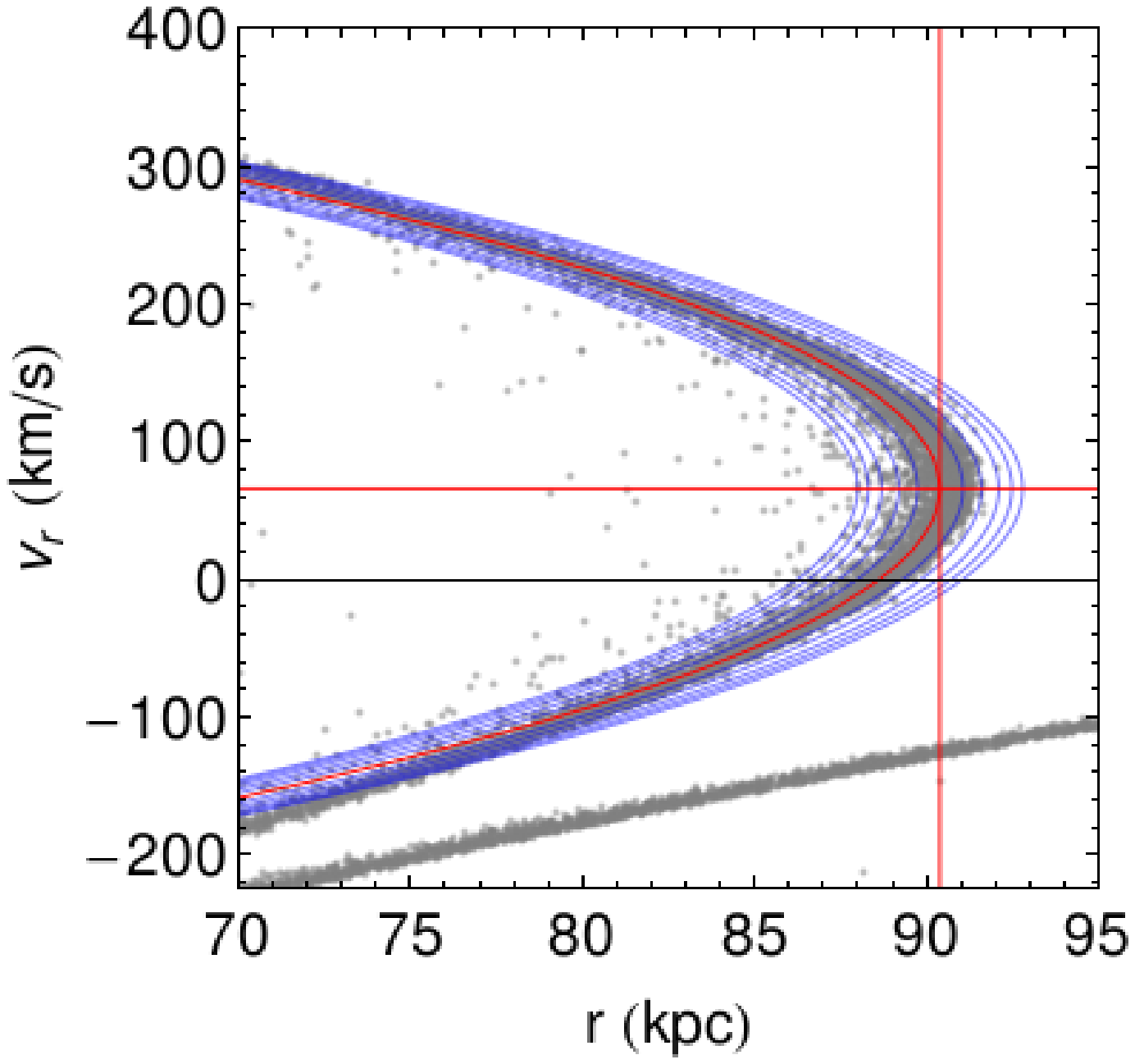} & \includegraphics[width=0.3\textwidth]{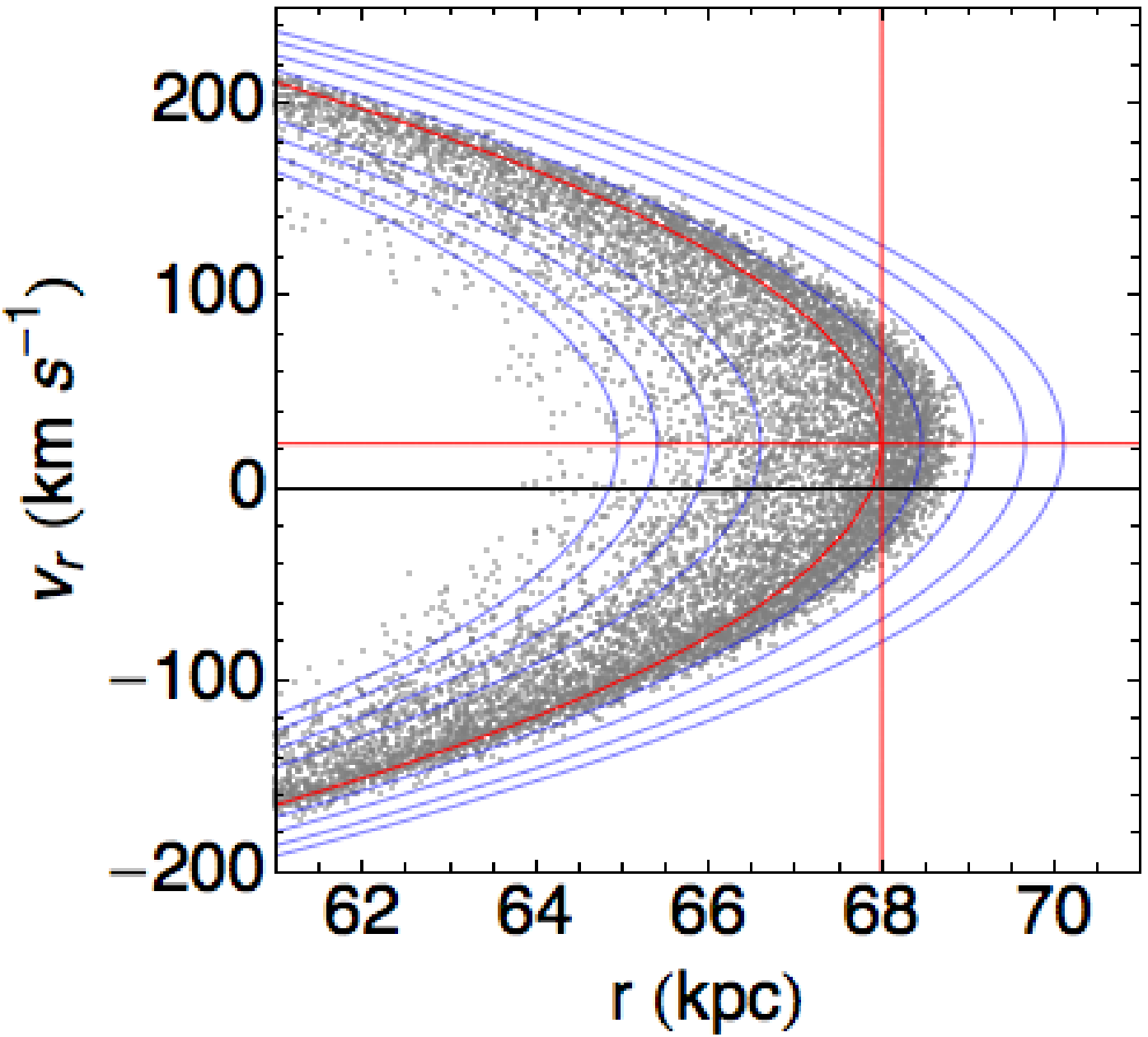}
\end{tabular}
\caption{Contours of the model phase space density distribution in the plane of galactocentric radius $r$ and radial velocity $v_r$ for sample parameters (left); with fitted parameters overlaid on a caustic from the series B simulation with incident angle $\theta=20^\circ$; and with fitted parameters overlaid on a caustic from the series C simulation with $\beta=30^\circ$ (right). In all panels the straight red lines cross at the caustic radius and expansion velocity $(r_s, v_s)$. In the center and right panels the red parabola denotes the fit of Equation \eqref{eq:rvrcurve} used to give $\kappa$, $v_s$, and $r_s$; a fit to the density profile of Equation \eqref{eq:generalDP} determined the caustic thickness $\delta_r$. }
\label{fig:phaseSpaceGaussian}
\end{figure*}

\begin{figure*}
\begin{tabular}{cc}
  \includegraphics[width=0.45\textwidth]{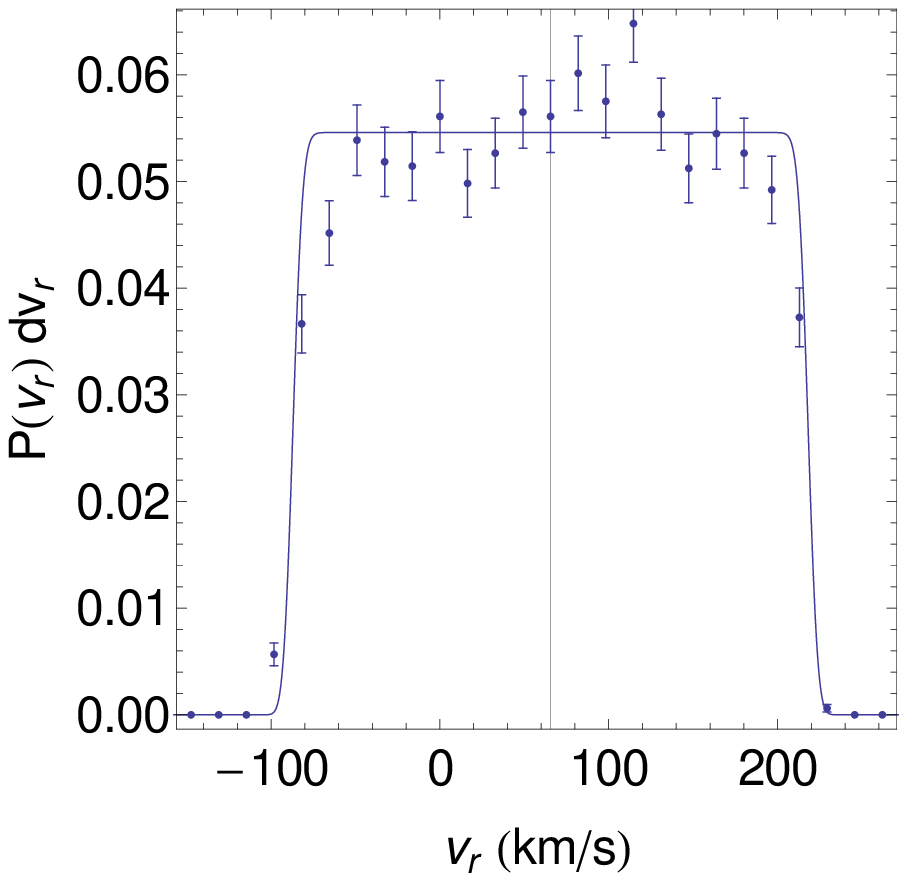} &   \includegraphics[width=0.45\textwidth]{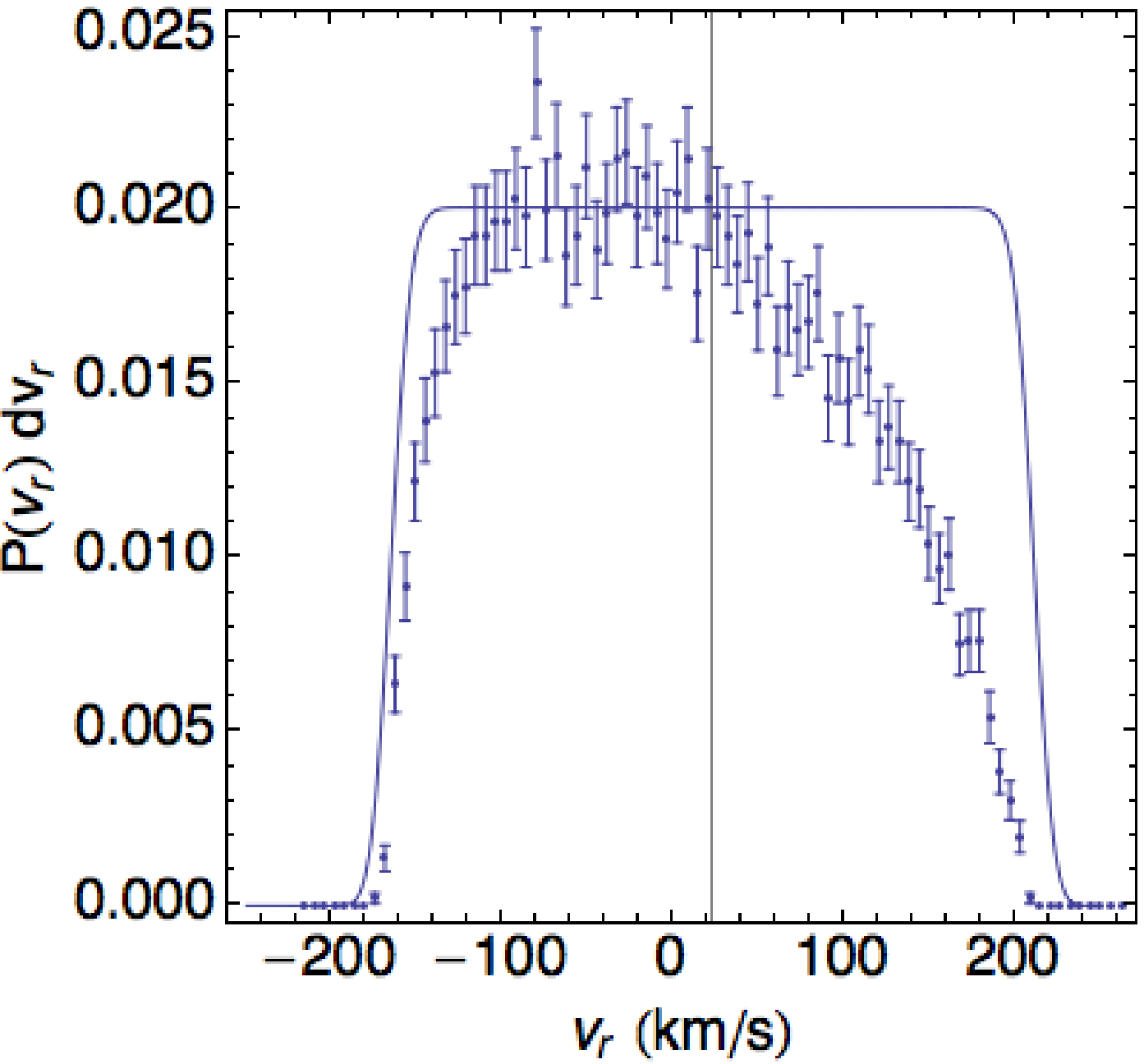} 
\end{tabular}
\caption{Probability distributions for the edges of two simulated caustics, compared to the phase space distribution integrated over $r>\sub{r}{min}$.  The normalization of the model is scaled by eye to match the data; the error bars indicate Poisson error on the number of counts per bin.  Left panel: a caustic from series B ($\theta=20^\circ$) with $r_s = 90.4$ kpc, $\delta_r = 0.6$ kpc and $\sub{r}{min} = 81$ kpc. Right panel: a caustic from series C  ($\beta=30^\circ$) with $r_s = 67.5$ kpc, $\delta_r = 0.68$ kpc and $\sub{r}{min} = 60.5$ kpc.}
\label{fig:phaseSpaceVelocityProjection}
\end{figure*}

We also compare the model phase-space density distribution of Equation \eqref{eq:phaseSpaceGaussian} with the simulated caustics, as shown in Figure \ref{fig:phaseSpaceGaussian}. The density contours of the model do not exactly track the distribution of mass in the caustic. This is because we obtained Equation \eqref{eq:phaseSpaceGaussian} by assuming that mass is evenly distributed along the stream, but this is not the case for real caustics.  However, both the density profile (obtained by collapsing along the $v_r$ axis) and the velocity profile (obtained by collapsing along the $r$ axis) match the simulated caustics quite well for the spherical potential, as shown for $v_r$ in Figure \ref{fig:phaseSpaceVelocityProjection}. There is a slight difference between the completely flat distribution in $v_r$ given by the model and the slightly peaked distribution seen in these two examples because the mass in the simulated stream is not necessarily evenly distributed.  The difference is more apparent in the case of the extremely flattened potential of Simulations C.  In the density distribution (Figure \ref{fig:phaseSpaceGaussian}), the caustic edge is sharper relative to the tail than can be represented in our model, thanks to the large spread perpendicular to the $r$--$v_r$. This means that the model must compromise on the caustic radius, moving it slightly too far back from the true edge. In the velocity distribution we see that the assumption that the material is evenly distributed is indeed not a very good one, since one side of the distribution now has slightly more material than the other. This asymmetry is also present in Figure \ref{fig:phaseSpaceGaussian}. However, the model contours in the $r$--$v_r$ plane and the width and edge sharpness of the model velocity distribution (Figure \ref{fig:phaseSpaceVelocityProjection}) are still fairly close to the simulation. Although the details of the phase space distribution are not perfectly encapsulated, our model can represent the main features of caustics formed from mergers quite far from the ideal case of a radial orbit in a spherical potential.

We have also not taken into account in our analytic model the distribution of the material in $v_\theta$ and $v_\phi$, instead assuming that all the material has zero angular momentum so $v_\theta=v_\phi=0$. In reality, even a satellite whose center of mass is on a perfectly radial orbit will contain material with a distribution of small angular momenta, and even in a spherical potential this will cover a distribution of orbital planes and tilt the distribution slightly out of the $(r,v_r)$ plane. In this simplest case, the tilt results in a linear relationship between $\theta$ and $v_\theta$ and between $\phi$ and $v_\phi$ near the caustic. The relationship is more complicated when angular momentum is added or the potential is not spherical, thanks to the precession of the orbits in these cases.  Unlike the caustic in the $r-v_r$ plane, the  behavior in these subspaces is much more dependent on the details of the particular interaction; we saw an example in the right-hand panel of Figure \ref{fig:phaseSpaceVelocityProjection}. In the $r-v_r$ plane the main effect is to blur the caustic, leading to an over-estimate of the width $\delta_r$ compared to the true stream thickness. By using the projected $r-v_r$ space as our full phase space, therefore, we are mainly losing the ability to measure $f_0$, which is related to the mass of the satellite galaxy. We will show in Section \ref{subsec:envfits} that this projection effect does not compromise our ability to measure the host galaxy's mass.

\subsection{Comparisons of model and simulated images}
\label{subsec:imageComparison}

\begin{figure}
 \includegraphics[width=0.48\textwidth]{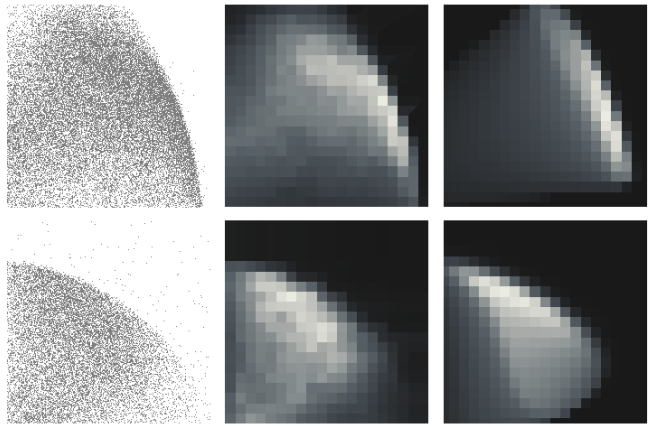}
\caption{The binned ``images'' (relative brightness) generated for views of two caustics from simulation D (left: points from Nbody simulation; center: binned surface density) can be adequately modeled by a projection of Equation \eqref{eq:generalDP} onto the sky plane using a spherical cone to model the geometry as described in the text (right). The caustic parameters $(r_s, \delta_r, \mathcal{A}, \alpha, \theta_s, \phi_s)$ used to generate the model image on the right in each case were obtained by a least-squares fit to the mock data in the center image.}
\label{fig:projSurfBright}
\end{figure}

In Section \ref{sec:images}, we showed how to derive a surface brightness distribution from our model.  Figure \ref{fig:projSurfBright} illustrates a test of how well model images fit two simulated caustics in simulation D. We binned the N-body realizations of the caustics (left column) to construct two surface-density distributions (center column), which were then least-squares fit to Equation \eqref{eq:sbxy} to obtain the parameters $(r_s, \delta_r, \mathcal{A})$, which describe the radial density distribution, and the parameters $(\theta_s,\phi_s,\alpha)$, which describe the orientation and angular extent of the shell. The best-fit models are shown in the right-hand column. Although the data clearly have smaller-scale complexity that is not accounted for in the model, the model is approximately consistent with the smooth features of the shell when comparing by eye. The surface-brightness fit is mainly used to provide an input geometry for fitting the kinematic data, from which $g_s$ is recovered, so a model that roughly describes the extent, variation, and orientation of the shell is sufficient.

One significant exception is seen in the top row of Figure \ref{fig:projSurfBright}.  Compared to the model, the caustic taken from the N-body simulation appears to have $r_s$ smaller toward the top of the frame. Correlations between the energy and orbital phase of the material in the stream, combined with the presence of angular momentum, will introduce a variation in the caustic radius with physical angle $r_s(\theta,\phi)$ which is not allowed for in the model. This variation can be comparable to $\delta_r$. This discrepancy can be addressed by allowing $r_s$ to vary, but since $\delta_r$ is mainly used as input into the kinematic models, this extra parameterization may not be needed.

\subsection{Tests of discrete line-of-sight velocity model}
\label{subsec:DLOScomparison}

In this section we describe two types of tests of the model projections developed in Section \ref{sec:vzmax}, to describe the maximum $v_z$ as a function of $R$ (Equation \ref{eq:vlosmax}) and the full projected distribution of the phase-space density in $(R, v_z)$ (Equation \ref{eq:projectedPhaseDensity}). First we test how well these two equations qualitatively match the $(R, v_z)$ distributions of the simulated caustics when the parameters of the model are determined by fitting the caustics in the $r$--$v_r$ plane (Section \ref{subsec:projModelComparison}). We also assess how the angle of the shell relative to the line of sight and the amount of angular momentum (Section \ref{subsec:projeffects}) affect the appearance of the projected distribution. Finally, we test how well the equation for $v_{z,\ \mathrm{max}}(R)$ can recover $g_s$ when fit to the simulated caustics (Section \ref{subsec:envfits}).

\subsubsection{Comparison of line-of-sight velocity model to simulations}
\label{subsec:projModelComparison}

\begin{figure*}
\begin{tabular}{cc}
\multicolumn{2}{c}{ \includegraphics[width=0.95\textwidth]{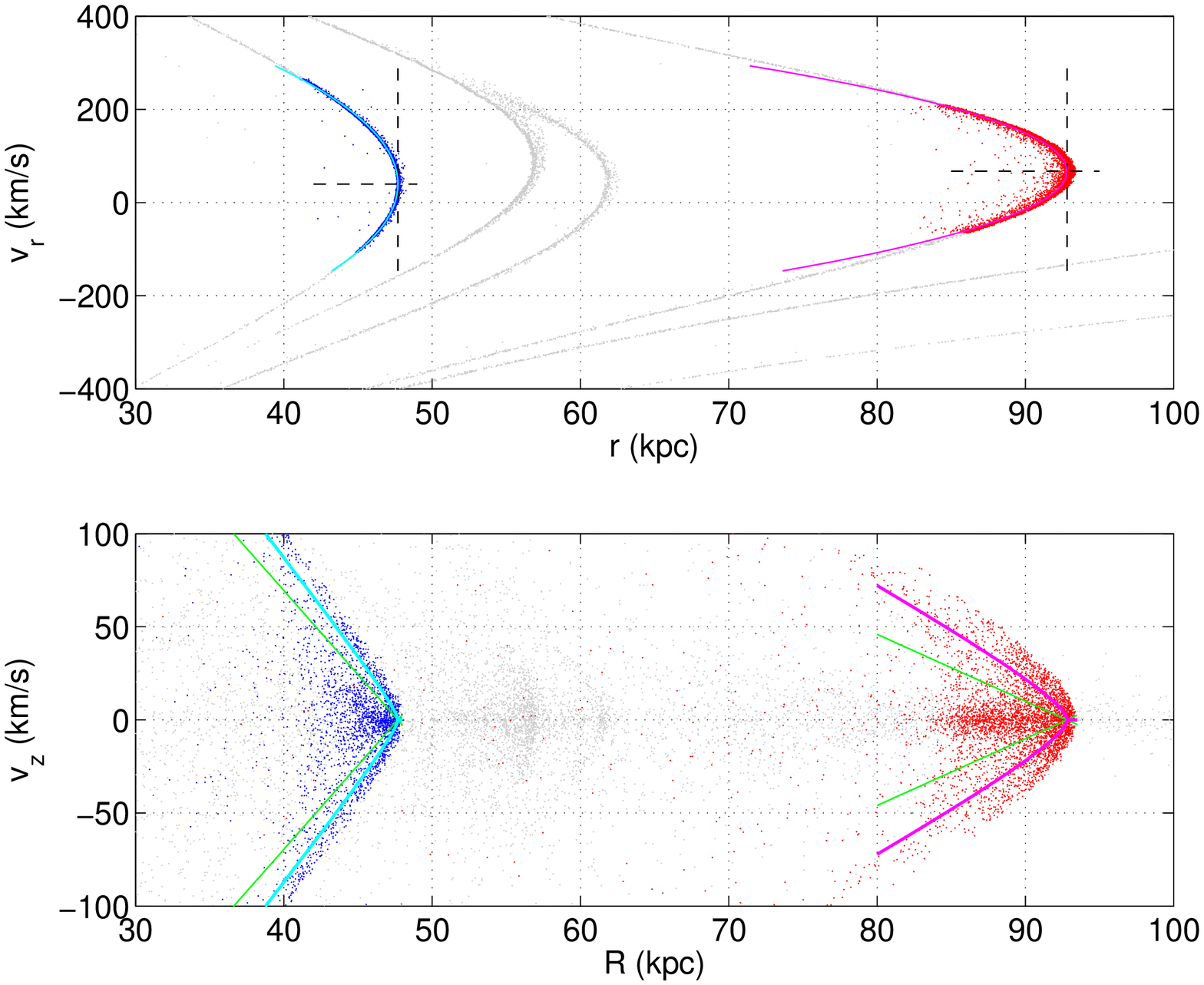}} \\
 \includegraphics[width=0.4\textwidth]{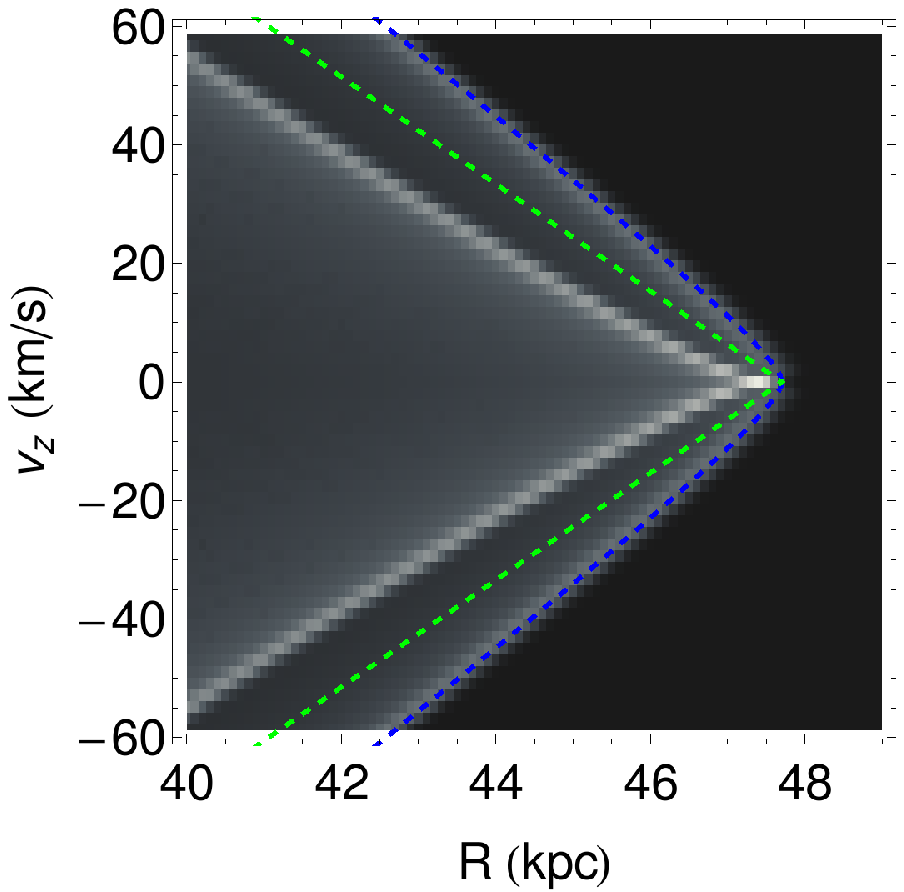} &  \includegraphics[width=0.4\textwidth]{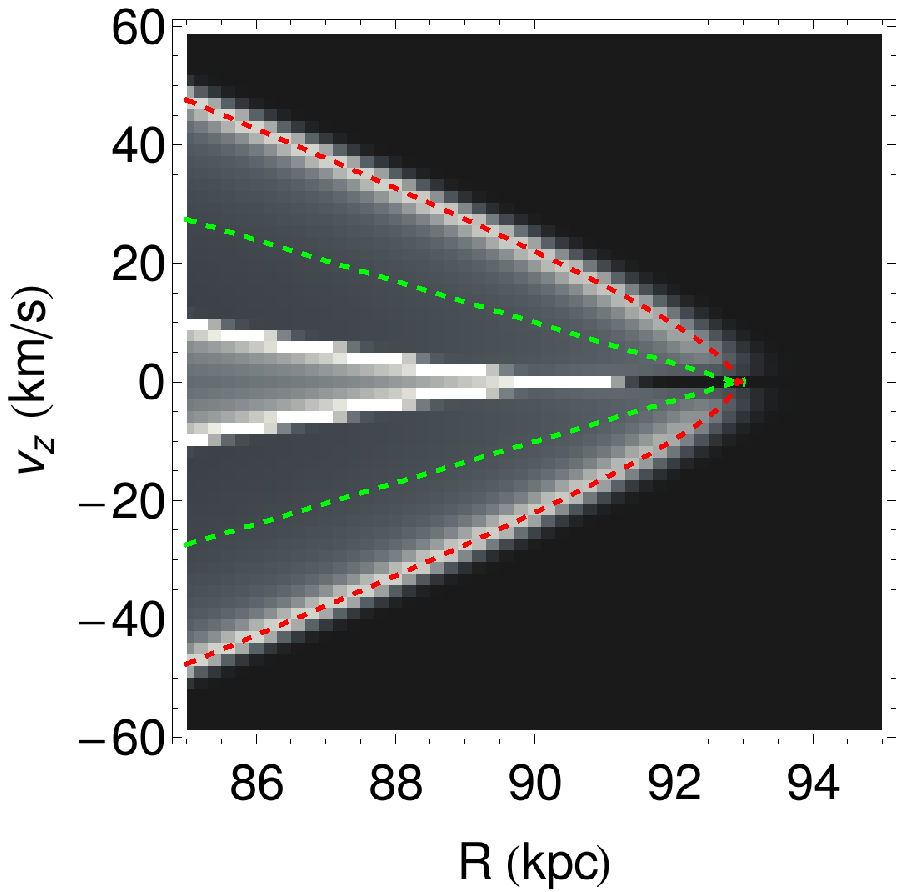} 
 \end{tabular}
\caption{Line-of-sight velocity profiles and phase space distributions for two caustics from simulation A. The galactocentric $(r,v_r)$ projection and the selected caustics are shown in the top panel (blue and red points) with fits to Equation \eqref{eq:rvrcurve} overplotted (cyan and pink lines). Dashed black lines cross at the point $(r_s,v_s)$ (see Equation \ref{eq:rvrcurve}). The middle panel shows the projected line-of-sight velocities $v_z$ versus projected distance $R$ in the plane of the sky overlaid with our derived expression for $v_{z,\ \mathrm{max}}$ (Equation \ref{eq:vlosmax}; cyan/pink) and the version derived by \citeauthor{Merrifield1998} that assumes $v_s=0$ (Equation \ref{eq:vlosMK}; green). The equations use values for $r_s$ and $v_s$ obtained from the fits in the top panel and $\kappa = 1/|2g(r_s)|$ calculated from the input potential as in Equation \eqref{eq:radialEnergy}. The bottom panel shows the same two equations (green and blue/red dashed, respectively) superposed on the projected density, $f(R,v_z)$ (Equation \ref{eq:projectedPhaseDensity}), for the same two caustics.}
\label{fig:TwoCausticsA}
\end{figure*}

We first consider simulation A, which satisfies all the assumptions used to derive the models. The top panel of Figure \ref{fig:TwoCausticsA} shows the system in the $(r,v_r)$ projection of phase space, with two of the four caustics identified in red and blue.  The simulation provides us access to the full phase-space information of the shell: as a first test of the model's ability to represent the observed quantities, we can get the parameters of $v_{z,\mathrm{max}}(R)$ from the complete phase space information and compare them in the observed space $(R,v_z)$. For this test we just analyze the inner- and outermost caustics to get the widest range of $g_s$ values, but the procedure would work the same for the other two as well; as seen in the top panel of Figure \ref{fig:TwoCausticsA} they also take on the same parabolic shape. To determine the parameters we fit Equation \eqref{eq:rvrcurve} to each caustic (solid cyan and magenta lines) to obtain values for $r_s$, $v_s$, and $g_s$; we also obtain  $\delta_r$ by fitting the radial density profile of each caustic (a histogram of the $r$ coordinates of the colored particles) with Equation \eqref{eq:generalDP}. We need $\delta_r$ because the spread around $r_s$ is not taken into account in our derivation of Equation \eqref{eq:vlosmax}, so the envelope should be shifted radially outward to the very edge of the distribution at $r_s + 2 \delta_r$. The parameters are given in Table \ref{tbl:ModelParams}.

\begin{table}
 \begin{tabular}{lll}
 Parameter & Value, Caustic 1 & Value, Caustic 2 \\
\hline
$r_s$ (kpc) & 92.8 & 47.7 \\
$\delta_r$ (kpc) & 0.32 & 0.15 \\
$v_s$ (km \unit{s}{-1}) & 67. & 40. \\
$|g_s|$ (km \unit{s}{-1} \unit{Myr}{-1}) & 1.2 & 4.0 
\end{tabular}
\caption{Parameters for the phase-space models of the two caustics shown in Figure \ref{fig:TwoCausticsA}, derived by fitting the full phase-space information as described in Section \ref{subsec:DLOScomparison}.}
\label{tbl:ModelParams}
\end{table}

\begin{figure*}
\includegraphics[width=0.95\textwidth]{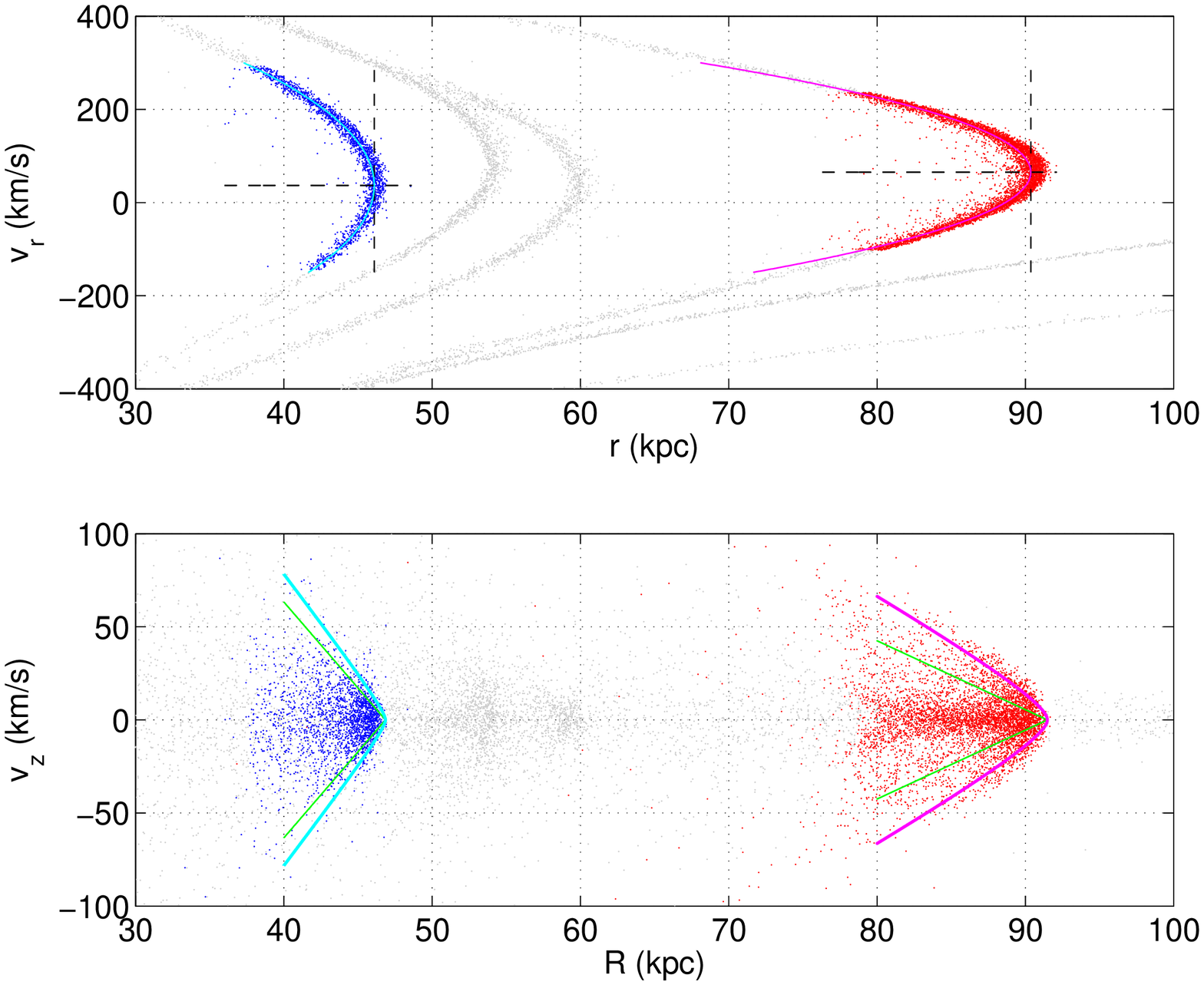}
\caption{As in Figure \ref{fig:TwoCausticsA}, but using the series B simulation with $\theta=20^\circ$. The nonzero angular momentum actually improves the agreement of the model and simulation in the $(R,v_z)$ plane somewhat.}
\label{fig:TwoCausticsB}
\end{figure*}

The center panel of Figure \ref{fig:TwoCausticsA} shows the same system in the observed space $(R,v_z)$. For both analyzed caustics it is clear that Equation \eqref{eq:vlosmax} (cyan/magenta), which allows a nonzero $v_s$, is a better description of the outer envelope than the relation derived by \citeauthor{Merrifield1998} which assumes $v_s=0$ (Equation \ref{eq:vlosMK}, green). However, some material still lies above the envelope described by the equation for $\sub{v}{max}$, especially for the outer caustic, even though we have adjusted for the spread of the material around $r_s$. This is a result of correlations in the perpendicular $\theta-v_\theta$ space that are specific to spherical potentials. The velocity $v_\theta$ varies linearly with $\theta$ along the stream thanks to the symmetry in this coordinate, is centered on zero near the caustic, and has a range of values comparable to $v_s$ in the range of interest in $R$. For this highly symmetric system, this correlation causes a rounding of the outer envelope and leads to a slight overestimate of $g$. If we examine a case where the angular momentum is nonzero (Simulation B, Figure \ref{fig:TwoCausticsB}), we see that the addition of angular momentum breaks the symmetry in the $\theta-v_\theta$ plane, leading, counterintuitively, to an improvement in the fit.

The lowest row of Figure \ref{fig:TwoCausticsA} shows the modeled projected density in the $R-v_z$ plane, Equation \eqref{eq:projectedPhaseDensity}, for each of the two modeled caustics. The model captures the interior structure in the density distribution that is seen in the N-body distribution (second row), including significant density enhancement in the interior of the outer (red) caustic in the projected space. In practice, the number of measurements in this space is likely to be fairly small, probably insufficient to distinguish the structure of the distribution by eye. Because the outer envelope is not filled with a constant density of points, a sparse sampling of measurements in this plane will not give a good estimate of $v_{z,\mathrm{max}}(R)$. Fitting Equation \eqref{eq:vlosmax} directly to an ensemble of measured line-of-sight velocities is therefore not a good way to determine the gravitational force. Instead, the measurements should be assigned probabilities for a given set of model parameters using Equation \eqref{eq:projectedPhaseDensity}. Then the parameters, including $g$, can be recovered by maximizing the total likelihood.

\begin{figure*}
\begin{tabular}{cc}
\includegraphics[width=0.45\textwidth]{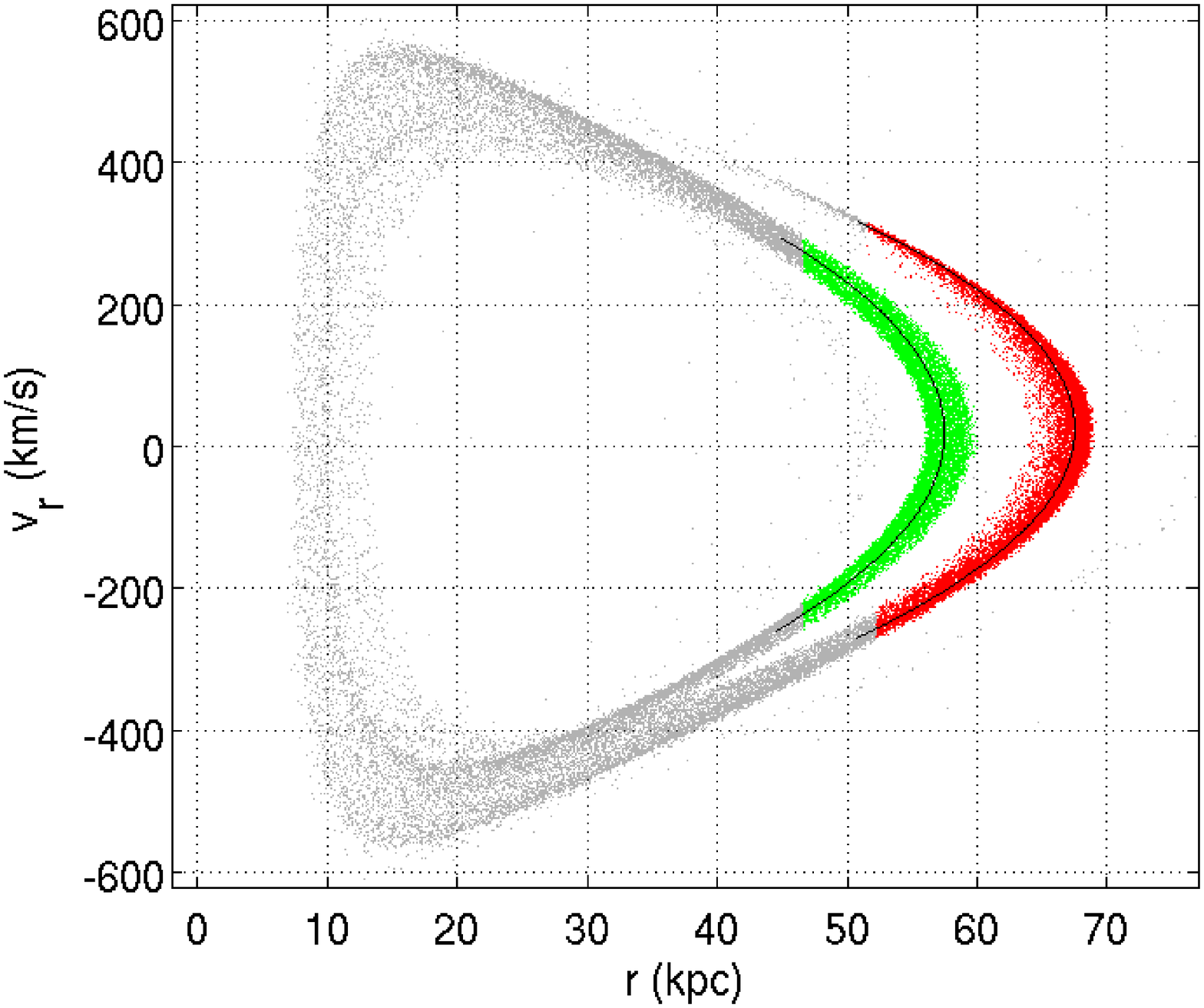} & \includegraphics[width=0.45\textwidth]{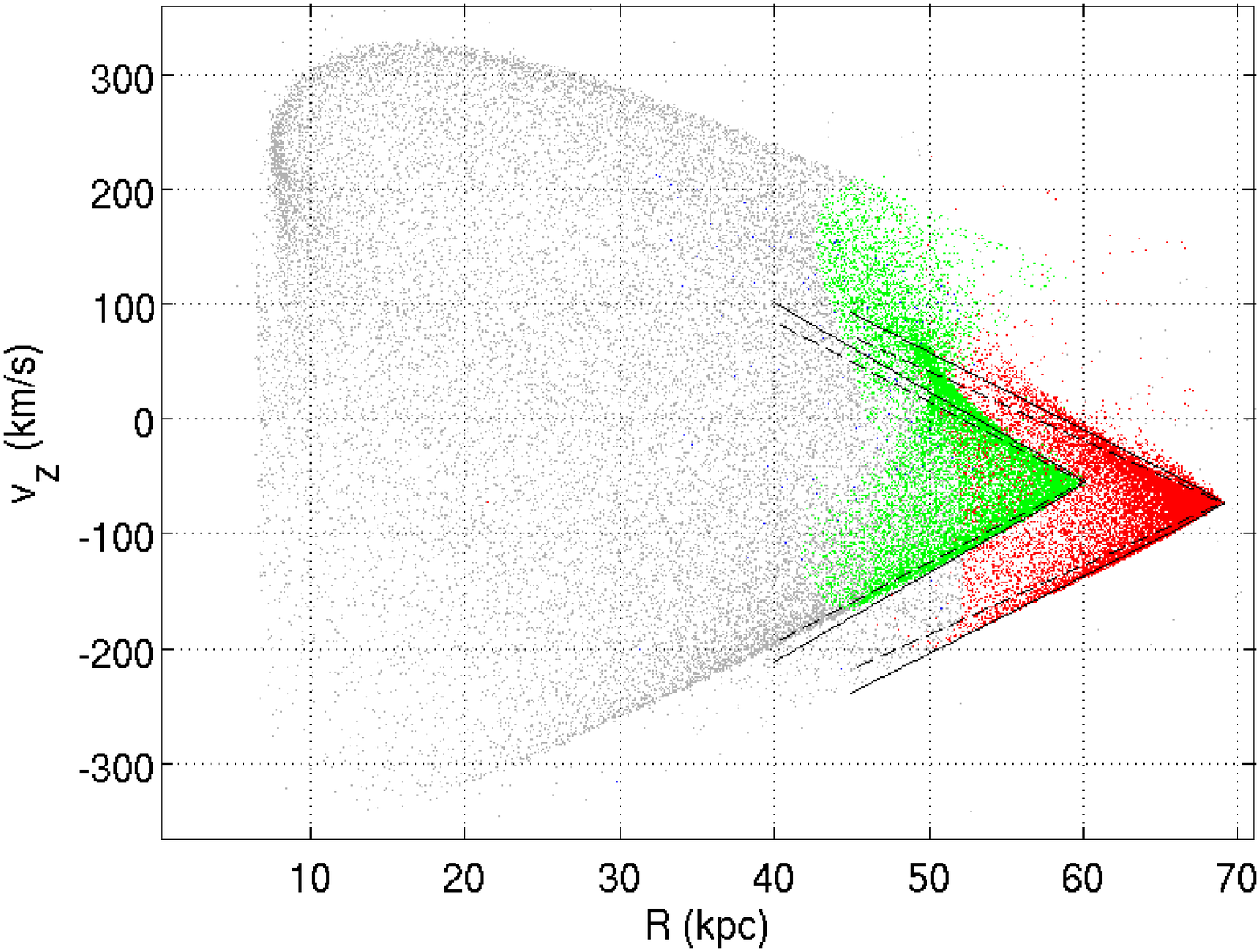} 
\end{tabular}
\caption{Two caustics from Simulation C (red and green points) with $\beta=30^\circ$ in the $(r,v_r)$ projection (left panel) and the $(R, v_z)$ projection (right panel). The caustics were selected and fit in the left panel (fits shown in black) to obtain the parameters used to model the envelope in the right panel: using our derived Equation \eqref{eq:vlosmax}, in solid black, and using the version with $v_s=0$ (Equation \ref{eq:vlosMK}), in dashed black.}
\label{fig:TwoCausticsC}
\end{figure*}

With Simulations C we identify the effects of a significantly flattened potential. Figure \ref{fig:TwoCausticsC} shows the $(r, v_r)$ and $(R, v_z)$ planes for the two caustics in the simulation with $\beta=30^\circ$. As before, we fit the shape in the former space (left panel) to predict the outer envelope shape in the latter space (right panel). However now the distribution in the $(R,v_z)$ plane is not centered around $v_z=0$ as it was in the previous examples. This is due to the satellite's angular momentum and the precession induced by the flattening of the potential, which give the caustic a nonzero bulk velocity along the line of sight. In the case of a spherical potential the displacement of the symmetry axis in $v_z$ has a simple dependence on $L$ and $v_s$, which we will discuss further in Section \ref{subsec:projeffects}. However for an axisymmetric potential, the displacement can be much larger and is potential-dependent, so in the right panel of Figure \ref{fig:TwoCausticsC} we have simply translated the envelopes by eye. Once this is done, our expression for the envelope (Equation \ref{eq:vlosmax}) describes the outer edge of the region near each caustic quite well; as before, ignoring $v_s$ is not as good a fit. 

\begin{figure*}
\begin{tabular}{cc}
 \includegraphics[width=0.45\textwidth]{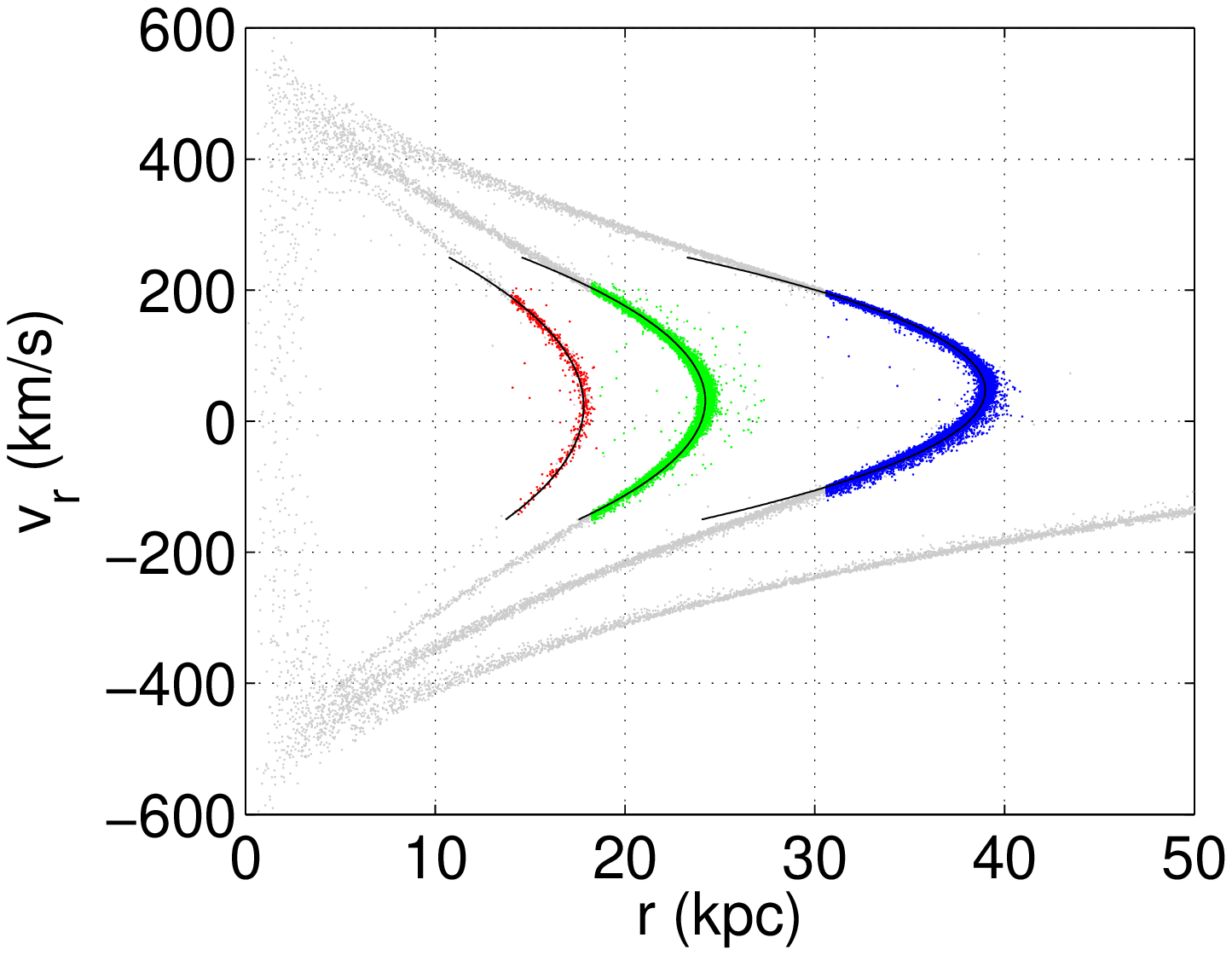} &\includegraphics[width=0.45\textwidth]{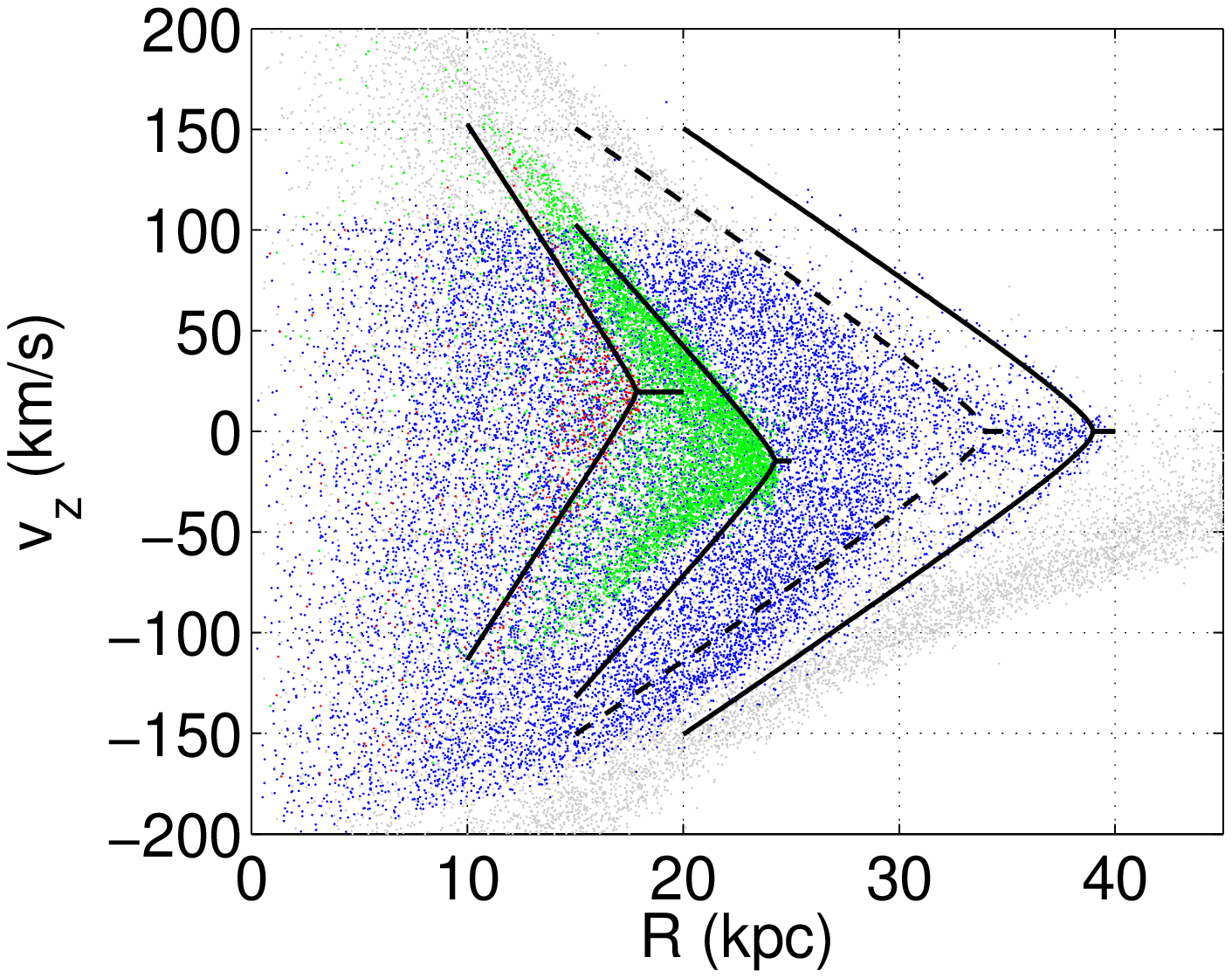} 
\end{tabular}
\caption{Views similar to Figure \ref{fig:TwoCausticsA}, but for simulation D. The red, green, and blue regions are the same as in Figure \ref{fig:simsxy}. The left panel shows the view in the $r-v_R$ plane with fits of Equation \eqref{eq:rvrcurve} to the colored regions as black lines. The right panel shows the $R-v_z$ plane with envelopes calculated using Equation \eqref{eq:vlosmax} and the values from the fits in the left panel; the two inner envelopes are translated by eye in the $v_z$ direction to fit the symmetry axis. The dashed line is a translation in $R$ of the envelope for the outermost caustic. See the text for further discussion.}
\label{fig:M31Projection}
\end{figure*}

Next we consider an example in which the potential has a realistic axisymmetric contribution from a disk: simulation D (Figure \ref{fig:M31Projection}). The caustics in simulation D combine three important differences from the highly symmetric situation in simulation A: they are not aligned exactly with the line of sight, they were formed with some small angular momentum, and the potential is not spherical. In the plane of the sky (leftmost panel), the outermost caustic, shown in blue, appears less sharp-edged than the next caustic, shown in green. As before, we fit the $r-v_r$ profiles of the caustics (middle panel) to obtain $r_s$ and $v_s$; in this case we also use the fitted value for $\kappa$ which is not far from the true value of $g(r_s)$.  In the projected space (right panel) we see again that the velocity profiles are not all symmetric around $v_z=0$, but the displacement can be adequately modeled by a constant, and the distributions are fairly symmetric around an offset velocity axis. We will show momentarily that these are primarily projection effects. Second, the outermost caustic is rotated enough to affect the velocity profile, since the bulk of the material does not satisfy the approximation $R \approx r$ (or $\theta_s \approx \pi/2$) used in the derivation. Nonetheless, the envelope fits the material that does satisfy this approximation, and when the envelope is translated in $R$ to encompass the bulk of the observed material (dashed black line) it is still a roughly accurate description. The difference between the very outer edge of the shell and the outer edge of the main part of the material is about 5 kpc, so the uncertainty in $R$ will not result in a large error in $g_s$. This suggests that Equation \eqref{eq:vlosmax}, with some modifications, should be able to recover $g_s$ for real shells.

\subsubsection{Projection effects}
\label{subsec:projeffects}
\begin{figure*}
\begin{tabular}{cc}
 \includegraphics[width=0.45\textwidth]{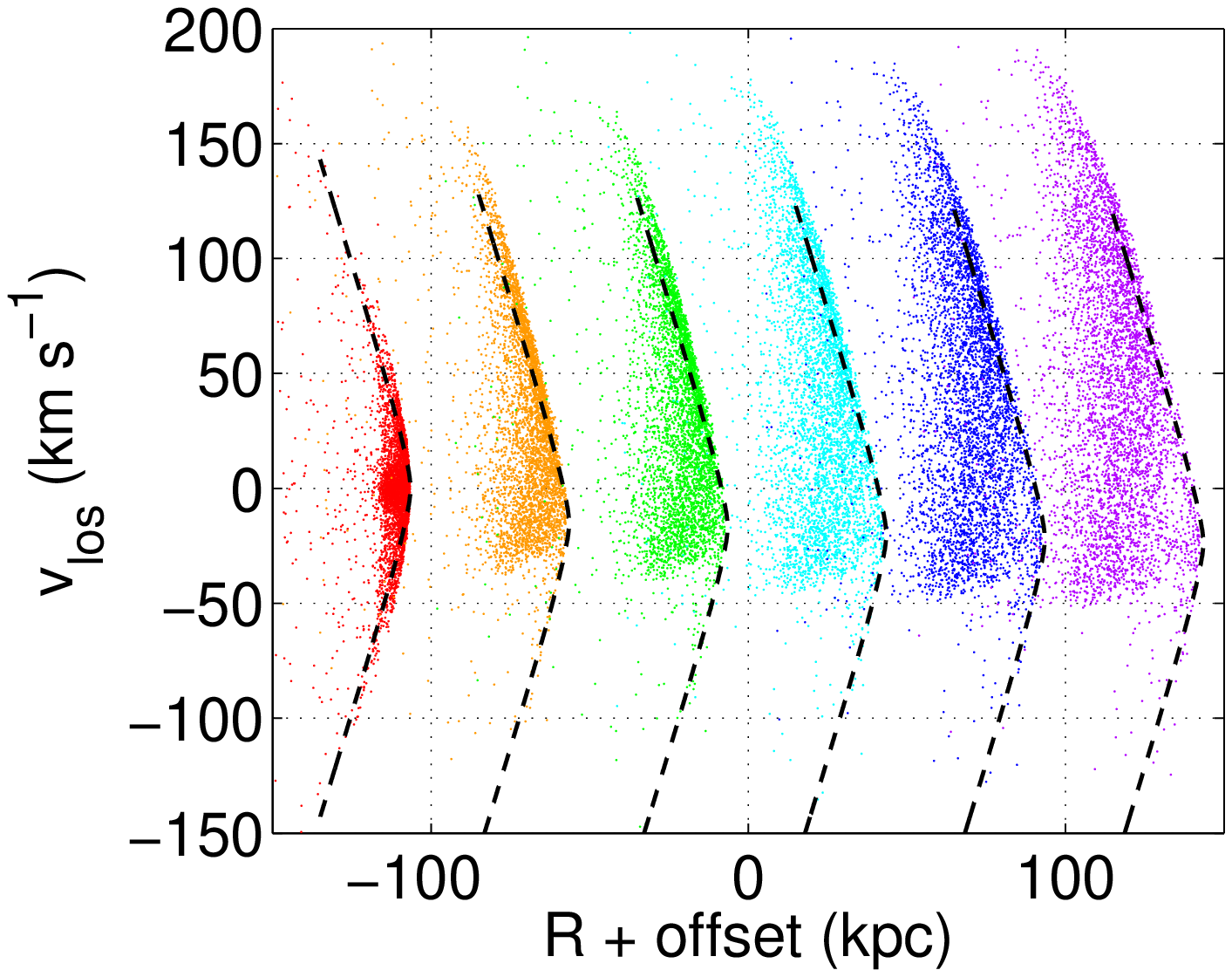} & \includegraphics[width=0.45\textwidth]{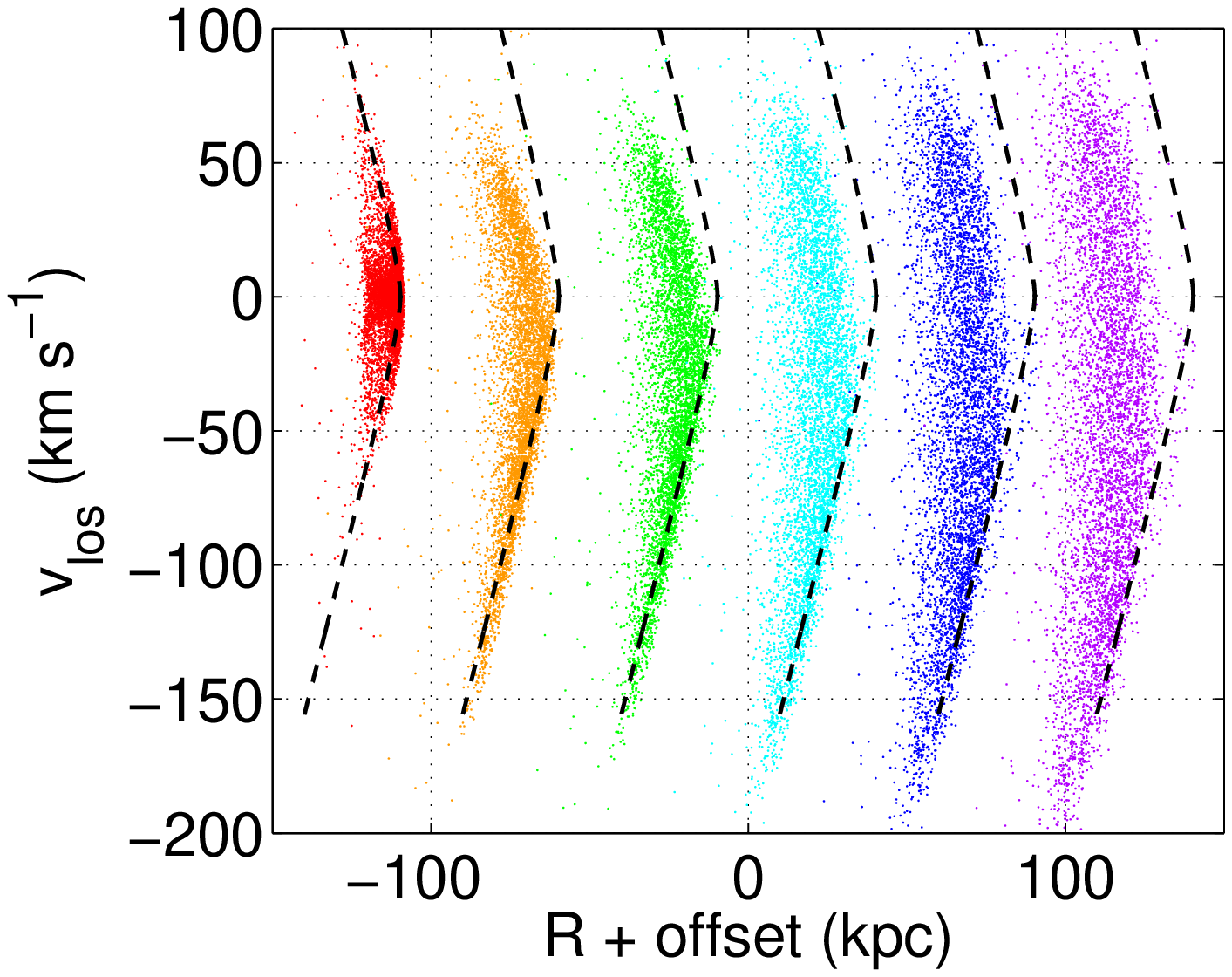}
\end{tabular}
\caption{Effect of shell orientation relative to the line of sight, $\theta_s$ (see Section \ref{subsec:projeffects}), on the $(R, v_z)$ plane, in simulations A (left) and B($\theta=20^\circ$) (right). The colors red through purple indicate $\theta_s = \{90, 63, 60, 53, 49, 45\}^{\circ}$. Each caustic is displaced along the $R$ axis by an arbitrary offset so they do not overlap. The dashed black lines are the envelope derived in Section \ref{subsec:envelope}, translated in $v_z$ by the amount defined in Equation \eqref{eq:vzoffset}. The $v_z$ shift and asymmetry are results of taking $\theta_s \ne 90^\circ$, but the envelope still fits the more fully populated side of the distribution.}
\label{fig:rotationAngles}
\end{figure*}

The orientation of the shell with respect to the line of sight can affect the view in the $(R,v_z)$ plane. In Figure \ref{fig:rotationAngles} we show the effect of rotating the outermost caustics from simulations A and B from a perfectly symmetric orientation ($\theta_s=90^\circ$) away from the observer ($\theta_s < 90^\circ$). All these examples have $\theta_s$ very different from $90^\circ$; at inclinations closer to $90^\circ$ there is not very much change to the distribution, but if the shells are nearly spherical (as these are) then even at relatively large inclinations they may still have a sharp edge in projection. We see that there are three main effects: the peak density is no longer on the $v_z$ axis, the distribution is no longer symmetric about the $v_z$ value defined by this point, and the internal structure seen in the distribution is smoothed out. 

The distribution is offset in $v_z$ because the average dot product with the shell expansion velocity $v_s$, and any net rotation in the line-of-sight direction, is now nonzero. In the figure we have modeled this shift by
\begin{equation}
\label{eq:vzoffset}
 \sub{v}{offset} = -\half\left(v_s + \frac{\sub{L}{los}}{r_s}\right) \cos \theta_s,
\end{equation} 
where the first term accounts for the shift from the shell velocity and the second for the rotation in the line-of-sight direction, $\sub{L}{los} \equiv \mathbf{L} \cdot \hat{z}$. As shown in the figure by the dashed lines, this model is a good predictor of the vertical shift for spherically symmetric potentials. In the case of simulation B, the angular momentum is in the $-z$ direction so that the two terms nearly cancel. This also causes the lower half of the distribution to be selected instead of the upper half for this case, even though in both simulations the viewing angles are the same.

The asymmetry about the $v_z-\sub{v}{offset}$ axis is a combined effect of the inclination angle and the finite angular extent of the shell---essentially, the line of sight now probes an asymmetric distribution of velocities. However, it is clear from the figure that the ``filled'' side of the distribution still has the same envelope as the symmetric distribution. Rotating the caustic in the other direction (toward the observer instead of away from them, $\theta_s > 90^\circ$) will ``fill'' the other side of the distribution instead, but the maximum value of $|v_z-\sub{v}{offset}|$ still follows the model. Additionally, for caustics inclined to the line of sight the projection smoothes out the interior structure of the distribution somewhat, which could be an advantage when sampling this plane with discrete sources since the sampling is then more uniform in the populated part.  In any case, the only asymmetry that the projection effect can produce is to cut out part of the filled area under the envelope. This may complicate efforts to fit the envelope, so an understanding of the geometry, especially the inclination angle $\theta_s$, is important since kinematic measurements are unlikely to populate the whole $(R,v_z)$ plane the way these simulations do.

\subsubsection{Recovery of the gravitational force from projected observables}
\label{subsec:envfits}
\begin{figure}
 \includegraphics[width = 0.45\textwidth]{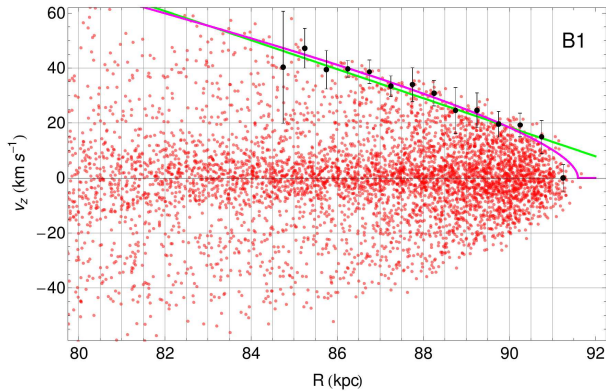}
\caption{Example of envelope data and fits for the outer caustic from simulation B($\theta=20^\circ$). The black points with error bars are obtained from the red points (the simulation particles) as described in Section \ref{subsec:envfits}; the green line is the fit to Equation \eqref{eq:vlosMK} and the pink line is the fit to Equation \eqref{eq:vlosmax}. This is one of the less well-fit cases, as indicated by the overshoot on the front edge of the distribution, which leads to a larger uncertainty on the recovered $g_s$.}
\label{fig:fakeEnvelopeData}
\end{figure}

Finally, we tested how well the model could recover the value of $g_s$ in the case of projected observables. For each caustic we need to fit the envelope of the points from the N-body simulation in the $(R,v_z)$ plane. To do so, we bin the points in $R$ and choose $\sub{v}{z,max}$ in each bin at the $\super{n}{th}$ percentile in $v_z$, where $n$ is usually between 95 and 100. We choose $n$ as large as possible in each case so that we get the edge of the envelope but are not contaminated by single outlier points above it. We estimate the errors on these $\sub{v}{z,max}$ points by looking at the difference in $v_z$ between roughly the $n$ and $n-2$ percentile. Finally, we add a point at the front edge of the distribution where $\sub{v}{z,max}$ goes to zero. This method is not meant to be representative of how real data would be analyzed (for one thing, it is not likely that we would have thousands of real velocity measurements to work with); it is just a system for obtaining the envelopes from the particle representations of the simulated caustics. An example is shown in Figure \ref{fig:fakeEnvelopeData}. We then carry out weighted least-squares fits to recover $g_s$ using both Equation \eqref{eq:vlosmax}, which we derived in Section \ref{sec:vzmax} allowing for a nonzero $v_s$, and Equation \eqref{eq:vlosMK}, which assumes $v_s=0$.  We compared the fit results to test which form gives results closer to the input value of $g_s$. In the case of Simulations C we compare the recovered value of $g_s$ to the radial derivative of the potential in the plane, since the debris in the caustics is roughly symmetric around $z=0$. 

\begin{figure}
 \includegraphics[width=0.45\textwidth]{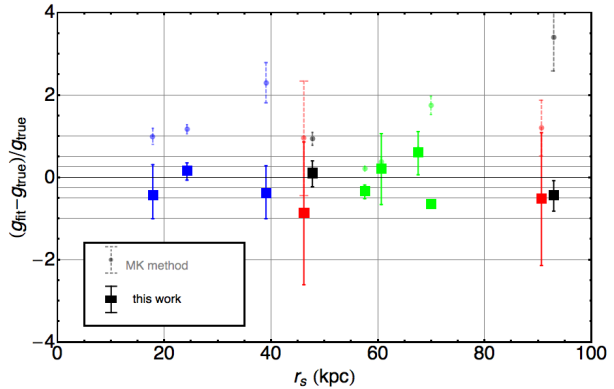}
\caption{Error on recovered values of the gravitational force, $g$, at the shell radius, $r_s$, for the caustics from simulations A-D pictured in Figures \ref{fig:TwoCausticsA}-\ref{fig:M31Projection} (black=A, red=B($\theta=20^\circ$), green=C($\beta=0^\circ$ and $\beta = 30^\circ$), blue=D). Each caustic was fit twice, once using Equation \eqref{eq:vlosMK} (the MK method; points with dashed error bars) and once using Equation \eqref{eq:vlosmax} (filled squares with solid error bars). The error bars are the projected 95 percent confidence intervals from the least-squares fits. For absolute relative errors less than 0.5, the gridlines are spaced in intervals of 0.25.}
\label{fig:gFromEnvelope}
\end{figure}

The results, shown in Figure \ref{fig:gFromEnvelope}, show that fitting Equation \eqref{eq:vlosMK} (open squares) consistently overestimates $g_s$ by large factors, up to a factor of 4 in some cases. Fits to Equation \eqref{eq:vlosmax} (filled circles) are consistently better at recovering $g_s$, although in a few cases the enormous error bars indicate that this can be a more difficult fit to converge.  For simulations C (green points) there are a few cases where there is a negligible difference between the fits of Equations \eqref{eq:vlosmax} and \eqref{eq:vlosMK}, and in some cases we see that the uncertainty has likely been underestimated by the fitting procedure. These are indications that our model is having some difficulty fitting the form of these caustics, which is unsurprising given the large difference in the velocity distribution that we saw in Figure \ref{fig:phaseSpaceVelocityProjection}, and the difficulty in determining the caustic radius and $\delta_r$, due to the spread perpendicular to the $(r, v_r)$ plane, that we discussed in Section \ref{subsec:phaseSpaceComparison}. Nevertheless, in most cases we can still recover $g_s$ within 50 percent using Equation \eqref{eq:vlosmax}, even with an extremely flattened potential. Most importantly, the degree of flattening contributed by the exponential disk in Simulation $D$, whose potential is fit to real data from M31, does not prevent the model from recovering $g_s$ to better than 50\%.

\section{An additional application of the model: spectroscopic line-of-sight velocity profiles}
\label{sec:spectra}

Spectroscopic measurements of integrated light use a slit or a set of fibers to record the line-of-sight velocity profile at a given location in the plane of the sky. We can calculate the expected velocity distribution in a slit or fiber from the phase-space distribution. As an example, we consider a single round fiber of the type that might be used in an integral-field unit. The center of the fiber is located at $(\sub{x}{fib},\sub{y}{fib})$ relative to the host galaxy center, and the fiber has a radius $\sub{\rho}{fib}$. The fiber position and radius are usually specified in angular units; we implicitly include a factor of the distance $D$ from the observer to the target, so that in the following analysis they have units of length. This is done so that $g(r_s)$, embedded in the phase space distribution, has the appropriate units. The supplied profiles can then be convolved using spectral simulator tools to produce the expected lineshape for fitting purposes.

For a circular fiber, the velocity profile is obtained by changing variables in Equation \eqref{eq:projectedPhaseDensity}, and then integrating over the line of sight and over two auxiliary variables $\xi$ and $\eta$ that span the area subtended by the fiber:  
\begin{eqnarray}
\label{eq:fiberProfile}
&& f(v_z) = 2 \mathcal{F}_0 \alpha \int_{-\sub{\rho}{fib}}^{\sub{\rho}{fib}} d\xi \int_{-\sqrt{\sub{\rho}{fib}^2 - \xi^2}}^{\sqrt{\sub{\rho}{fib}^2 - \xi^2}} d\eta \int_{\zmin(\xi,\eta)}^{\zmax(\xi,\eta)} dz  \nonumber \\
&& \exp\left\{-\left[r_s -  \sqrt{R^2 + z^2} - \kappa (\frac{R^2 v_z}{r z} + \frac{z v_z}{r}-v_s)^2\right]^2/2 \delta_r^2\right\}, \nonumber \\
\end{eqnarray}
where
\begin{equation}
 R \equiv \sqrt{(\sub{x}{fib}+\xi)^2 + (\sub{y}{fib}+\eta)^2}
\end{equation} 
and
\begin{equation}
 r \equiv \sqrt{(\sub{x}{fib}+\xi)^2 + (\sub{y}{fib}+\eta)^2 + z^2}.
\end{equation}  
A different spectroscopic geometry, e.g. a slit instead of a fiber, would simply have a different integration range for $\xi$ and $\eta$.

We assume that the fiber is small enough that $\sub{\rho}{fib} \lesssim \delta_r$. For most shells this probably sufficient; for example, a 4" fiber has a fiber radius of about 0.5 kpc at a distance of around 25 Mpc, comparable to $\delta_r$ for the shells in our simulations (see Table \ref{tbl:ModelParams}). This assumption allows us to replace the limits of the line-of-sight integral with their central values $\super{\sub{z}{min}}{max}(\sub{x}{fib},\sub{y}{fib})$, and approximate the $\xi$ and $\eta$ integrals with $\pi \sub{\rho}{fib}^2$ times the central value of the integrand. Then Equation \eqref{eq:fiberProfile} reduces to a single integral over the line of sight:
\begin{eqnarray}
\label{eq:fiberProfileApprox}
 f(v_z) &=&  2\pi \sub{\rho}{fib}^2 \mathcal{F}_0 \alpha \int_{\zmin}^{\zmax} dz \exp\left\{-\left[r_s -  \sqrt{\sub{R}{fib}^2 + z^2} \right. \right. \nonumber \\
&& \left. \left. - \kappa \left(\frac{\sub{R}{fib}^2 v_z}{z\sqrt{\sub{R}{fib}^2 + z^2}} + \frac{z v_z}{\sqrt{\sub{R}{fib}^2 + z^2}}-v_s \right)^2\right]^2/2 \delta_r^2 \right\}, \nonumber \\
\end{eqnarray} 
where
\begin{equation}
 \sub{R}{fib} \equiv \sqrt{\sub{x}{fib}^2 + \sub{y}{fib}^2}.
\end{equation} 

\begin{figure*}
\includegraphics[width=\textwidth]{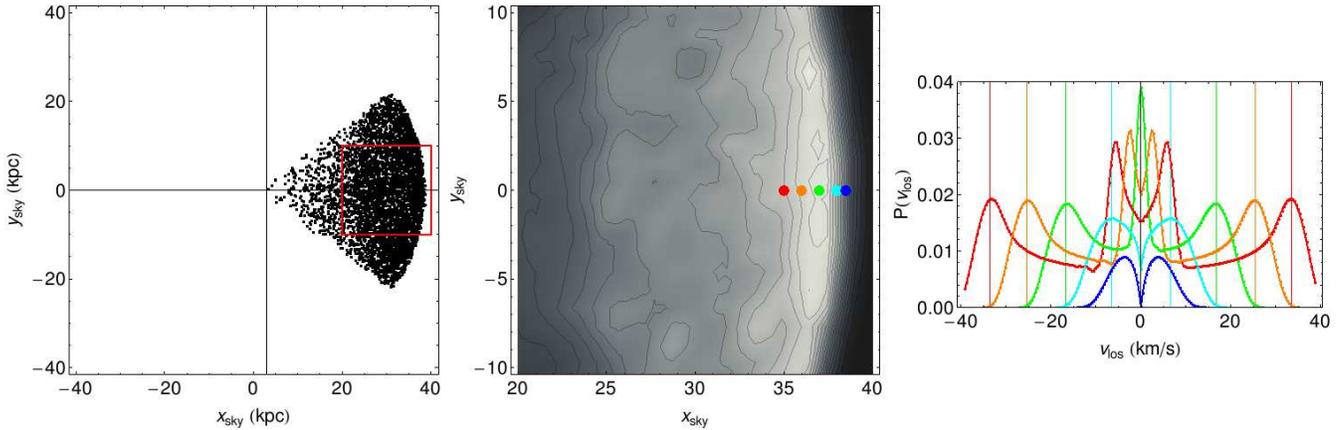}
\caption{Equation \eqref{eq:fiberProfile} can be implemented numerically to predict the velocity distribution for different lines of sight through a shell. Here we illustrate using a particle realization (left) of a shell modeled on the one shown in green in Figure \ref{fig:M31Projection}. The different lines of sight (center; colored circles) are overlaid on a surface-brightness map of the model convolved with a Gaussian of $2\sub{r}{fib}$; the red box in the right-hand panel indicates the area shown in the center panel.  This example uses a fiber radius of 0.5 kpc, equivalent to 4 arcsec at 25 Mpc, shown to scale in the colored circles of the center panel. The different lines of sight produce different velocity profiles (right) based on their distance from the shell edge. Farther from the edge the profile has two pairs of peaks, while close to the shell edge only one pair of peaks is visible.  The colored vertical lines mark the outer peak locations predicted by Equation \eqref{eq:vlosmax} as described in the text.}
 \label{fig:SingleFiberLinesOfSight}
\end{figure*}

For some lines of sight this distribution has four symmetric peaks, while others closer to the shell have a double peak  (Figure \ref{fig:SingleFiberLinesOfSight}). The shape of the profile depends on the distance from the shell edge, the shell thickness, the orientation of the shell relative to the line of sight, and the angular span of the shell. Four peaks will be observed when the line of sight probes radii farther from the shell edge, where the two values of $\pm|v_r-v_s|$ are widely separated (see Figure \ref{fig:angles}). As long as $|\theta_s-\pi/2| <\alpha$, most lines of sight will intersect the shell edge twice, once on each side of $z=0$. Each intersection produces a pair of peaks in the velocity distribution on either side of $v_z=0$, corresponding to $v_z$ of the incoming and outgoing streams. The signs of $v_z$ for these two streams are reversed on opposite sides of $z=0$, producing a symmetric distribution with four peaks.

For lines of sight closer to $r_s$, $|v_r-v_s|$ becomes smaller and each bifurcated pair of peaks will merge into a single one, with the profile remaining symmetric about $v_z=0$. The distance from the shell edge where this occurs depends on the shell thickness, but also on fiber size and spectral resolution. An asymmetric, two-peaked distribution is produced for lines of sight that intersect the shell edge only once instead of twice.

The spectral line shapes shown in Figure \ref{fig:SingleFiberLinesOfSight} are similar to those presented in \citet{EbrovaIvana2012}, but include the broadening and changes in peak position induced by the range of energies of the stars in the shell. Their velocity profiles can be obtained from our expression by taking the limit $\delta_r \to 0$, or equivalently using Equation \eqref{eq:phaseSpaceSingleE} for the distribution function instead of Equation \eqref{eq:phaseSpaceGaussian}.

The locations of the centers of the outer peaks are given by $\pm\sub{v}{max}(R)$ as in Equation \eqref{eq:vlosmax}, but since the velocity profile convolves the physical density of the material with its shape in $(r,v_r)$ space, we should replace $r_s$ in the equation with $\rmax$, the radius of peak density defined in Equation \eqref{eq:rmax}. Thanks to the stream thickness these two radii differ slightly. The outer peak locations computed this way are marked in the left-hand panel of Figure \ref{fig:SingleFiberLinesOfSight}; the colors match the respective lines of sight marked in the center panel. The interior peaks' locations can likewise be computed by letting $v_s \to -v_s$ in Equation \eqref{eq:vlosmax} and again using $r_s \to \rmax$.  So one way to measure $g_s$ from spectra is similar to that discussed in Section \ref{sec:vzmax} for discrete measurements, except that since the spectra probe the integrated light they will automatically measure the correct $\sub{v}{max}$. However, the caveats about potential flattening and angular span discussed above still apply: for this reason it is important to understand the geometry of the shell by fitting the surface brightness profile in order to judge whether Equation \eqref{eq:vlosmax} is applicable.

\begin{figure*}
\begin{tabular}{cc}
\includegraphics[width=0.7\textwidth]{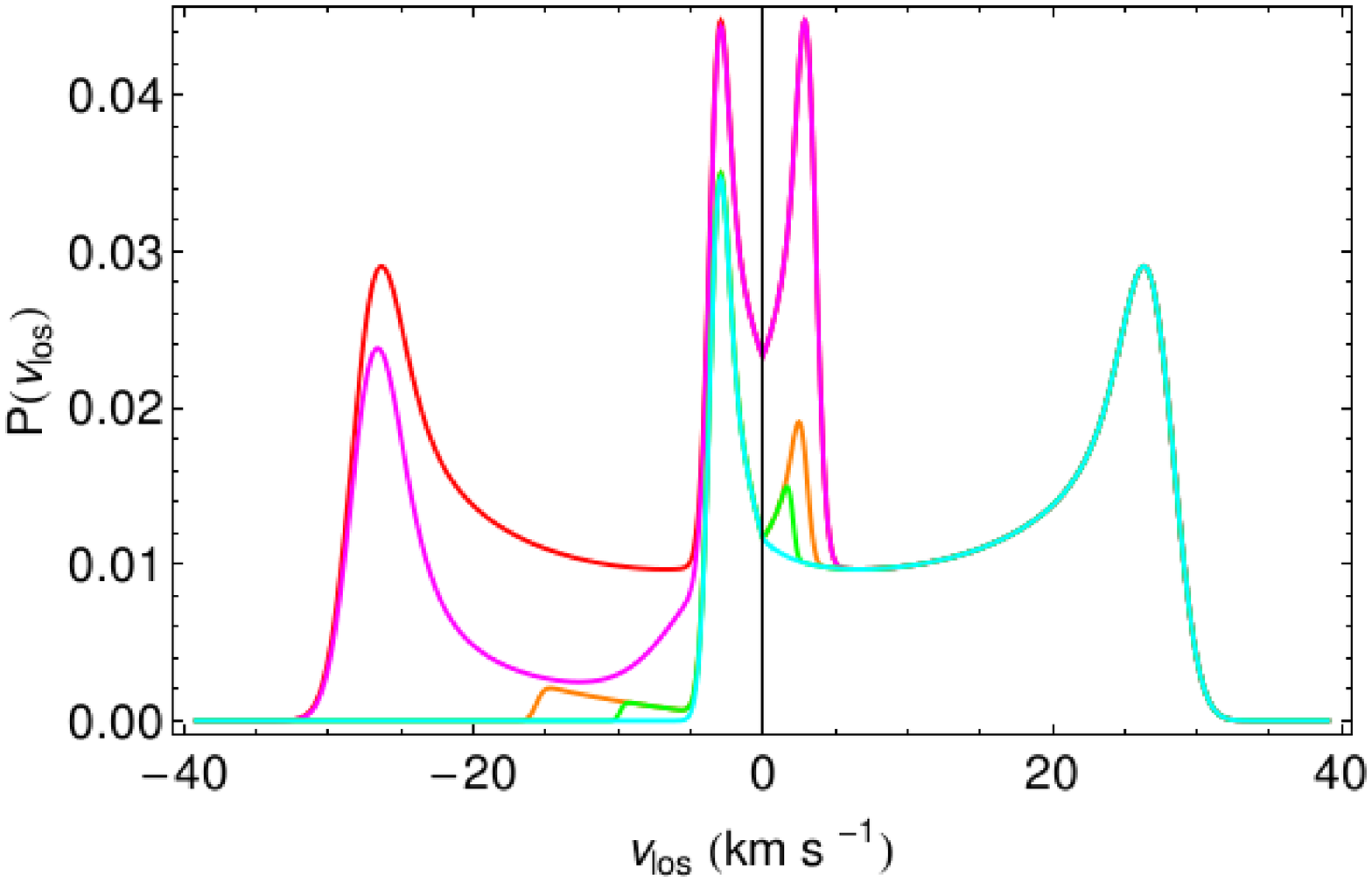} & \includegraphics[width=0.3\textwidth]{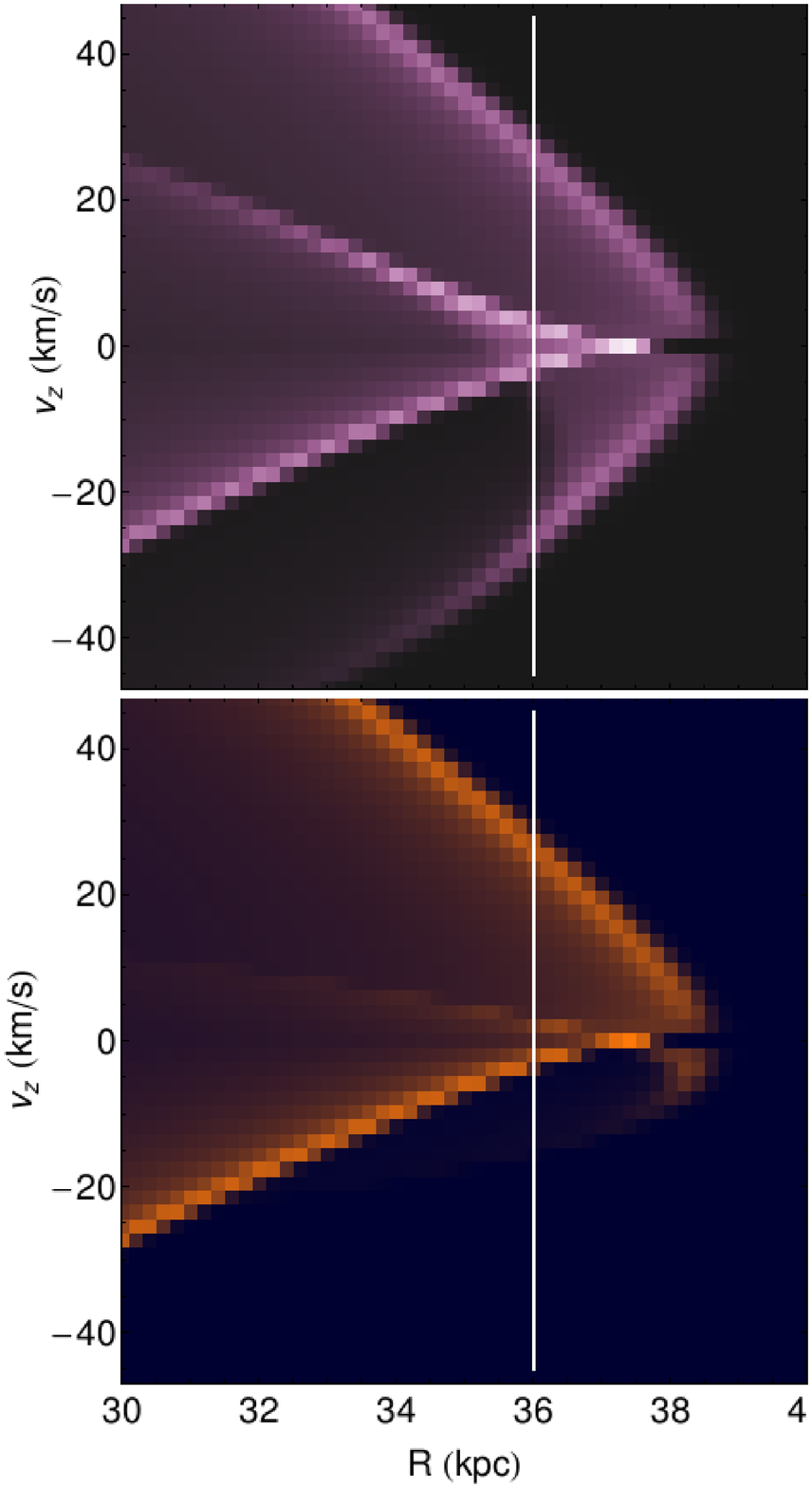}
\end{tabular}
\caption{The effect of the inclination of a shell with opening angle $\alpha=34^\circ$ relative to the line of sight on the integrated velocity profile is shown in the left panel for the inclination angles $\theta_s = \{90, 75, 63, 60, 53\}^{\circ}$ ($\theta_s$ decreasing away from the observer) in \{red, magenta, orange, green, cyan\}; the colors in the plot correspond with those in Figure \ref{fig:rotationAngles}. For $\theta_s<90-\alpha=56^\circ$ the infalling and outgoing streams in the shell are intersected by the line of sight only once instead of twice, and the profile is given by the cyan curve. For $\theta_s>90^\circ$ the effect is the same, but mirrored about the $\sub{v}{los}$ axis. The red profile is normalized; the others are scaled by the same normalization to show the reduction relative to the $\theta_s=90^\circ$ case. On the right are shown the phase-space density distributions for $\theta=75^\circ$ (top) and $\theta=63^\circ$ (bottom), with the line of sight marked in white.}
\label{fig:SingleFiberVaryThetaC}
\end{figure*}

The inclination and angular span of the shell have an effect on the integrated velocity profile just as they do on the distribution in $(R,v_z)$. In certain cases the limited angular span of visible material restricts the velocity profile so that only one of the two pairs of peaks (one of the two intersections with $r_s$) is probed. For $\theta_s$ near $\pi/2$, the line of sight is roughly symmetric around $z=0$ so the two pairs of peaks are both the same height, but for inclinations much different from this, lines of sight can intersect only some of the density maxima and cut off others, as shown in Figure \ref{fig:SingleFiberVaryThetaC}. Knowing the geometry of the shell in advance from image fitting is therefore a great advantage when deciding where to place fibers, since it is much easier to measure $\sub{v}{z,max}$ if the distinctive four-peaked velocity profile can be identified.

\begin{figure}
\includegraphics[width=0.5\textwidth]{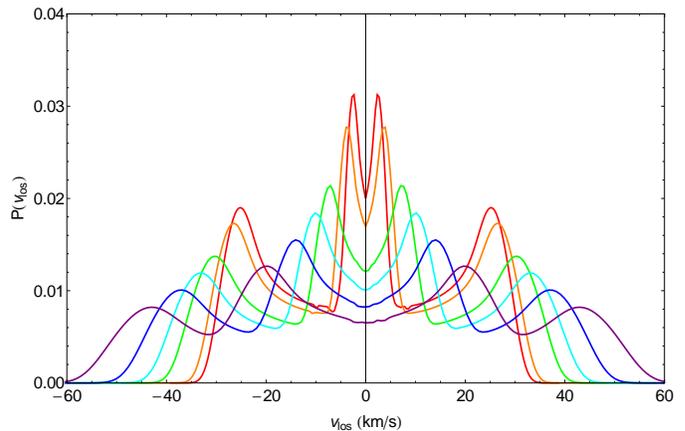}
\caption{Changing the value of $g(r_s)$, for an otherwise constant set of shell parameters and constant line of sight (along the symmetry axis of the shell near its edge), alters the velocity profile. The colors from red to purple are the values $g(r_s)=\{1.2,1.5,2.5,3.4,4.9,7.6\}$ km \unit{s}{-1} \unit{Myr}{-1}, corresponding to the gravitational force at $r=\{90,80,60,50,40,30\}$ kpc in the potential of simulations A-C and the range $r=10-35$ kpc in the potential of simulation D. Stronger gravity results in a broader profile and lower peaks. The shell is the same model as in Figure  \ref{fig:SingleFiberLinesOfSight}; the orange profiles in the two figures correspond.}
 \label{fig:SingleFiberVaryG}
\end{figure}

 The profile in even a single fiber is sensitive to $g(r_s)$ (Figure \ref{fig:SingleFiberVaryG}), so $g_s$ can in principle be recovered by fitting the full velocity profiles of even a handful of lines of sight, even if there are too few to get the variation of the peak widths and fit Equation \eqref{eq:vlosmax}.  Fitting the lineshape also allows the combination of multiple fibers or the use of a few slits to improve the data quality.  We will assess the quantitative ability of this strategy to recover $g_s$ in future work.

\section{Discussion and Conclusions}
\label{sec:conclusion}

In this paper we have presented a simple analytical model that successfully describes tidal shells with a small number of parameters, some of which are linked to the masses of the two interacting galaxies.  The model includes self-consistent expressions for the radial density profile (Section \ref{subsec:udp}) and $(r, v_r)$ phase-space distribution (Section \ref{subsec:PDF}) of material near the edge of a single tidal shell. Given an image of a shell (Section \ref{sec:images}), the surface-brightness model jointly constrains the fine-grained phase-space density of the satellite material, $f_0$, and the gravitational force at the shell edge through the parameter $\kappa$, which is related to $g_s$ as discussed in Section \ref{subsec:udp}.  The gravitational force $g_s(r_s)$ can be independently measured if line-of-sight velocities are also obtained. These can be either the velocities of individual point sources in a shell (Section \ref{sec:vzmax}) or velocity profiles obtained from integrated-light spectra (Section \ref{sec:spectra}). If multiple shells are observed around a galaxy, each one gives an independent measurement of the gravitational force at the radius of its edge.

The model makes three assumptions about the two interacting galaxies: that the interaction orbit is radial, that the potential is spherical at the shell radius with negligible tides, and that the satellite originally had a Maxwellian velocity distribution. To test these assumptions we performed a series of N-body simulations (Section \ref{sec:sims}) that formed tidal caustics while relaxing one assumption at a time. Using these simulations we demonstrate in Section \ref{subsec:assumptionTests} that the assumptions of zero angular momentum, zero tides, and spherical symmetry in the potential are sufficient for caustics that appear as fan-like shells on the sky. The choice of a satellite velocity distribution affects the distribution of mass along the stream, which is assumed to be constant in our model. In Section \ref{subsec:phaseSpaceComparison} we show that variations in the bulk velocity profile from the slightly nonuniform density along simulated streams are minor. 

We further test the model by comparing images (Section \ref{subsec:imageComparison}) and line-of-sight velocity distributions (Sections \ref{subsec:DLOScomparison} and \ref{subsec:projModelComparison}) generated by the model with the caustics from our simulations. With the exception of extremely flattened potentials, the model succeeds in reproducing the main features of caustics and permits recovery of $g_s$ within 50 percent by fitting the shape of the outer envelope of line-of-sight velocities (Section \ref{subsec:envfits}). 

We also include a discussion of a further application of the model to calculate velocity profiles for integrated-light spectra (Section \ref{sec:spectra}). Although we have not yet compared these profiles to simulations, they are likely a more promising route to the potential than pointwise line-of-sight velocity measurements because each line of sight samples the complete velocity distribution at a given location in the shell, and each resulting velocity profile is individually sensitive to $g_s$.

One possible caveat to this work is that even for high-mass-ratio mergers like those studied here, the use of a static potential is not necessarily justified during the disruption of the satellite at pericenter, where the enclosed mass of the host galaxy can be comparable to the satellite mass. Previous work \citep[e.g.,][]{1996A&A...310..757S} has shown that including this response results in slightly thicker, more massive caustics at slightly smaller radii, but does not destroy the shells. In addition, \citet{2008ApJ...674L..77M} performed Simulation D with a live representation of M31's potential and found that the same shells-and-stream system formed in that case. The halo response can prevent the use of the shell \emph{radii} to measure the enclosed mass, since it alters the absolute energies of the stream stars during stripping, but the phase space curvature that we measure in this work is sensitive to the force at \emph{apocenter} where the halo back-reaction is minimal. This suggests that shells are not fundamentally different in live host galaxies and that our results should extend to these cases.

Finally, each shell observed around a galaxy provides its own mass estimate at its own radius, without having to assume anything about the relationship between different shells. In galaxies with many shells, therefore, these independent estimates would produce a rough dynamical mass profile of the host galaxy at otherwise inaccessible locations outside the luminous disk or ellipsoid. By using their distinctive morphology as a clue to their symmetry, shells will open for us a new window into the shapes and masses of external galaxies.

\section{Acknowledgements}
The authors gratefully acknowledge support from the European Research Council under ERC-Starting Grant GALACTICA- 240271. RES thanks Ed Bertschinger, Roya Mohayaee, David Martinez-Delgado, Kyle Westfall, \& Aaron Romanowsky for helpful discussions and feedback. The N-body simulations used in the examples in this paper used code provided by Will Farr, and were run on either the Kapteyn computer network or the MIT Kavli Institute High-Performance Computing Cluster. 

\bibliography{TidalCaustics}

\appendix

\section{Potentials used in simulations}
\label{appx:potls}

Simulation A and Series B (see Table \ref{tbl:sims}) use the spherical isochrone potential with total mass $M$ and scale radius $b$ described by 
\begin{equation}
\super{\sub{\Phi}{iso}}{sph}(r) = -\frac{G M}{b + \sqrt{r^2 + b^2}},
\end{equation}
where $r$ is the spherical radial coordinate and $G$ the gravitational constant.

\begin{figure}
\includegraphics[width=0.45\textwidth]{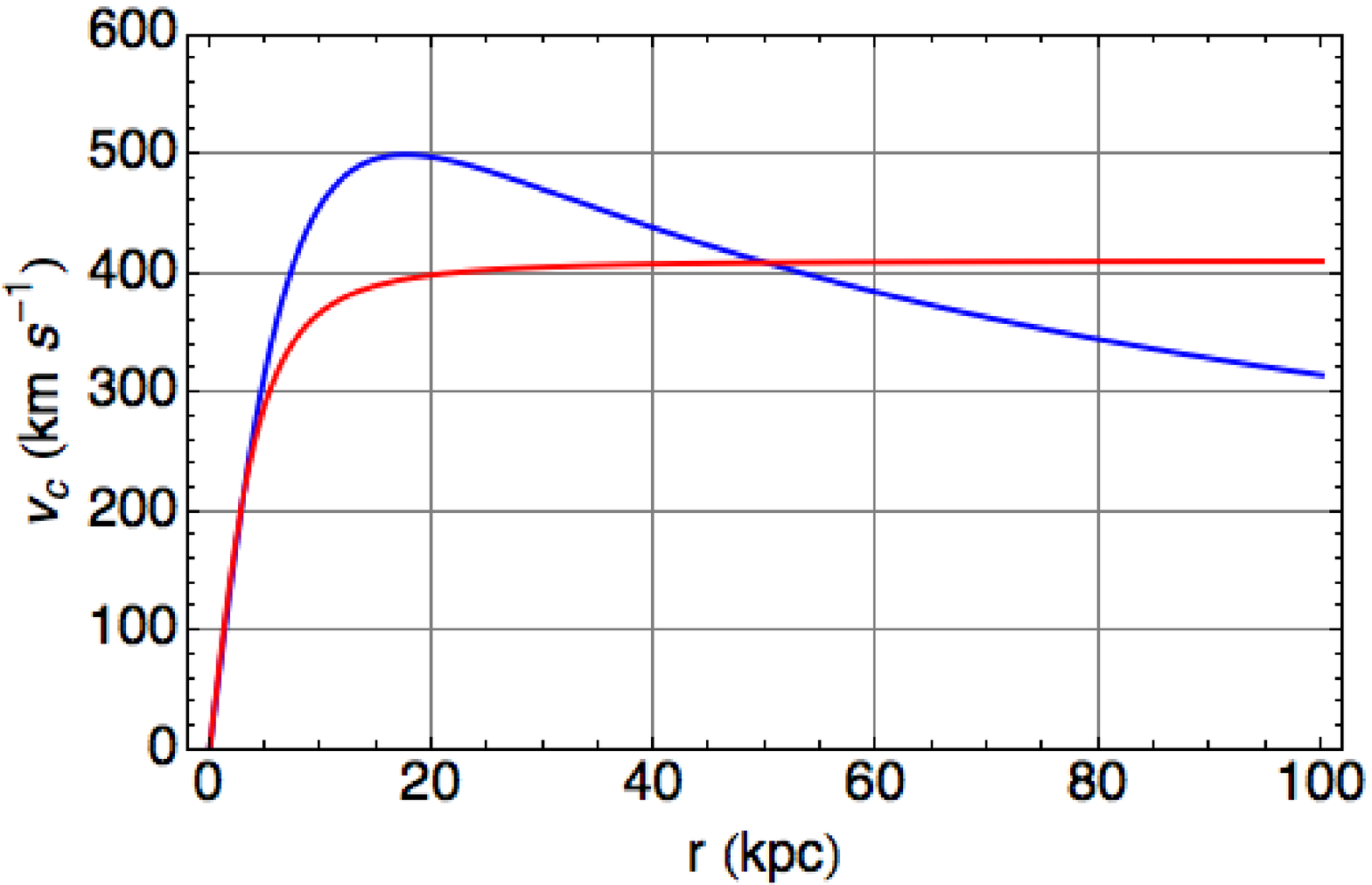}
\caption{Rotation curves for the isochrone potential used in Simulations A and B (blue) and the logarithmic potential used in Simulations C (red). The logarithmic curve is shown for the case $q=1$, or equivalently in the plane for $q \neq 1$, and is the best fit to the isochrone curve in the displayed region.}
\label{fig:rotationCurveFit}
\end{figure}

\begin{figure*}
\begin{tabular}{ccc}
\includegraphics[width=0.3\textwidth]{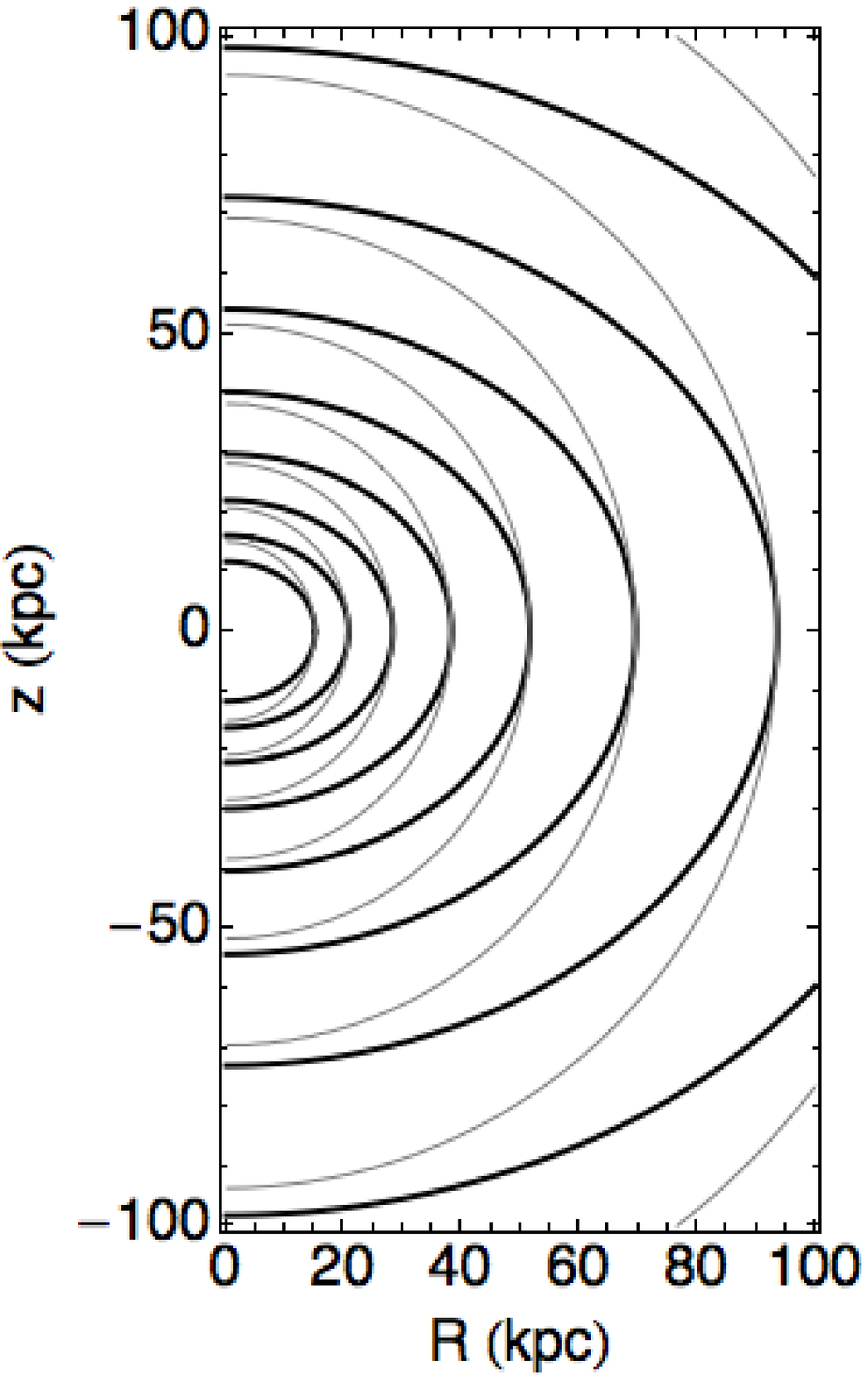} & \includegraphics[width=0.3\textwidth]{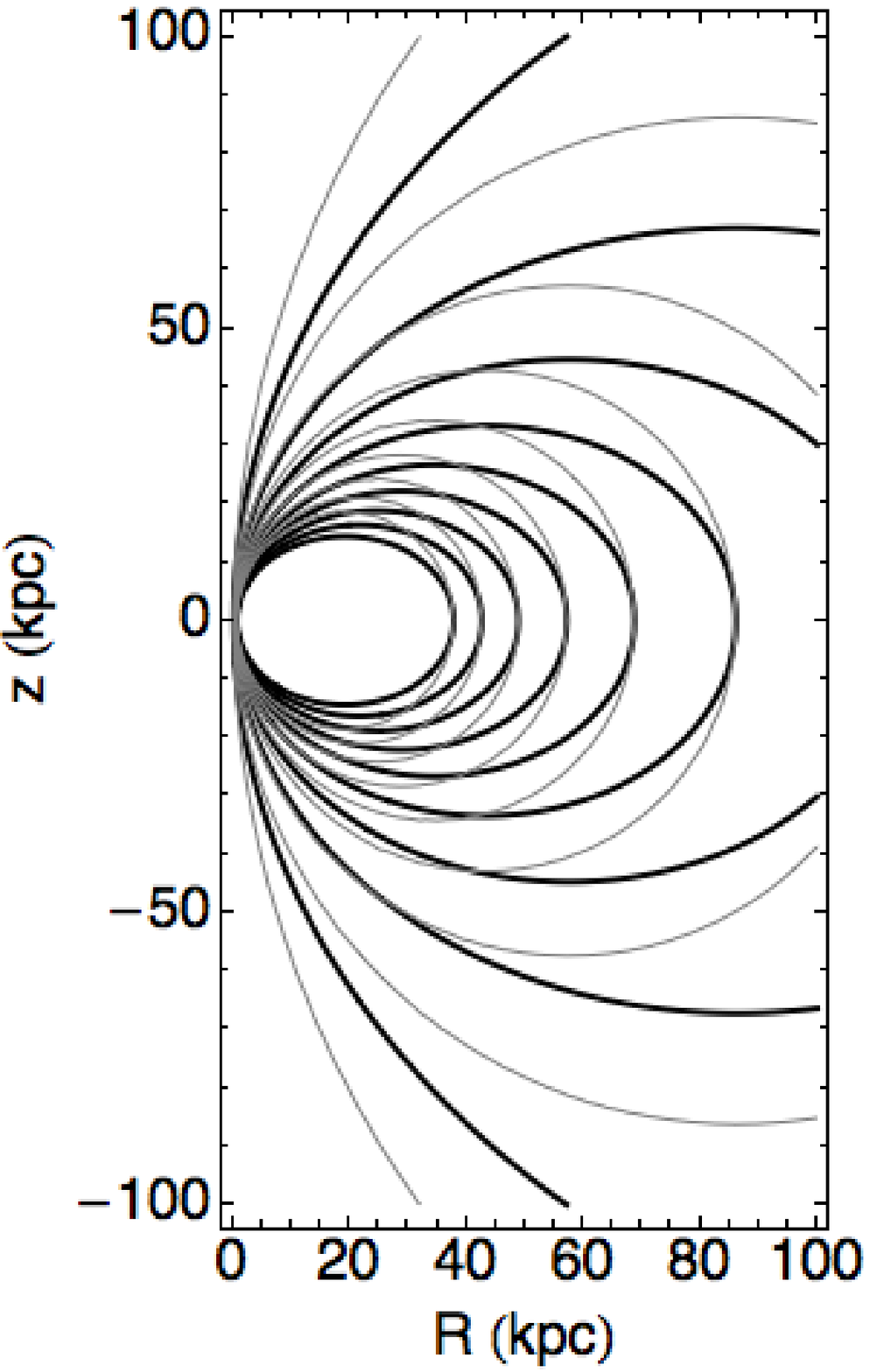} & \includegraphics[width=0.3\textwidth]{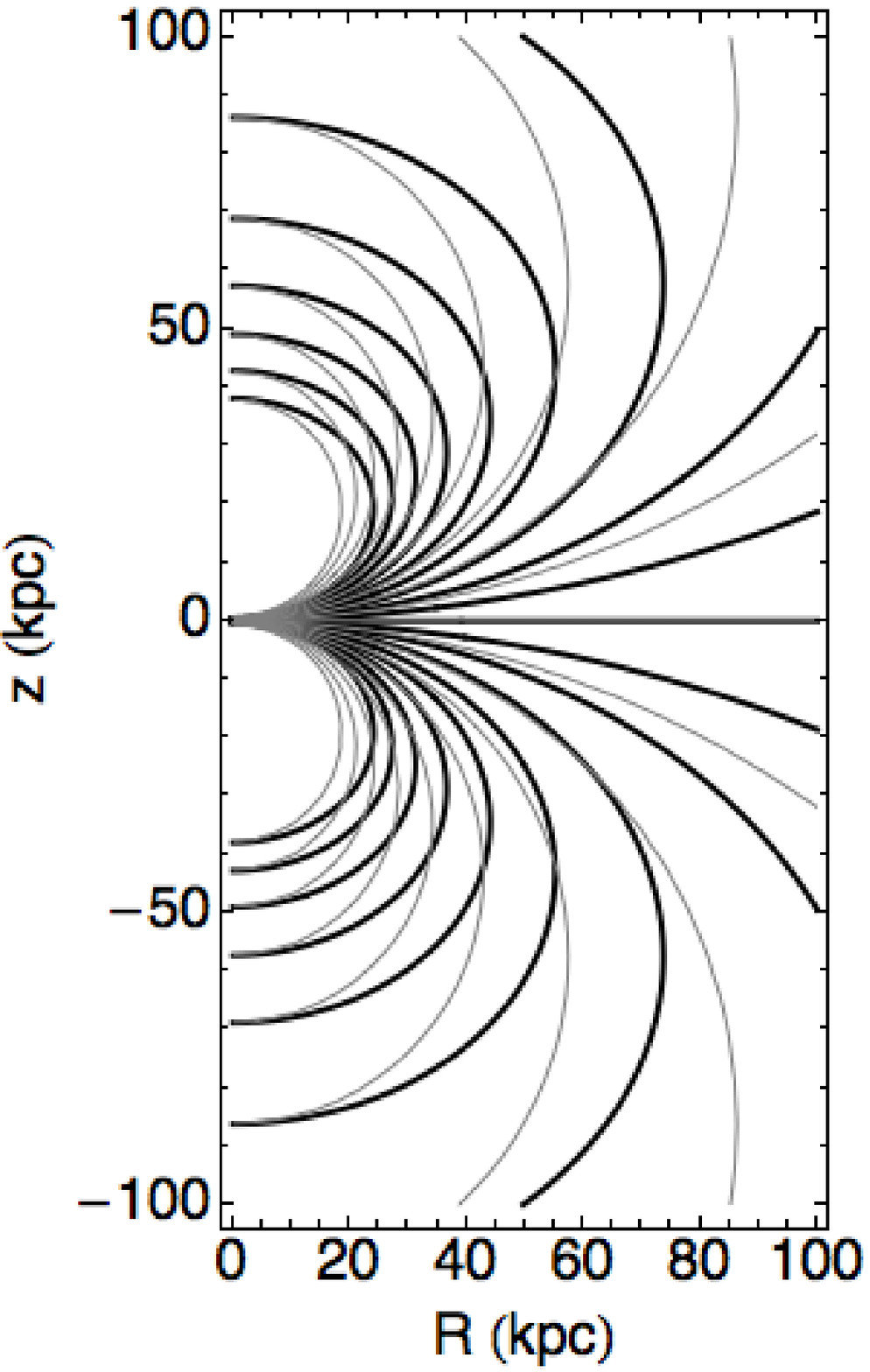}
\end{tabular}
\caption{Contours of the potential (left), $R$ force (center), and $z$ force (right) for the flattened logarithmic potential used in this work. Black lines show $q=0.78$ while grey lines show $q=1$ for comparison. In the left panel, contour spacing is $5\times 10^4$ $(\textrm{km } \unit{s}{-1})^2$; in the center and right panels it is 0.5 km \unit{s}{-1} \unit{Myr}{-1}. In the rightmost panel the contours are negative above the plane and positive below it.The contours in each panel have the same range for both $q$ values.}
\label{fig:FlattenedContours}
\end{figure*}

Series C uses a flattened axisymmetric cored logarithmic potential,
\begin{equation}
\super{\sub{\Phi}{log}}{axi}(R,z) = \half v_c^2 \ln\left(r_0^2 + R^2 + \frac{z^2}{q^2}\right) + \Phi_0
\end{equation}
where $q$ is the flattening parameter, $v_c$ is the asymptotic circular velocity, $r_0$ is the core radius and $\Phi_0$ is chosen so that the potential is negative for all $R<1000$ kpc in the plane. In the limit $q\to 1$ one obtains a spherical potential, whose rotation curve we fit to the spherical isochrone potential used in Simulations A and B in the region $0\leq r\leq100$ kpc (Figure \ref{fig:rotationCurveFit}). 
We take $q=0.78$, which produces the degree of flattening shown in the potential and force contour plots in Figure \ref{fig:FlattenedContours}. 

\begin{figure}
\includegraphics[width=0.45\textwidth]{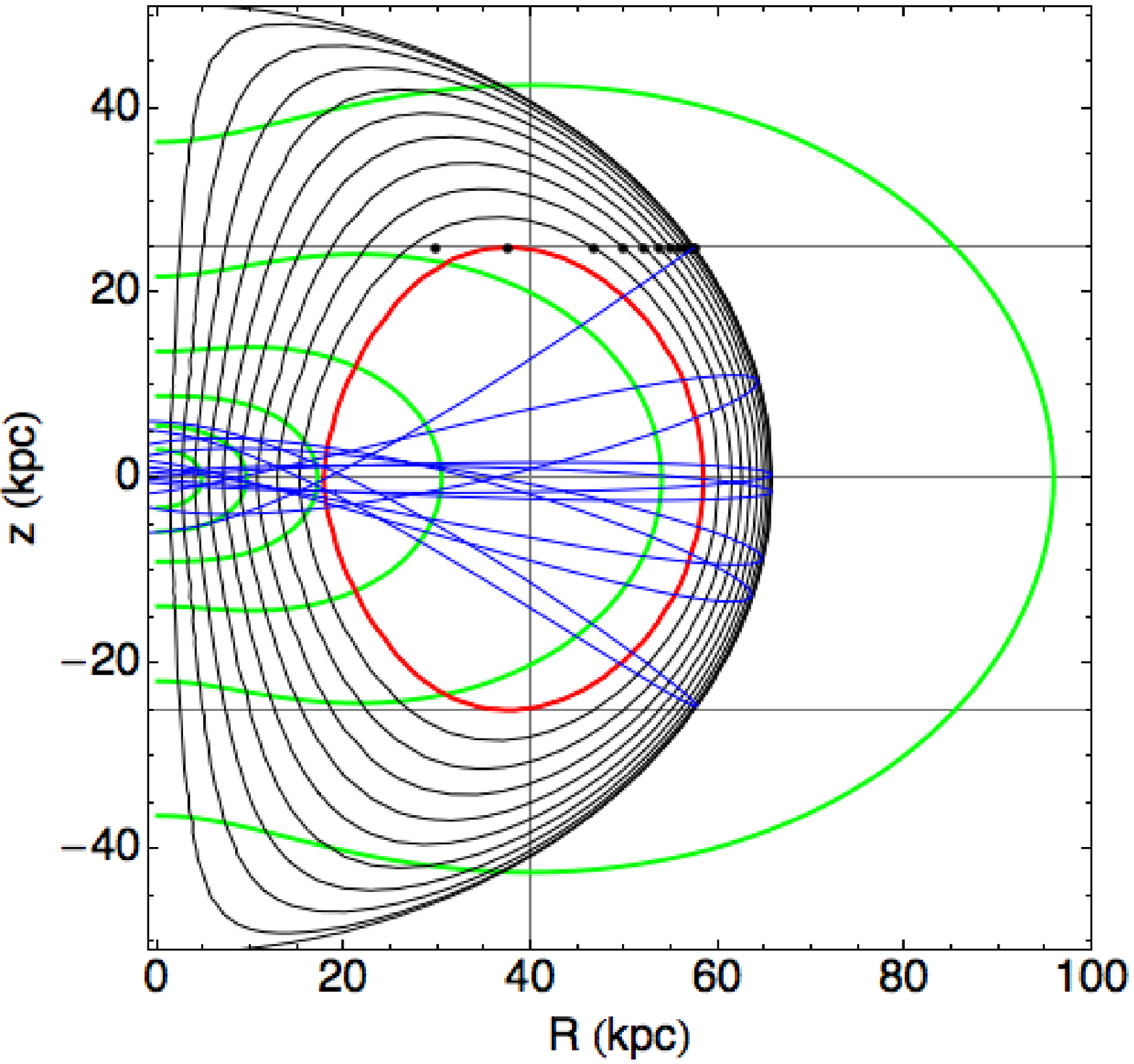}
\caption{Generation of initial conditions for the axisymmetric logarithmic potential. The satellite COM is placed on the zero-velocity curve corresponding to $L_z = \eta \sub{L}{circ}$ (black curves) for $\eta$ from zero up to $\sub{\eta}{max}$ (red curve) at $\sub{z}{max}=25$ kpc (black dots). The single black dot interior to the one at $\sub{\eta}{max}$ is the shell orbit with the same $\sub{z}{max}$. The green contours show the log of the density distribution; the COM orbit for $\eta=0$ is shown in blue as an example.}
\label{fig:logAxiICs}
\end{figure}

For each simulation in Series C, we place the center of mass of the satellite galaxy on the zero-velocity curve for an orbit with $\sub{z}{max}=25$ kpc and angular momentum $L_z = \eta\sub{L}{circ}(40\textrm{ kpc})$. The maximum value of $\eta$ for this $\sub{z}{max}$ is $\sub{\eta}{max}=0.71$ for this orbit family. We increase $\eta$ from 0 in steps of $\sub{\eta}{max}/10$, and for each value of $\eta$ the zero-velocity curve determines $\sub{R}{max}$ (Figure \ref{fig:logAxiICs}). The satellite COM is then placed at $(\mathbf{x}_0; \mathbf{v}_0)=(-\sub{R}{max}, 0, \sub{z}{max}; 0, \eta \sub{L}{circ}/\sub{R}{max}, 0)$ in Cartesian coordinates.

Simulation D uses the three-component potential described in Table \ref{tbl:ModelParams} and the references therein. The components include a Hernquist bulge with scale radius $r_b$ and total mass $M_b$:
\begin{equation}
\Phi(r) = -\frac{G M_b}{r_b + r},
\end{equation}
an exponential disk with surface density $\Sigma_d$, scale length $R_d$ and scale height $z_d$ (here we give the density distribution since the potential requires numerical integration):
\begin{equation}
\rho(R,z) = \frac{\Sigma_d}{2 z_d} \exp\left(-\frac{R}{R_d} - \frac{|z|}{z_d} \right),
\end{equation}
and a spherical NFW halo with density $\rho_h$ and scale radius $r_h$:
\begin{equation}
\Phi(r) = - 4\pi G \rho_h r_h^2 \left(\frac{r_h}{r}\right) \ln \left( \frac{r}{r_h} + 1 \right).
\end{equation}
In \citet{fardal:2007aa}, the halo density is given as a multiple of the critical density: $\rho_h = \delta_c \rho_c$.

\section{Derivation of the normalized cold phase space density}
\label{appx:phaseSpaceNorm}
Near a caustic, the phase-space distribution function (DF) can be approximated by a line, a quadratic equation relating the galactocentric radius $r$ and the radial velocity $v_r$:
\begin{equation}
 f(r, v_r) = \mathcal{F}_0 \delta\left[r_s - r - \kappa \left(v_r - v_s\right)^2\right].
\end{equation}
Here we have suppressed notation indicating that there is no variation in $v_\theta$ or $v_\phi$ (two trivial delta functions) and also notation limiting the angular extent of the shell to some total solid angle $\Omega_s$ of the sphere. We will deal with this second notion by adjusting the limits of integration in angle.
To determine the normalization $\mathcal{F}_0$ we impose the normalization condition
\begin{equation}
 \int_0^{r_s} r^2 dr \int_{\Omega_s} d\Omega \int_{-\infty}^{\infty} dv_r dv_\theta dv_\phi  f(r, v_r) = 1
\end{equation}
The integrals over the $v_\theta$ and $v_\phi$ delta functions are equal to unity, and the integral over the solid angle spanned by the shell, $\Omega_s$, simply contributes a prefactor. Thus this expression simplifies to
\begin{equation}
\mathcal{F}_0 \Omega_s \int_0^{r_s} r^2 dr \int_{-\infty}^{\infty} dv_r \delta\left[r_s - r - \kappa \left(v_r - v_s\right)^2\right] = 1.
\end{equation}
We now change variables to the argument of the delta function,
\begin{equation}
 u=r_s - r - \kappa \left(v_r - v_s\right)^2 \equiv \Delta_r -\kappa \left(v_r - v_s\right)^2 ,
\end{equation}
to integrate over $v_r$. Note that the quantity $\Delta_r\equiv r_s-r$ is always positive since the caustic only includes material at $r<r_s$. 

This change of variables has two solutions for $v_r$,
\begin{equation}
 v_{r,\pm} = v_s \pm \sqrt{\frac{\Delta_r - u}{\kappa}}
\end{equation}
To understand how to do the integral we therefore make a preliminary auxiliary change to the variable $x = v_r-v_s$, so that the integrand is symmetric in $x$:
\begin{equation}
 I_{v_r} = \int_{-\infty}^{\infty} dx \delta(\Delta_r - \kappa x^2).
\end{equation}
In terms of u, 
\begin{equation}
 x_\pm = \pm \sqrt{\frac{\Delta_r - u}{\kappa}} \qquad \textrm{and} \qquad u = \Delta_r - \kappa x^2
\end{equation}
Now we break the integral up into positive and negative $x$ pieces, so that it is clear which branch of the solution to take on each piece:
\begin{equation}
  I_{v_r} = \int_{-\infty}^0 dx\ \delta(\Delta_r - \kappa x_-^2) + \int_0^{\infty} dx\ \delta(\Delta_r - \kappa x_+^2) 
\end{equation}
 We note from the change of variables that at $x=\pm \infty$, $u=-\infty$, while at $x=0$, $u$ takes on its maximum (positive) value of $\Delta_r$. So the zero point will still be included in both pieces of the integral, allowing the delta functions to be evaluated. To complete the change of variables we need the Jacobian $dx/du$:
\begin{equation}
 \frac{dx_\pm}{du} = \mp \frac{1}{2 \sqrt{\kappa(\Delta_r-u)}}
\end{equation}
so that
\begin{eqnarray}
  I_{v_r} &=& \int_{-\infty}^{\Delta_r} \frac{du}{2 \sqrt{\kappa(\Delta_r-u)}} \ \delta(u) - \int_{\Delta_r}^{-\infty}\frac{du}{2 \sqrt{\kappa(\Delta_r-u)}} \ \delta(u) \nonumber \\
&=&\int_{-\infty}^{\Delta_r} \frac{du}{\sqrt{\kappa(\Delta_r-u)}} \ \delta(u).
\end{eqnarray}
Now we can evaluate the integral directly thanks to the delta function:
\begin{equation}
 I_{v_r} = \frac{1}{\sqrt{\kappa(r_s-r)}} .
\end{equation}
Note that this integral gives the functional form of the density distribution, which has the characteristic $1/\sqrt{r}$ falloff behind the caustic. This shows that we have chosen an appropriate form for the DF, since it gives the right density distribution when integrated over velocity.

Replacing the evaluated integral in the normalization condition gives
\begin{equation}
 \mathcal{F}_0 \Omega_s \int_0^{r_s} \frac{r^2 dr}{\sqrt{\kappa(r_s-r)}} = 1.
\end{equation}
The remaining integral can be evaluated quite easily to give the definition of $\mathcal{F}_0$:
\begin{equation}
\frac{ 16 \mathcal{F}_0 \Omega_s r_s^{5/2}}{15 \sqrt{\kappa}} = 1, \qquad \textrm{or} \qquad \mathcal{F}_0 = \frac{15\sqrt{\kappa}}{16 r_s^{5/2} \Omega_s}.
\end{equation}
Thus, the complete expression for the DF is
\begin{equation}
  f(r, v_r) = \frac{15\sqrt{\kappa}}{16 r_s^{5/2} \Omega_s}\delta\left[r_s - r - \kappa \left(v_r - v_s\right)^2\right],
\end{equation}
where we have suppressed the trivial delta functions on $v_\theta$ and $v_\phi$.

\section{Derivation of the phase-space density; verification of the density distribution}
\label{appx:phaseSpaceGaussian}
The phase-space density is obtained by averaging over an ensemble of cold distributions, each with a slightly different caustic radius:
\begin{equation}
 f(r,v_r) \propto \int_{-\infty}^{\infty}  \exp\left(-\frac{x^2}{2 \delta_r^2}\right) \delta\left[r_s + x - r - \kappa \left(v_r - v_s\right)^2\right] x \equiv \mathcal{I}.
\end{equation}
To perform the integral we change variables to the quantity inside the delta function,
\begin{equation}
u = r_s + x - r - \kappa (v_r-v_s)^2, 
\end{equation}
 which has the same endpoints and a Jacobian of 1 ($du = dx$). Then the integral becomes trivial:
\begin{equation}
 \mathcal{I} = \int_{-\infty}^{\infty} \exp\left\{-\frac{\left[u-r_s+r+\kappa(v_r-v_s)^2\right]^2}{2 \delta_r^2}\right\} \delta(u) du,
\end{equation}
so that
\begin{equation}
  \mathcal{I} = \exp\left\{-\frac{\left[r_s-r-\kappa(v_r-v_s)^2\right]^2}{2 \delta_r^2}\right\}.
\end{equation}

We can verify that this expression for the phase density gives the same functional form as Equation \eqref{eq:generalDP} for the density $\rho$ by integrating over $v_r$, although the normalization will be different since the phase density is normalized to 1. The density will be proportional to the integral of $\mathcal{I}$ over $v_r$:
\begin{equation}
 \rho \propto \int_{-\infty}^{\infty}  \exp\left\{-\frac{\left[r_s-r-\kappa(v_r-v_s)^2\right]^2}{2 \delta_r^2}\right\} dv_r.
\end{equation}
We change variables to make the exponent Gaussian in the variable of integration,
\begin{equation}
 u = \frac{r_s-r-\kappa(v_r-v_s)^2}{\sqrt{2} \delta_r},
\end{equation}
and determine the Jacobian:
\begin{equation}
 dv_r = \mp \frac{\delta_r}{\sqrt{2\kappa (r_s - r - \sqrt{2} \delta_r\ u)}},
\end{equation}
where the upper sign is for the part of the integral where $v_r>v_s$ (the upper half of the parabola) and the lower sign is for the part when $v_r<v_s$ (the lower half). This implies that we must break the integral into two parts with the following endpoints in $u$, taking into account the sign of the Jacobian:
\begin{eqnarray}
 -\infty < v_r < v_s &\to& -\infty < u < \frac{r_s-r}{\sqrt{2}\delta_r} \nonumber \\
v_s \le v_r < \infty & \to & \frac{r_s-r}{\sqrt{2}\delta_r} \le u > -\infty \nonumber
\end{eqnarray}
So the two parts of the integral are identical; thus we can say
\begin{equation}
 \rho \propto \int_{-\infty}^{(r_s-r)/\sqrt{2}\delta_r} \frac{e^{-u^2} du}{\sqrt{r_s - r - \sqrt{2} \delta_r\ u}}.
\end{equation}
This integral can be evaluated as a combination of Bessel functions depending on the sign of $r_s-r$. Behind the caustic, this quantity is positive; in front it is negative, so we get a piecewise solution with the same domains as Equation \eqref{eq:generalDP}. Performing the integral gives the result:
\begin{equation}
  \rho \propto \sqrt{|r_s-r|} e^{-\frac{(r_s-r)^2}{4 \delta_r^2}} \mathcal{B}\left[\frac{(r_s-r)^2}{4 \delta_r^2}\right],
\end{equation}
which is the same functional form as Equation \eqref{eq:generalDP}.

\section{Equations for a rotated cone}
\label{appx:rotatedCone}

A cone of height $h$ and opening angle $\alpha$, with its point at the origin and its axis of symmetry along the $z$ axis has the parametric equations (in Cartesian coordinates)
\begin{eqnarray}
x &=& (h-u) \cos \vartheta \tan \alpha \\
y &=& (h-u) \sin \vartheta \tan \alpha \\
z &=& h-u
\end{eqnarray}
where the parameter $ u \in [0,h] $ describes the distance from the base of the cone and the parameter $\vartheta \in [0,2\pi]$ describes the azimuthal location on the cone.  If the cone is rotated so that its axis of symmetry points along the unit vector 
\begin{equation}
\hat{n} = \sin \theta_s \cos \phi_s \hat{x} + \sin \theta_s \sin \phi_s \hat{y} + \cos \theta_s \hat{z}
\end{equation}
then the equations become
\begin{eqnarray}
\label{eqs:rotatedConeX}
x &=& (h-u) \left\{  \cos \vartheta \left[ 1 - \cos^2 \phi_s \left(1-\cos \theta_s\right) \right] \tan \alpha \right.  \nonumber \\
&& -  \sin \vartheta \left[ \cos \phi_s \sin \phi_s \tan \alpha \left(1-\cos \theta_s\right)\right] \nonumber \\
&& + \left. \sin \theta_s \cos \phi_s  \right\}\\
\label{eqs:rotatedConeY}
y &=& (h-u) \left\{ - \cos \vartheta \left[ \cos \phi_s \sin \phi_s \tan \alpha \left(1 - \cos \theta_s\right)\right] \right. \nonumber \\
&& + \sin \vartheta \left[ 1 - \sin^2 \phi_s \left(1 - \cos \theta_s\right) \right] \tan \alpha \nonumber \\
&& + \left. \sin \theta_s \cos \phi_s \right \} \\
z &=& (h - u) \left[ - \cos\left(\vartheta - \phi_s\right) \sin \theta_s \tan \alpha + \cos \theta_s \right]
\label{eqs:rotatedConeZ}
\end{eqnarray}
To determine \zmin\ and \zmax\ as a function of $x$ and $y$, we solve the system of the $x$ and $y$ parametric equations to obtain the parameters $u$ and $\vartheta$.  The system can have zero, one, or two solutions depending upon the values of $x$ and $y$: zero solutions for lines of sight that do not intersect the cone, one solution for lines of sight that intersect the cone once, and two solutions for lines of sight that intersect the cone twice.  In cases where there is one solution, the other limit can safely be taken to be $\pm \infty$, where the plus sign is taken if the bulk of the cone is in front of the intersection point (the solution is the lower bound of $z$), and the minus sign is taken if the cone is behind the intersection (the solution is the upper bound of $z$).  The system can always be solved analytically and we present the solution here.  First we will define some auxiliary quantities to make the notation simpler.  We extract the $\theta_s$, $\phi_s$, and $\alpha$ dependence of Equations (\ref{eqs:rotatedConeX}-\ref{eqs:rotatedConeY}) into coefficients that need only be calculated once for a given cone:
\begin{eqnarray}
A_x &=& \tan \alpha \left[1- \cos^2 \phi_s \left(1-\cos \theta_s\right)\right] \\
B_x &=& \tan \alpha \cos \phi_s \sin \phi_s \left(1 - \cos \theta_s\right) \\
C_x &=& \sin \theta_s \cos \phi_s \\
A_y &=& B_x \\
B_y &=& \tan \alpha \left[1- \sin^2 \phi_s \left(1-\cos \theta_s\right)\right] \\
C_y &=& \sin \theta_s \sin \phi_s, 
\end{eqnarray}
so that Equations (\ref{eqs:rotatedConeX}-\ref{eqs:rotatedConeY}) become
\begin{eqnarray}
x &=& (h-u) \left(A_x \cos \vartheta - B_x \sin \vartheta + C_x \right) \\
y &=& (h-u) \left(-A_y \cos \vartheta + B_y \sin \vartheta + C_y \right).
\end{eqnarray}
We can solve for $\vartheta$ by dividing the two equations, noting that if $x=0$ or $y=0$ then we are at the point of the cone ($u = h$) and $\vartheta$ is degenerate.  We obtain an equation for $\vartheta$ in terms of the ratio $\eta \equiv x/y$:
\begin{equation}
\label{eq:varthetaRoots}
\mathcal{S} \sin \vartheta - \mathcal{C} \cos \vartheta + \mathcal{K} = 0,
\end{equation}
where 
\begin{eqnarray}
\mathcal{C} &\equiv& A_x + \eta A_y\\
\mathcal{S} &\equiv& B_x + \eta B_y\\
\mathcal{K} &\equiv& -C_x + \eta C_y.
\end{eqnarray}

Equation \eqref{eq:varthetaRoots} can be recast as a quadratic equation for either $\sin \vartheta$ or $\cos \vartheta$.  To avoid using inverse trigonometric functions (and the associated difficulties in choosing the right branch) we simply solve for both and use them in the rest of the solution:
\begin{eqnarray}
\left(\cos \vartheta\right)_{\pm} &=& \frac{\mathcal{K} \mathcal{C} \pm \mathcal{S} \mathcal{D}}{\mathcal{C}^2 + \mathcal{S}^2} \\
\left(\sin \vartheta\right)_{\pm} &=& \frac{\mathcal{K} \mathcal{S} \pm \mathcal{C} \mathcal{D}}{\mathcal{C}^2 + \mathcal{S}^2}
\end{eqnarray}
where $\mathcal{D}^2$ is the discriminant
\begin{equation}
\mathcal{D}^2 \equiv \mathcal{C}^2 + \mathcal{S}^2 - \mathcal{K}^2.
\end{equation}
As usual, if $\mathcal{D}^2 < 0$ the point is outside the cone and the equation has no real roots; otherwise it has two real roots.  When two solutions exist for $\vartheta$, one or both of them may lead to a value of $u$ outside its allowed range.

Having determined $\vartheta$ either Equation \eqref{eqs:rotatedConeX} or Equation \eqref{eqs:rotatedConeY} can be used to determine the quantity $h-u$ that is necessary to find the limits on $z$:
\begin{eqnarray}
\left(h-u\right)_{\pm} &=& \frac{x}{A_x \left(\cos \vartheta\right)_{\pm}  - B_x \left(\sin \vartheta\right)_{\pm} + C_x} \nonumber \\
&=& \frac{y}{-A_y \left(\cos \vartheta\right)_{\pm} + B_y \left(\sin \vartheta\right)_{\pm} + C_y}
\end{eqnarray}
The quantity $h-u$ should be in the range $(0,h]$---if it is not, that value of $\vartheta$ is discarded as a root and one of the limits in $z$ goes to $\pm \infty$. 

Finally, the limits $z_{\pm}$ are given by plugging in the valid solutions for $u$ and $\vartheta$:
\begin{eqnarray}
z_{\pm} &=& \left(h-u\right)_{\pm} \left\{ \cos \theta_s - \sin \theta_s \tan \alpha \left[ \left(\cos \vartheta\right)_{\pm} \cos \phi_s  \nonumber \right. \right. \\
&&\left. \left. + \left(\sin \vartheta\right)_{\pm} \sin \phi_s \right] \right\}
\end{eqnarray}
By inspection, we see that \zmin\ is not always equal to $z_-$ and \zmax\ is not always $z_+$ since both $\left(\cos \vartheta\right)_{\pm}$ and $\left(\sin \vartheta\right)_{\pm}$ can take any sign: the roots must be compared and the smaller assigned to \zmin.  If one root is out of range in $u$, its out-of-range $z$ value can still be calculated (for this work the height $h$ of the cone is arbitrary because the density function effectively cuts off the cone along a spherical segment) and compared to the value of the in-range root to determine whether the out-of-range limit is \zmin\ or \zmax.

In constructing a fitting routine, the partial derivatives of $\Sigma$ with respect to the projection parameters $\alpha, \theta_s, \phi_s$ and to the density profile parameters $r_s, \delta_r, \kappa, f_0$ may be needed.  In the following, we define the shorthand
\begin{eqnarray}
\cosdiffmax & \equiv & \cosmax \cos \phi_s + \sinmax \sin \phi_s \nonumber \\
\sindiffmax & \equiv & \sinmax \cos \phi_s - \cosmax \sin \phi_s \nonumber 
\end{eqnarray}
where $\cosmax$ and $\sinmax$ are taken to be the roots that lead to the value of $\zmax$, and likewise
\begin{eqnarray}
\cosdiffmin & \equiv & \cosmin \cos \phi_s + \sinmin \sin \phi_s \nonumber \\
\sindiffmin & \equiv & \sinmin \cos \phi_s - \cosmin \sin \phi_s \nonumber 
\end{eqnarray}
for the roots leading to the value of $\zmin$.  A similar notation is used to denote the appropriate root of $(h-u)$.  We also use the shorthand 
\begin{equation}
\rho(\zmax) \equiv \rho(\sqrt{x^2 + y^2 + \zmax^2})
\end{equation}
in the following, since $x$ and $y$ are understood to be constant.  With these definitions, the derivatives with respect to the projection parameters are:
\begin{eqnarray}
\frac{\partial \Sigma}{\partial \alpha} &=& \frac{\sin \theta_s \sec^2 \alpha}{\Upsilon} \left\{ \rho(\zmin) \cosdiffmin \hmumin \right. \nonumber \\
&&    \left. - \rho(\zmax) \cosdiffmax \hmumax \right\} \\
\frac{\partial \Sigma}{\partial \theta_s} &=& \frac{1}{\Upsilon} \left\{ \rho(\zmin) \left[ \hmumin \sin \theta_s   \right.\right. \nonumber \\
&&\qquad \qquad  \left. + \cosdiffmin \cos \theta_s \tan \alpha \right]  \nonumber \\
&&  -\rho(\zmax) \left[ \hmumax \sin \theta_s \right. \nonumber \\
&& \qquad \qquad  + \left. \left. \cosdiffmax \cos \theta_s \tan \alpha \right] \right\} \\
\frac{\partial \Sigma}{\partial \phi_s} &=& \frac{\sin \theta_s \tan \alpha}{\Upsilon} \left\{ \rho(\zmin) \hmumin \sindiffmin \right. \nonumber \\
&& \left. - \rho(\zmax) \hmumax \sindiffmax \right\}
\end{eqnarray}

The derivatives with respect to the profile parameters are all integrals of the derivatives of the density, of the form
\begin{equation}
\frac{\partial \Sigma}{\partial \pi_i} = \frac{1}{\Upsilon} \int_{\zmin}^{\zmax} \frac{\partial \rho}{\partial \pi_i} dz
\end{equation}
for a given parameter $\pi_i$, since the limits of the integral and the variable being integrated over do not depend on any of the $\pi_i$ and the function $\rho$, although defined as a piecewise function in Equation \eqref{eq:generalDP}, is continuous over the entire integration range.  To compactly write derivatives of $\rho$ with respect to the parameters, we expand the definition of $\mathcal{B}$ to include other Bessel functions:
\begin{equation}
\mathcal{B}_{n}(u) \equiv \left\{ \begin{array}{cc}
\frac{\pi}{2} \left[ \mathcal{I}_{(2n + 1)/4}(u) +  \mathcal{I}_{-(2n + 1)/4}(u) \right]  & r \le r_s  \\
 (-1)^{n+1} \frac{\pi}{2}\left[ \mathcal{I}_{(2n + 1)/4}(u) -  \mathcal{I}_{-(2n + 1)/4}(u) \right]  &  r > r_s 
\end{array}
\right.
\end{equation}
with $u \equiv (r-r_s)^2/4\delta_r^2$ as before.  The definition used in Equation \eqref{eq:generalDP} is equivalent to $\mathcal{B}_0$ in this new notation.  With this simplification, the derivatives $\partial \rho/\partial \pi_i$ are:
\begin{eqnarray}
\frac{\partial \rho}{\partial r_s} &=& \frac{f_0}{\sqrt{2 \pi \kappa}} e^{-u} \frac{\sqrt{|r-r_s|}}{r-r_s} \times \nonumber \\
&& \times \left\{  u\left[ 2 \mathcal{B}_0 (u) - \mathcal{B}_2(u) - \mathcal{B}_1(u) \right] -  \half \mathcal{B}_0 (u)\right\} \\
\frac{\partial \rho}{\partial \delta_r} &=& \frac{f_0}{\sqrt{2 \pi \kappa}} e^{-u}\sqrt{|r-r_s|} \frac{u}{\delta_r} \times \nonumber \\
&& \qquad \times \left[2 \mathcal{B}_0 (u) -  \mathcal{B}_2(u) - \mathcal{B}_1(u) \right] \\
\frac{\partial \rho}{\partial f_0} &=& \frac{\rho}{f_0}\\
\frac{\partial \rho}{\partial \kappa} &=& -\frac{\rho}{2\kappa}
\end{eqnarray}

\end{document}